\PassOptionsToPackage{hyphens}{url}
\documentclass[format=acmsmall, review=false, screen=true, dvipsnames]{acmart_arXiv}

\acmJournal{POMACS}
\acmYear{2018}
\acmMonth{0}
\acmPrice{15.00}
\copyrightyear{2018}
\setcopyright{acmcopyright}
\acmDOI{}

\usepackage[ruled]{algorithm2e} 

\SetAlFnt{\small}
\SetAlCapFnt{\small}
\SetAlCapNameFnt{\small}
\SetAlCapHSkip{0pt}
\IncMargin{-\parindent}

\usepackage{etoolbox}
\usepackage{datetime}
\usepackage{enumitem}
\usepackage{booktabs}
\usepackage{multirow}
\usepackage{color}
\usepackage{array}
\usepackage{textcomp}
\usepackage[counterclockwise, figuresright]{rotating}
\usepackage{subcaption}
\usepackage{setspace}
\usepackage{colortbl}

\usepackage[binary-units=true,parse-numbers=false,per-mode=symbol, range-units=single, range-phrase={--}]{siunitx}

\usepackage{dblfloatfix}
\newcolumntype{L}[1]{>{\raggedright\let\newline\\\arraybackslash\hspace{0pt}}m{#1}}
\newcolumntype{C}[1]{>{\centering\let\newline\\\arraybackslash\hspace{0pt}}m{#1}}
\newcolumntype{R}[1]{>{\raggedleft\let\newline\\\arraybackslash\hspace{0pt}}m{#1}}

\renewcommand\footnotetextcopyrightpermission[1]{}

\newcommand{\todo}[1][]{}
\newcommand{\chI}[0]{}
\newcommand{\chII}[0]{}
\newcommand{\chIII}[0]{}
\newcommand{\chIV}[0]{}
\newcommand{\chV}[0]{}
\newcommand{\chVI}[0]{}
\newcommand{\chVII}[0]{}
\newcommand{\chVIII}[0]{}
\newcommand{\chIX}[0]{}
\newcommand{\chX}[0]{}
\newcommand{\chXI}[0]{}
\newcommand{\chXII}[0]{}

\newcommand{\sg}[0]{}
\newcommand{\sph}[0]{}

\newcommand{\figscale}{.7}

\begin{document}
\sloppy

\title[Tolerating Early Retention Loss and Process Variation in 3D NAND Flash Memory]{Improving 3D NAND Flash Memory Lifetime \\
by Tolerating Early Retention Loss and Process Variation}

\author{Yixin Luo}
\affiliation{%
  \institution{Carnegie Mellon University}
}
\email{yixinluo@cs.cmu.edu}

\author{Saugata Ghose}
\affiliation{%
  \institution{Carnegie Mellon University}
}
\email{ghose@cmu.edu}

\author{Yu Cai}
\affiliation{%
  \institution{Carnegie Mellon University}
}
\email{yucaicai@gmail.com}

\author{Erich F. Haratsch}
\affiliation{%
  \institution{Seagate Technology}
}
\email{erich.haratsch@seagate.com}

\author{Onur Mutlu}
\affiliation{%
  \institution{ETH Z{\"u}rich \& Carnegie Mellon University}
}
\email{omutlu@ethz.ch}

\renewcommand{\shortauthors}{Y. Luo et al.}

\authorsaddresses{}




\begin{abstract}

Compared to planar \chX{(i.e., two-dimensional)} NAND flash memory, 3D NAND flash memory uses a new
flash cell design, and vertically stacks dozens of silicon layers in a
single chip. This allows 3D NAND flash memory to increase storage density
using a much less aggressive manufacturing process technology than planar 
\chX{NAND flash memory}.
The circuit-level and structural changes in 3D NAND flash memory
significantly alter how different error sources affect the reliability
of the memory.

\chVI{In this paper,} through experimental characterization of real, state-of-the-art 3D NAND flash
memory chips,
we find that 3D NAND flash memory exhibits \emph{three} new error sources
that were not previously observed in planar NAND \sg{flash memory}:
(1)~\emph{layer-to-layer process variation}, 
\sg{a new phenomenon specific to the 3D nature of the device,
where the average error rate of each 3D-stacked layer in a chip is
significantly different;}
(2)~\emph{early retention loss}, \sg{a new phenomenon where the number of
errors due to charge leakage increases \emph{quickly within several hours}
after programming;} and
(3)~\emph{retention interference}, \sg{a new phenomenon where the rate 
at which charge leaks from a flash cell is dependent}
on the \sg{data} value stored in the neighboring cell.

Based on our experimental results, we develop new analytical models \chVI{of}
layer-to-layer process variation and retention loss 
in 3D NAND flash memory. 
Motivated by our new findings and models, we develop four new techniques to
mitigate process variation and early retention loss in 3D NAND flash memory.
Our first technique,
\chVI{\underline{La}yer \underline{V}ariation \underline{A}ware \underline{R}eading (LaVAR)},
reduces the effect of layer-to-layer process variation by \chVI{fine-}tuning
the read reference voltage \chVII{separately} for each layer.
Our second technique, \chVI{\chX{\underline{L}ayer-\underline{I}nterleaved}
\underline{R}edundant \underline{A}rray of \underline{I}ndependent \underline{D}isks
(LI-RAID)}, \sg{uses information about layer-to-layer
process variation to \chVII{intelligently group pages} under the RAID 
error recovery technique \chVII{in a manner} that \chVII{reduces} the likelihood that the
recovery of a group \chVIII{fails} significantly \chVII{earlier than} the recovery of other groups.}
\chVI{Our third technique, \underline{Re}tention \underline{M}odel
\underline{A}ware \underline{R}eading (ReMAR), reduces retention errors in 3D
NAND flash memory by tracking the retention time of the data using our \chVII{new} retention
model and adapting the read reference voltage to data age. Our
fourth technique, \underline{Re}tention Interference Aware
\underline{N}eighbor-Cell \underline{A}ssisted \underline{C}orrection (ReNAC),
adapts the read reference voltage to the
amount of retention interference \chVII{a page has experienced, in order} to re-read the data after a
read operation fails.
These four techniques are complementary, and can be combined together
to significantly improve flash memory reliability.  Compared to a
state-of-the-art
baseline, our techniques, when combined, improve flash memory lifetime
by 1.85$\times$.  Alternatively, if a NAND flash
\chX{vendor} wants to keep the lifetime of the 3D NAND flash memory device
constant, our techniques reduce the storage overhead required to hold error
correction information by 78.9\%.}

\end{abstract}

\maketitle


\section{Introduction}
\label{sec:introduction}

Solid-state drives (SSDs), which consist of NAND flash memory chips, are a
popular data storage medium in modern computer systems.
Traditionally,
NAND flash memory has employed a \emph{planar} (i.e., two-dimensional)
architecture, where the entire chip resides on a single layer of silicon. In
planar NAND flash memory, a flash cell is made using a
\emph{\chI{floating-gate} transistor}, where data is represented by the amount of
charge stored in the transistor's floating gate.  The amount of charge stored
in \sg{the floating gate} determines the \emph{threshold voltage} of the 
\sg{flash cell} transistor
(i.e., the voltage at which the transistor turns on).

For planar NAND flash memory, to continually increase the SSD capacity and decrease
the cost-per-bit of the SSD, flash vendors have \sg{been aggressively scaling}
NAND flash memory to smaller manufacturing process \chX{technology nodes}. This,
however, comes at the cost of \chVI{lower}
reliability~\cite{mielke.irps08, cai.date12, cai.procieee17}. 
Due to a
combination of manufacturing process \chVI{technology} limitations and \chVI{reduced}
reliability \sph{of planar NAND flash memory}, it has become increasingly difficult for \chX{vendors} to
continue to scale \sph{\sg{the density of planar NAND flash memory 
chips}}~\chVI{\cite{park.jssc15, cai.book18, grossi.bookchapter16}}.

To overcome this scaling challenge, 3D NAND flash memory
has recently been introduced~\cite{park.jssc15, kang.isscc16,
im.isscc15}. 
\chI{Although}
\chX{3D NAND flash memory is already being deployed \chVII{at large} scale in new computer
systems, there is a lack of available knowledge on the error
characteristics of real 3D NAND flash \sg{memory} chips, which makes it harder to estimate
the reliability characteristics of systems that employ such chips.}
Previous publicly-available experimental
studies on NAND flash memory errors using real flash memory chips
(e.g.,~\chVII{\cite{mielke.irps08, cai.date12, cai.date13, cai.iccd12, cai.itj13,
cai.iccd13, cai.sigmetrics14, cai.hpca15, cai.dsn15, cai.hpca17,
parnell.globecom14, luo.jsac16, cai.procieee17, cai.book18}}) have \chVI{mostly} been on planar NAND \sg{flash
memory} devices.\footnote{\chVI{With the exception of our very recent prior
work~\cite{luo.hpca18}, which examined two specific \chVII{important} aspects of 3D NAND flash
memory reliability: temperature and self-recovery effects.}}

We identify that 3D NAND flash memory has \emph{three} fundamental
differences from the most recent generation (i.e., \chI{\SIrange{10}{15}{\nano\meter}}) of
planar NAND flash memory, which lead to \chVI{new} error
characteristics \chVI{for 3D NAND flash memory} that we observe experimentally\chVI{:}
\chI{(1)}~3D NAND flash memory
currently uses a different flash cell architecture than planar NAND flash
\sg{memory}. 
Instead of using a floating-gate transistor, a cell in 3D NAND flash memory
consists of a \emph{charge trap transistor}~\cite{samsung.whitepaper14}, which
stores charge within an insulator. \chI{(2)}~Unlike planar NAND flash
memory, 3D NAND flash memory vertically stacks
\emph{multiple layers} of silicon together within a single chip.
Modern 3D NAND flash \sg{memory} chips typically contain \numrange{24}{96}
stack layers~\cite{techinsights.tr16, park.jssc15, kang.isscc16, kim.jssc18,
anandtech.web17,
im.isscc15}. Due to the high layer count, 3D NAND flash memory can
\chVI{provide high}
storage density without needing to scale the process technology as
aggressively as was done for planar NAND flash \sg{memory}. (3) While modern planar NAND flash \sg{memory}
uses a manufacturing process \chX{technology node} as small as
\chI{\SIrange{10}{15}{\nano\meter}}~\cite{techinsights.tr16, lee.isscc16}, 3D NAND
flash memory currently uses a much larger manufacturing process \chVI
{technology node}
(e.g., \chI{\SIrange{30}{50}{\nano\meter}}~\cite{samsung.whitepaper14}).

\textbf{Our goal in this work} is to (1)~identify and
understand the \emph{new} error characteristics of 3D NAND flash memory
(i.e., those that did \chVI{\emph{not}} exist previously in planar NAND \sg{flash memory}),
and (2)~develop new techniques to mitigate prevailing 3D
NAND flash \sg{memory} errors. We aim to achieve these goals via rigorous experimental
characterization of real, state-of-the-art 3D NAND flash memory chips
\chI{from a major flash vendor}.
Based on our comprehensive characterization and
analysis, we identify \emph{three new error characteristics} that were \chVI{\emph{not}}
previously observed in planar NAND flash memory, but are fundamental to the new
architecture of 3D NAND \sg{flash memory}:

\begin{enumerate}[leftmargin=15pt]

\item 3D NAND flash \chIII{memory} exhibits \emph{layer-to-layer process
variation}, a new phenomenon specific to the 3D nature of the device, where the
average error rate of each 3D-stacked layer in a
chip is significantly different \chVI{from one another} (Section~\ref{sec:variation}).
\chII{We are the \emph{first} to provide detailed experimental characterization
results of layer-to-layer process variation in real flash devices in open literature.}
\chII{Our results show that the raw bit error rate in the middle layer can be
6$\times$ \chIII{the \chVI{raw bit} error rate in} the top layer.}

\item 3D NAND flash memory experiences
\emph{early retention loss}, a new phenomenon where the number of
errors due to charge leakage increases \chIII{\emph{quickly within several hours}} after
programming, but then increases at a much slower rate (Section~\ref{sec:retention}).
\chI{We are the \emph{first} to perform an extended-duration observation
of early retention loss.  While \chVI{a prior study~\cite{choi.svlsi16}} examines the impact of early 
retention loss over only the first 5~minutes after data is written,
we examine the impact of early retention loss over the course of 24~days.}
\chII{Our results show that the retention error rate in \chIII{a 3D NAND flash memory
block} quickly increases by an order of magnitude \chIII{within $\sim$3 hours
after programming}.}

\item 3D NAND flash \chI{memory} experiences \emph{retention
interference}, a new
phenomenon where the rate at which charge leaks from a flash cell is
dependent on the amount of charge stored in neighboring flash cells (Section~\ref{sec:retention:interference}).
\chII{Our results show that charge leaks at a lower rate (i.e., the retention
loss speed is slower) when the \chX{\chVI{vertically}-\chXI{adjacent}} cell is
in a state that holds more charge (i.e., a higher-voltage state).}

\end{enumerate}

Our experimental observations indicate that we must revisit the error models
and the error mitigation mechanisms devised for planar NAND flash \chVI
{memory}, as they are
no longer accurate for 3D NAND flash \chVI{memory} behavior. To this end, we
develop \emph{new analytical model\chI{s}} 
of (1)~the layer-to-layer process variation in 3D NAND flash memory
(Section~\ref{sec:model:variation}), and 
(2)~retention loss in 3D NAND flash memory (Section~\ref{sec:model:retention}).
Our models estimate the
raw bit error rate (RBER), threshold voltage distribution, and the 
\emph{optimal read reference voltage} (i.e., the voltage at \chIV{which
\sg{the} RBER} is minimized when applied during a read operation) for each flash
\chI{page}. \chIII{Both models are useful for developing
techniques to mitigate raw bit errors in 3D NAND flash memory.}

We propose \emph{four new techniques} to
mitigate the unique layer-to-layer process variation and early retention loss
errors observed in 3D NAND flash memory. 
\chX{Each technique makes use of our new analytical models of
layer-to-layer process variation and retention loss in 3D NAND flash memory.}
Our first technique, \chVI{\underline{La}yer \underline{V}ariation
\underline{A}ware \underline{R}eading (LaVAR)}, reduces process variation by
fine-tuning the read reference voltage independently for each layer.
Our second technique, \chVI{\chX{\underline{L}ayer-\underline{I}nterleaved}
\underline{R}edundant \underline{A}rray of \underline{I}ndependent \underline{D}isks
(LI-RAID)}, \chIII{improves reliability by
changing how pages are grouped under \chIV{the} RAID
\chIV{error recovery technique}.  \chVI{LI-RAID uses}
information about
layer-to-layer process variation to reduce the likelihood that the RAID
recovery of a group could fail significantly earlier during the flash lifetime
than \sg{the} recovery of other groups.}
Our third technique, \underline{Re}tention \underline{M}odel
\underline{A}ware \underline{R}eading (ReMAR), reduces retention errors in 3D
NAND flash memory by tracking the retention time of the data using our retention
model and adapting the read reference voltage to data age. Our
fourth technique, \underline{Re}tention Interference Aware
\underline{N}eighbor-Cell \underline{A}ssisted \underline{C}orrection (ReNAC),
adapts the read reference voltage to the
amount of retention interference \chX{and re-reads} the data after a
read operation fails, \chVII{in order} \chVI{to correct the cells affected by retention
interference}.
These four techniques are complementary, and can be combined together
to significantly improve flash memory reliability.  Compared to a
state-of-the-art
baseline, our techniques, when combined, improve flash memory lifetime
by 1.85$\times$.  Alternatively, if a NAND flash
\chX{vendor} wants to keep the lifetime of the 3D NAND flash memory device
constant, our techniques reduce the storage overhead required to hold error
correction information by 78.9\%.

This paper makes the following \textbf{key contributions}:

\begin{itemize}[topsep=0pt,partopsep=0pt,noitemsep,leftmargin=10pt]

\item \chI{It presents \chVI{the first} \emph{comprehensive experimental
characterization}
\chX{of} real, state-of-the-art 3D NAND flash memory chips,
and \sg{provides} an \emph{in-depth analysis} of layer-to-layer process
variation, early retention loss, and retention interference, \chX{which are} \emph{three
new error characteristics} inherent to 3D NAND flash memory.}

\item It develops \emph{new analytical models} for (1)~layer-to-layer process
variation and (2)~early retention loss, \chVI{which can be used to estimate
the raw bit error rate, mean and standard deviation of the threshold voltage
distribution of each state, and the optimal read reference voltages}.

\item It develops \emph{four new mechanisms}, LaVAR, LI-RAID, ReMAR, and ReNAC,
to mitigate the three new error characteristics we have identified in 3D NAND
flash memory. \chI{It evaluates these techniques, and shows that, when applied
together, they improve 3D NAND flash memory lifetime by \chVI{1.85$\times$},
or reduce the storage overhead for error correction by 78.9\% if we keep the 
lifetime constant, compared to a
\chI{state-of-the-art baseline}.}

\end{itemize}


\section{Background}
\label{sec:background}

In this section, we first provide necessary background on the basics
of NAND flash memory (Section~\ref{sec:background:flash}).
Next, we briefly discuss the different known sources of errors within planar
NAND flash memory (Section~\ref{sec:background:errors}).
\sg{For an extended background on NAND flash memory, we refer the reader
to our prior works~\cite{cai.procieee17, cai.arxiv17, cai.book18}.}

\subsection{NAND Flash Memory Basics}
\label{sec:background:flash}

\begin{figure}[h]
  \centering
  \begin{subfigure}[b]{0.43\columnwidth}%
    \includegraphics[width=\linewidth, trim=205 140 130 190, clip]{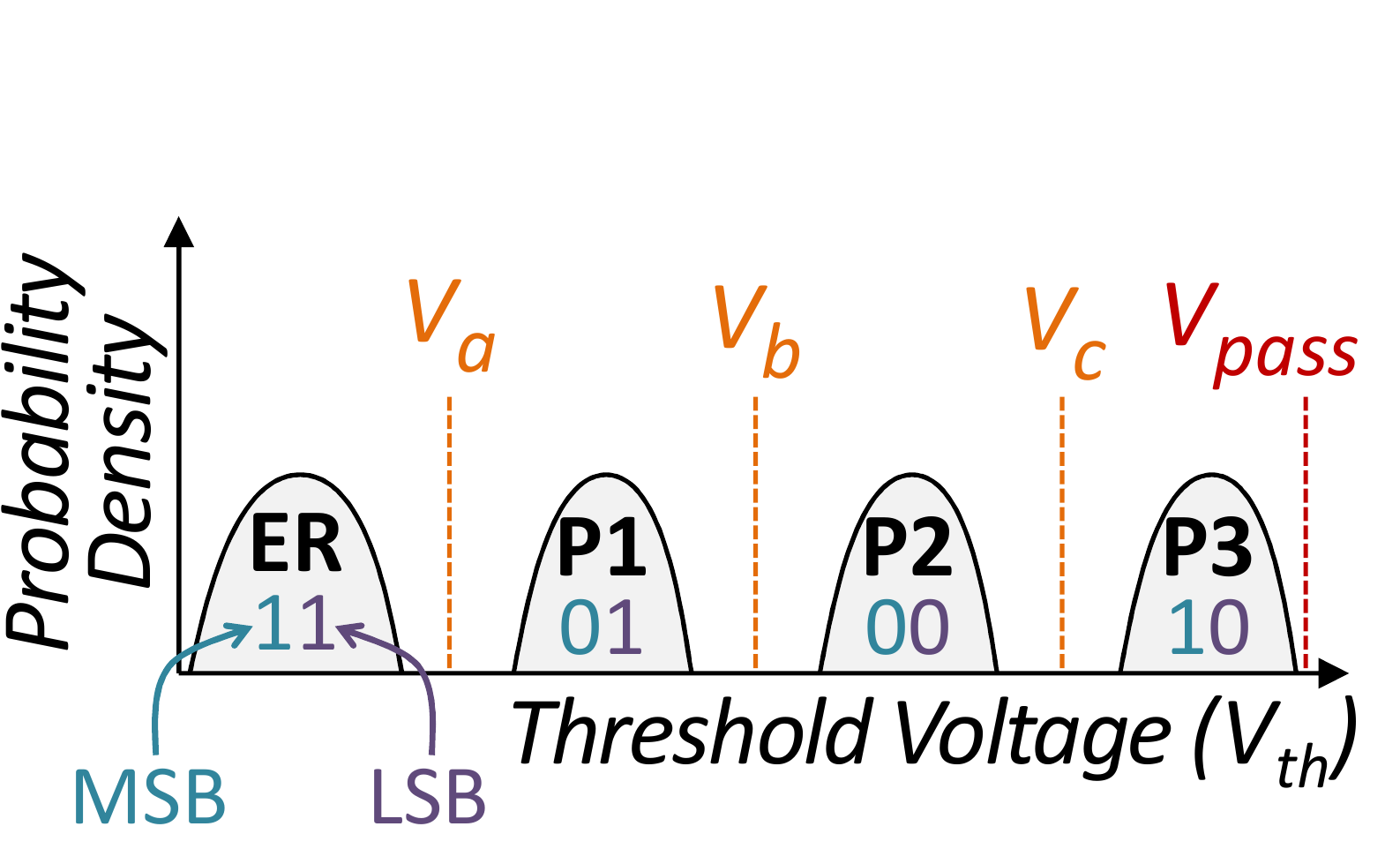}%
	\caption{}%
\label{fig:mlc}%
  \end{subfigure}%
  \hfill%
  \begin{subfigure}[b]{0.55\columnwidth}%
    \includegraphics[width=\linewidth, trim=195 243 165 155, clip]{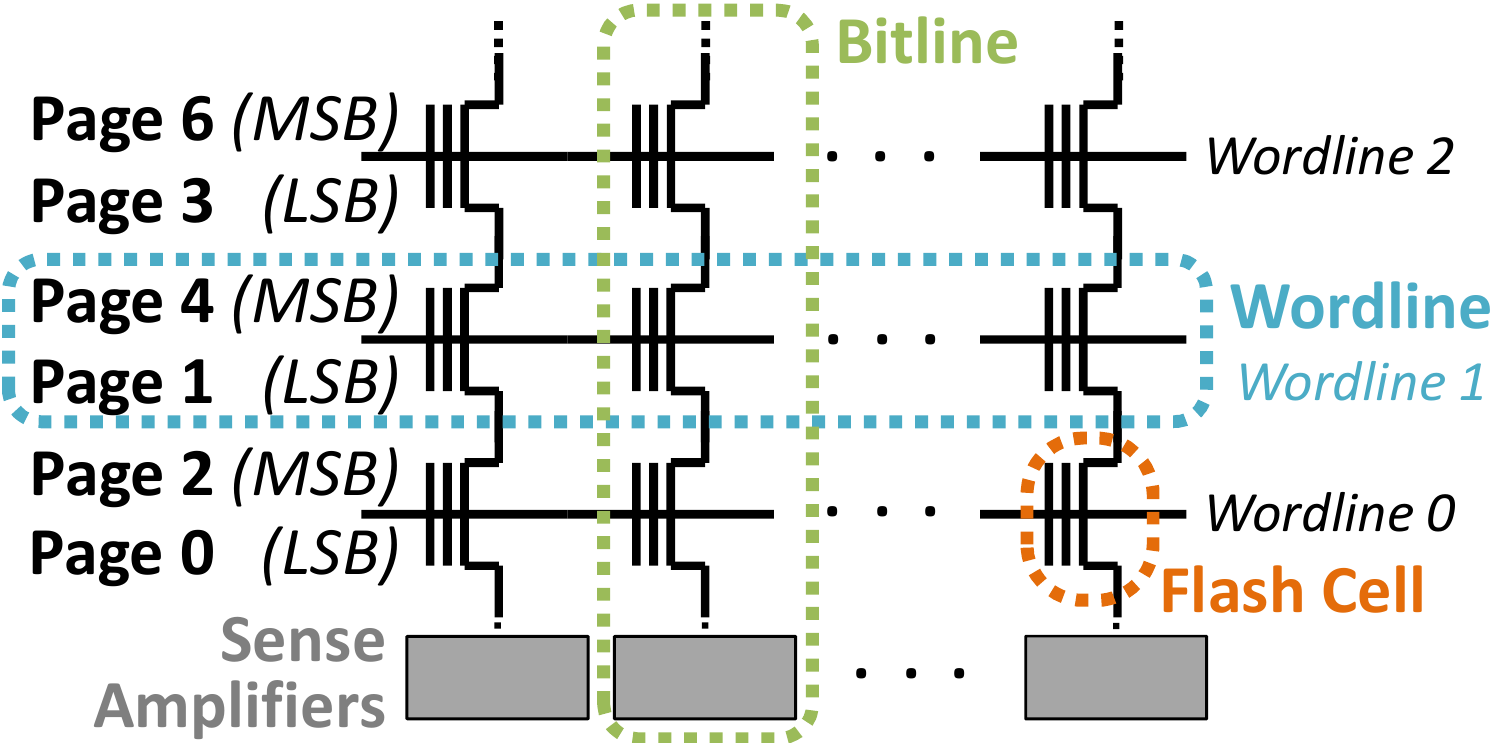}%
	\caption{}%
\label{fig:abl}%
  \end{subfigure}%
  \caption{(a)~Threshold voltage distribution and read reference voltages for MLC NAND flash memory;
        (b)~Internal \chVI{organization of a flash block}.}%
\end{figure}

In NAND flash memory, each flash cell consists of a transistor that can store 
charge. A flash cell represents a certain data value based on the \emph{threshold 
voltage} ($V_{th}$) of its transistor, \chV{which is \sg{determined} by the amount of
charge stored in it.}  In \emph{multi-level cell}
(MLC) flash memory, each cell stores two bits of data.  A threshold voltage
window \chVI{(i.e., \emph{state})} is
assigned for each possible two-bit value.
Figure~\ref{fig:mlc} shows the four possible states (i.e., ER, P1, P2, P3) in
MLC \chX{NAND} flash memory, along with their
corresponding bit values.  As a result of manufacturing process variation, the
threshold voltage of cells programmed to the same state follow a
Gaussian-like distribution across the voltage window of the
state~\cite{cai.date13, parnell.globecom14, luo.jsac16, cai.procieee17}, depicted as a
\chVI{probability density} curve
in Figure~\ref{fig:mlc}.

A NAND flash memory chip contains thousands of \emph{flash blocks}, which are
two-dimensional arrays of flash cells.  Figure~\ref{fig:abl} shows the internal
\chVI{organization} of a flash block.  Each block contains dozens of rows
(i.e., \emph{wordlines}) of flash cells, where each row typically contains
64K to 128K cells.  All of the cells on the same wordline are read and
programmed \chV{together as a group}.  
MLC \chX{NAND} flash memory partitions the two bits of \chV{each} flash
cell \chV{in a wordline} across two \emph{pages}, which are the unit of data 
programmed at a time (typically \SI{8}{\kilo\byte}).  The \emph{least significant bits} (LSBs) 
of all cells in one wordline form the \chVI{\emph{LSB page}} of that wordline, and the
\emph{most significant bits} (MSBs) of these cells form the \chVI{\emph{MSB page}}.
The sources and drains
of cells across different wordlines in the same block are connected in series
to form a \emph{bitline}.

Reads and writes \chV{to the flash memory} are managed by an SSD 
\emph{controller}.
The controller reads a page from a flash block by applying a \emph{read reference 
voltage} ($V_{ref}$) to 
the wordline that holds
the page.  A cell switches on only if $V_{th} > V_{ref}$.  Figure~\ref{fig:mlc}
shows the three read reference voltages ($V_a$, $V_b$, and $V_c$) that are used
to distinguish between each state.
A \emph{sense
amplifier} is attached to each bitline to detect if the cell is switched on. 
In order to detect the state of a particular cell on the bitline, the controller 
applies a \emph{pass-through voltage} ($V_{pass})$ to the wordlines of
\emph{all} unread cells in the \chV{flash} block.  This turns on the unread cells,
allowing
the value of the cell \chVI{that is} being read to propagate through the
bitline to the
sense amplifier.  To guarantee that all unread cells are \chV{\emph{on}}, $V_{pass}$ is set
to the maximum possible threshold voltage~\cite{cai.dsn15, cai.procieee17}.

Before new data can be written (i.e., \emph{programmed}) to a \chV{flash}
page, the controller must first
\emph{erase} the \chVI{\emph{entire block} (i.e., 512 to 1024 pages)} that the
page belongs to, due to wiring
constraints.  After erase, all of the \chV{cells in the erased block are
reset to} the ER~state. \chV{To program a flash cell,} the controller sends the
data to be programmed to the flash chip,
which repeatedly pulses a high programming voltage on a cell to increase a
cell's threshold voltage until the cell reaches its target state.  \chVI{This
iterative programming approach is called \emph{incremental step pulse programming} 
(ISPP)~\cite{mielke.irps08, bez.procieee03, suh.jssc95, wang.ics14}.}
Each pair of erase
and program operations is referred to as a \emph{program/erase} (P/E)
\emph{cycle}.

\subsection{Errors in NAND Flash Memory}
\label{sec:background:errors}

As \chX{vendors} work to increase the density of NAND flash memory, they use
aggressive manufacturing process technology scaling to reduce the size of a
flash \chVI{cell}.  As a result, each cell has a smaller capacity to store charge, 
and the cells move closer to each other.  These changes \chVI{reduce} the 
reliability of the NAND flash memory, \chVI{thereby} increasing the probability
of flash memory errors \chV{in} newer generations of planar (i.e.,
two-dimensional) NAND flash memory.
Errors occur when the cell threshold voltage ($V_{th}$) unintentionally changes
or is read incorrectly,
which can \chVI{alter the cell} state observed by the controller.
\chVI{Errors} can be induced by a range of sources\chVI{~\cite{mielke.irps08,
cai.date12, cai.date13, cai.iccd12, cai.iccd13, cai.sigmetrics14, cai.hpca15,
cai.dsn15, cai.procieee17, cai.hpca17, cai.itj13, luo.hpca18, cai.book18}},
which we divide into
four categories: process variation errors, retention errors, write-induced
errors, and read-induced errors.  We briefly describe each error source below,
and refer the reader to the prior work cited below for detailed explanations of
each error source. 
\chVI{A comprehensive treatment of different types of NAND flash memory
errors and mitigation mechanisms for them can be found in our recent
\chVII{survey papers}~\cite{cai.procieee17, cai.book18}.}

\emph{Process variation errors} occur as a result of the fabrication process.
Within a single chip, different flash cells have different attributes, due to 
the lithography limitations of modern manufacturing process
technologies~\cite{prabhu.trust11, cai.date12}.
As a result, there is inherent variation among the cells, and some cells have a
higher error rate than other cells.

\chVI{\emph{Retention errors}~\cite{cai.iccd12, cai.hpca15, cai.itj13} are
a
type of error that increase and accumulate over time after a flash cell is
programmed. A retention error occurs}
because charge leaks \chVI{out of} the transistor over
time.  As charge leaks from a cell, \chVII{the cell's} \chVI{threshold voltage ($V_{th}$)}
decreases.  In planar NAND flash memory, retention errors are the dominant 
source of all flash memory errors~\cite{cai.date12, cai.iccd12, cai.hpca15, cai.itj13},
\chVII{if aggressive refresh techniques~\cite{cai.iccd12, cai.itj13, luo.msst15}
are not employed.}

\emph{Write-induced errors} occur during program or erase operations.
\chV{\emph{P/E cycling errors} (or program/erase variation errors)}~\cite{cai.date13,
parnell.globecom14, luo.jsac16} \chVI{\chVII{are errors} that occur immediately
after erasing and programming a flash page. These errors} occur because of
the inaccuracy of each
program and erase operation. \chVI{This inaccuracy causes some cells to be programmed
into a state other than its desired target \chVII{state}.} As more P/E cycles
take place over the lifetime of a flash cell, the repeated stress causes more
electrons to become trapped within the transistor, which is known as
\emph{wearout}. \chVI{Wearout increases the inaccuracy during program and
erase operations, thereby increasing the number of P/E cycling errors.}
\chVI{\emph{Cell-to-cell program interference errors}~\cite{cai.iccd13,
cai.sigmetrics14}
are another type of write-induced error that \chVII{increases}} the threshold
voltage of a cell and thereby \chVII{increases} the RBER, when an 
\emph{adjacent}
cell in \chVII{\emph{another}} wordline is being programmed. \chVI{Since parasitic} capacitance
coupling exists between cells within close proximity of each \chVI{other, when} a
high programming voltage is applied on one cell, the capacitance coupling adds
charge to the transistors of the adjacent cells, increasing the program
interference errors.

\emph{Read-induced errors}
occur during read operations.
\chVI{\emph{Read errors}~\chXI{\cite{joe.ted11, compagnoni.edl09, ghetti.ted09}} are a type of
read-induced error where two reads to a flash cell may return different
data values.  A read error \chX{occurs} when the read reference voltage
is close to the cell's threshold voltage. \chVI{Such an error occurs}
when random fluctuations on the bitline cause}
the sense amplifier to \chVI{detect the wrong data}.
\chVI{\emph{Read disturb errors}~\cite{parnell.globecom14, cai.dsn15} are
another type of read-induced error where reading a page in a flash block may
\chVII{change the values stored in (i.e.,}
increase the RBER\chVII{)} of \chX{\emph{other}} pages in the same block. This type of error} occurs
\chVI{due to}
the \chVI{application of the} \emph{pass-through voltage} ($V_{pass}$) to
unread cells.  When one cell on a bitline is being read, \chVII{applying}
$V_{pass}$ \chVII{to the unread cells} can induce
a weak programming effect on the \emph{unread} cells, slowly transferring electrons
into the \chVII{unread cells'} transistors and increasing the \chVI{threshold
voltage} of the
unread cells.

To mitigate these errors, SSDs use error-correcting codes
(ECC) on the data.  ECC has a fixed \emph{error correction capability}: it can
correct only a limited number of errors, beyond which the data is no longer
correctable.  When a flash page is uncorrectable, we say that the SSD has
reached the end of its \emph{lifetime}.

\section{Architectural Differences Between 3D NAND and Planar NAND}
\label{sec:background:3d}

3D NAND flash memory (or 3D NAND) has three \emph{fundamental} differences from the most recent
generation (i.e., \SIrange{10}{15}{\nano\meter}) of planar NAND flash memory:
(1)~the flash cell design,
(2)~the organization of flash cells within a chip, and
(3)~the manufacturing process technology \chX{node}.

\begin{figure}[h]
\centering
\includegraphics[trim=0 280 290 0,clip,width=\figscale\linewidth]
{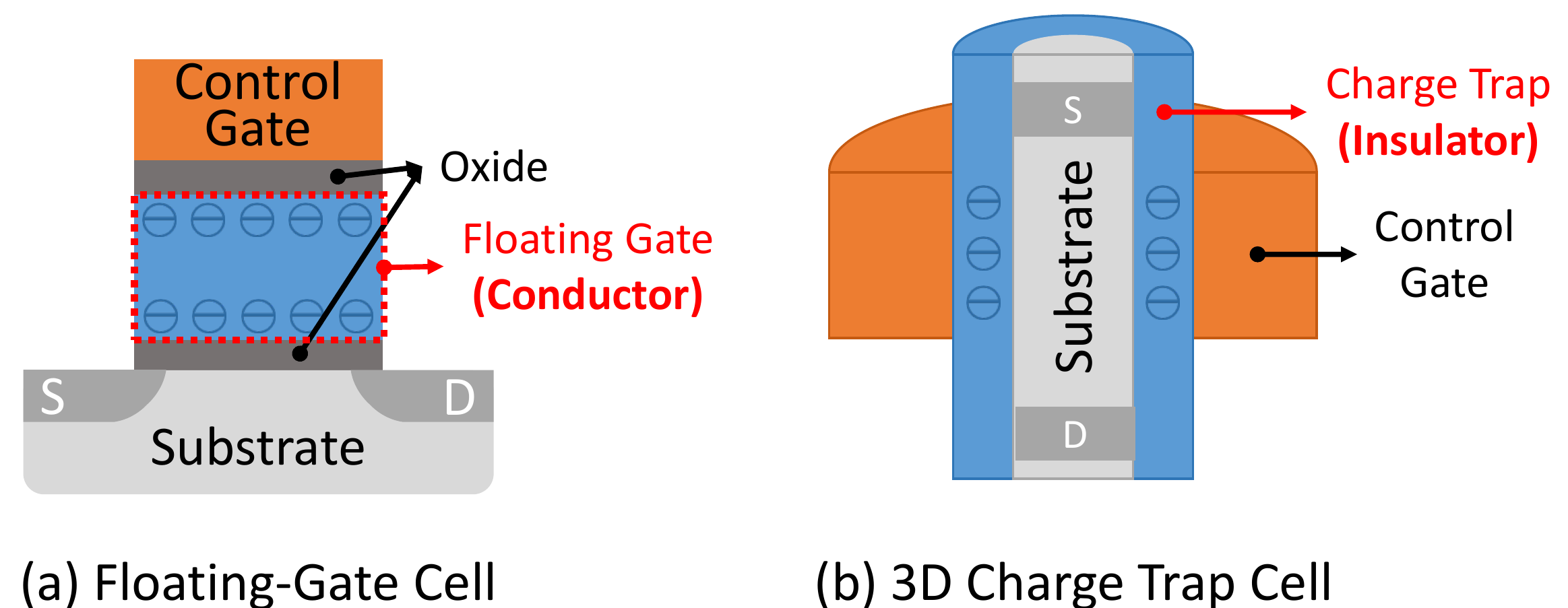}
\caption{The design of (a)~a floating-gate cell, and (b)~a 3D charge trap cell.}
\label{fig:cell}
\end{figure}

\textbf{Flash Cell Design.} 
In both planar and 3D NAND flash memory, each flash cell consists of a
transistor that can store charge, where the \chVI{amount of} charge determines
the
threshold voltage of the cell (i.e., the voltage at which the cell turns on).
The vast majority of planar NAND flash memory uses a
\emph{floating-gate transistor} (FG) for each cell.  Figure~\ref{fig:cell}a
illustrates the design of a floating-gate cell.  A \emph{control gate} sits at
the top of the transistor.  Read, program, and erase operations all apply a
voltage onto the control gate to turn on the cell or to add charge to the
transistor.  
A \emph{floating gate} sits in the middle of the transistor.  The floating gate
is a conductor that
stores the transistor's charge, and is sandwiched by \emph{oxide} layers.
The oxide layers minimize the amount of charge that leaks out of the floating gate.
At the bottom of the cell is the \emph{substrate}, which has two
terminals on either end, marked \emph{source} (S) and \emph{drain} (D).
When the voltage applied on the control gate is higher than the voltage of the
charge stored in the floating gate, an electrical \emph{channel} forms between
the source and drain, connecting them together.  The floating gate voltage can
be increased or decreased by applying a large positive or negative voltage, 
respectively, to the control gate, which induces Fowler-Nordheim tunneling~\cite{fowler.royal28}
of electrons through the oxide.

Instead of floating-gate transistors, \chVI{most existing} 3D NAND \chVI{flash
memory designs use a \emph{charge trap transistor}}
(CT) for each cell.  Figure~\ref{fig:cell}b illustrates the design of a 
charge trap cell.  The substrate, and therefore the channel between source
and drain, sits vertically in the center of the cell.  A \emph{charge trap layer}
wraps around the substrate.  The charge trap layer takes the place of the
floating gate, storing the transistor's charge.  However, unlike the floating gate,
the charge trap layer is an insulator.  The control gate still exists in a
charge trap cell, but it now wraps around the charge trap layer.

\begin{figure}[h]
\centering
\includegraphics[trim=0 40 140 0,clip,width=.8\linewidth]{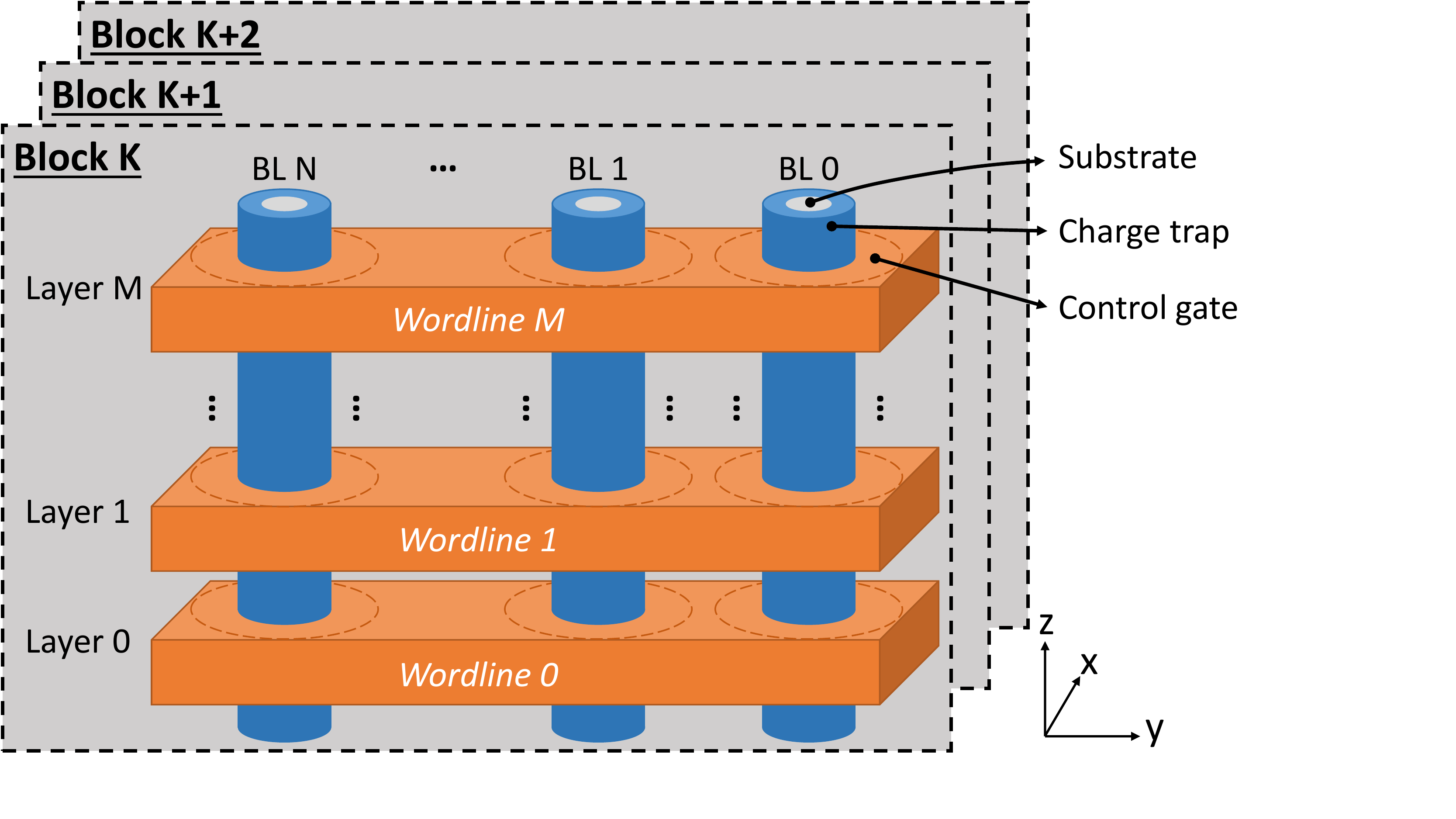}
\caption{3D NAND \chVI{flash memory} organization.}
\label{fig:organization}
\end{figure}

\textbf{Flash Chip Organization.} 
Figure~\ref{fig:organization} illustrates the
physical organization of flash cells in 3D NAND flash memory. The
charge trap transistor design allows the bitline (BL in
Figure~\ref{fig:organization}) of a block to stand \emph{vertically}
(i.e., along the z-axis) in the chip.  In other words, the bitline now connects 
together one charge trap cell from \emph{each layer} of the chip, as the
cells are stacked on top of each other.  Note that all of the cells along the 
z-axis share the same charge trap insulator, akin to how transistors are 
connected together on a bitline in planar NAND flash memory.
The control gates of cells in the same layer, along the y-axis, are connected
together to form a wordline.  In this
figure, we show a simple example where the cells in the same y--z plane 
form a flash block. In reality, to form larger flash blocks,
multiple stacks of flash cells are connected together to form longer bitlines,
thus increasing the number of wordlines within a block.
Multiple such flash blocks are aligned along the x-axis to
form a flash chip.

\textbf{Manufacturing Process Technology.} 
Compared with the most recent generation of planar NAND flash memory
(i.e., \SIrange{10}{15}{\nano\meter}), 3D NAND flash memory uses a much larger manufacturing process 
technology \chX{node} (e.g., \SIrange{30}{50}{\nano\meter})~\cite{samsung.whitepaper14}.  Because 3D NAND
flash memory has a large number \chX{of layers}
(typically \chVI{\numrange{24}{96}}~\chVII{\cite{techinsights.tr16,
park.jssc15, kang.isscc16, kim.jssc18, anandtech.web17,
im.isscc15}}), it can reach the same storage
\chVI{density} of
the most recent planar NAND flash memory generation while using much larger
flash cells.


\section{Characterization of 3D NAND Flash Memory Errors}
\label{sec:errors}

\chI{\textbf{Our goal} is to identify and understand new error characteristics
in 3D NAND flash memory, through rigorous experimental characterization of
real, state-of-the-art 3D NAND flash memory chips.}
We use the observations and \chVI{analyses obtained from such} characterization to
(1)~compare how the reliability of a 3D NAND flash memory chip differs from that of
a planar NAND flash memory chip,
(2)~develop a model of how \chI{each new} error source affects the error rate of
3D NAND flash memory,
(3)~understand if and how these reliability characteristics will change with
future generations of 3D NAND flash memory, and
(4)~develop mechanisms that can mitigate \chI{new error sources} in 3D NAND flash memory.

For our characterization, we use the methodology discussed in
Section~\ref{sec:methodology}. First, we perform a detailed
characterization and analysis of three error characteristics that are
drastically different in 3D NAND flash memory than in planar NAND flash \chVI{memory}:
\chV{layer-to-layer} process variation
(Section~\ref{sec:variation}), \chV{early retention loss}
(Section~\ref{sec:retention}), and retention interference
(Section~\ref{sec:retention:interference}).
\chI{In addition to identifying \emph{new} error sources in 3D NAND flash 
memory, we use our methodology to corroborate and quantify 3D NAND error
characteristics that are a result of error sources that were \chVI{\emph{previously}}
identified in planar NAND flash memory, including
retention loss~\cite{cai.book18, cai.procieee17, cai.iccd12, cai.hpca15, choi.svlsi16, park.jssc15, cai.itj13},
P/E cycling~\cite{cai.book18, cai.procieee17, cai.date13, parnell.globecom14, luo.jsac16, park.jssc15},
program interference~\cite{cai.book18, cai.procieee17, cai.iccd13, cai.sigmetrics14, park.jssc15, cai.hpca17},
read disturb~\cite{cai.book18, cai.procieee17, parnell.globecom14, cai.dsn15}, and
process variation~\cite{prabhu.trust11, cai.date12}.
We summarize our findings for these error types in
Section~\ref{sec:summary}, and provide detailed results on our characterization
of these previously-identified error sources in Appendix~\ref{sec:appendix}.}

\subsection{Methodology}
\label{sec:methodology}

We experimentally characterize several real, state-of-the-art 3D MLC
NAND flash memory chips from a single vendor.\footnote{The trends we observe
from the characterization are expected be similar for 3D charge trap flash
\chVI{memory}
manufactured by different vendors, as their 3D flash \chX{memory organizations} are
similar in design.}$^{,}$\footnote{We normalize the actual \chVI{number of}
stacked layers of the chips
and leave out the exact process technology
to protect the anonymity of the flash vendor and to avoid revealing proprietary
information.} We use a NAND flash characterization platform similar to prior
work~\chVII{\cite{cai.fccm11, cai.date12, cai.date13, cai.iccd12, cai.itj13,
cai.iccd13, cai.sigmetrics14, cai.hpca15, cai.dsn15, cai.hpca17,
parnell.globecom14, luo.jsac16, cai.procieee17, cai.book18, luo.hpca18}},
which allows us to issue \emph{read-retry} commands directly to the
flash chip. The read-retry command~\cite{cai.procieee17, cai.date13}
allows us to fine-tune the read
reference voltage used for each read operation. The smallest amount by
which we can change the read reference voltage is called a
\emph{voltage step}. We conduct all experiments at room temperature
(\SI{20}{\celsius}).

We use two metrics to evaluate 3D NAND \chVI{flash memory} reliability.
\chVI{First,} we show the
\emph{raw bit error rate} (RBER), which is the rate at which errors
occur in the data \emph{before error correction}. We show the RBER for
when we read data using the \emph{optimal read reference voltage}
($V_{opt}$), which is the read reference voltage that generates the
fewest errors in the data.\footnote{We show RBER at the optimal read
reference voltage to accurately represent the reliability of NAND
flash memory, as SSD controllers tune the read reference voltage to a
near-optimal point to extend the NAND flash lifetime~\cite{cai.hpca15,
papandreou.glsvlsi14, luo.jsac16, cai.procieee17}.}

Second, we show how the various error sources change the {\em
threshold voltage distribution}. These \chVI{changes} (i.e., shifting
and widening) in threshold voltage distribution directly \chVI{lead} to raw
bit errors in the flash memory. To obtain the distribution, we first
use the read-retry command to sweep over all possible voltage values,
to identify the threshold voltage of each cell.\footnote{\chVI{We}
refer to prior work for more detail \chVII{on} the methodology
to obtain the threshold voltage distribution~\cite{parnell.globecom14,
luo.jsac16, cai.date13}.}  Then, we use this data to calculate the probability
density of each state at every possible threshold voltage value.
As part of our analysis, we fit the threshold voltage
distribution of each state to a Gaussian distribution.  We
use the \emph{mean} of the Gaussian model to represent how the
distribution shifts as a result of errors, and we
use the \emph{standard deviation} of the model to represent how the
distribution widens. Throughout this paper, we present normalized voltage
values, as the actual voltage values are proprietary to NAND flash memory
\chX{vendors}. A normalized voltage of $1$ represents a single \chV{fixed} voltage step.

\begin{figure}[h]
\centering
\includegraphics[trim=0 10 0 10,clip,width=\figscale\linewidth]{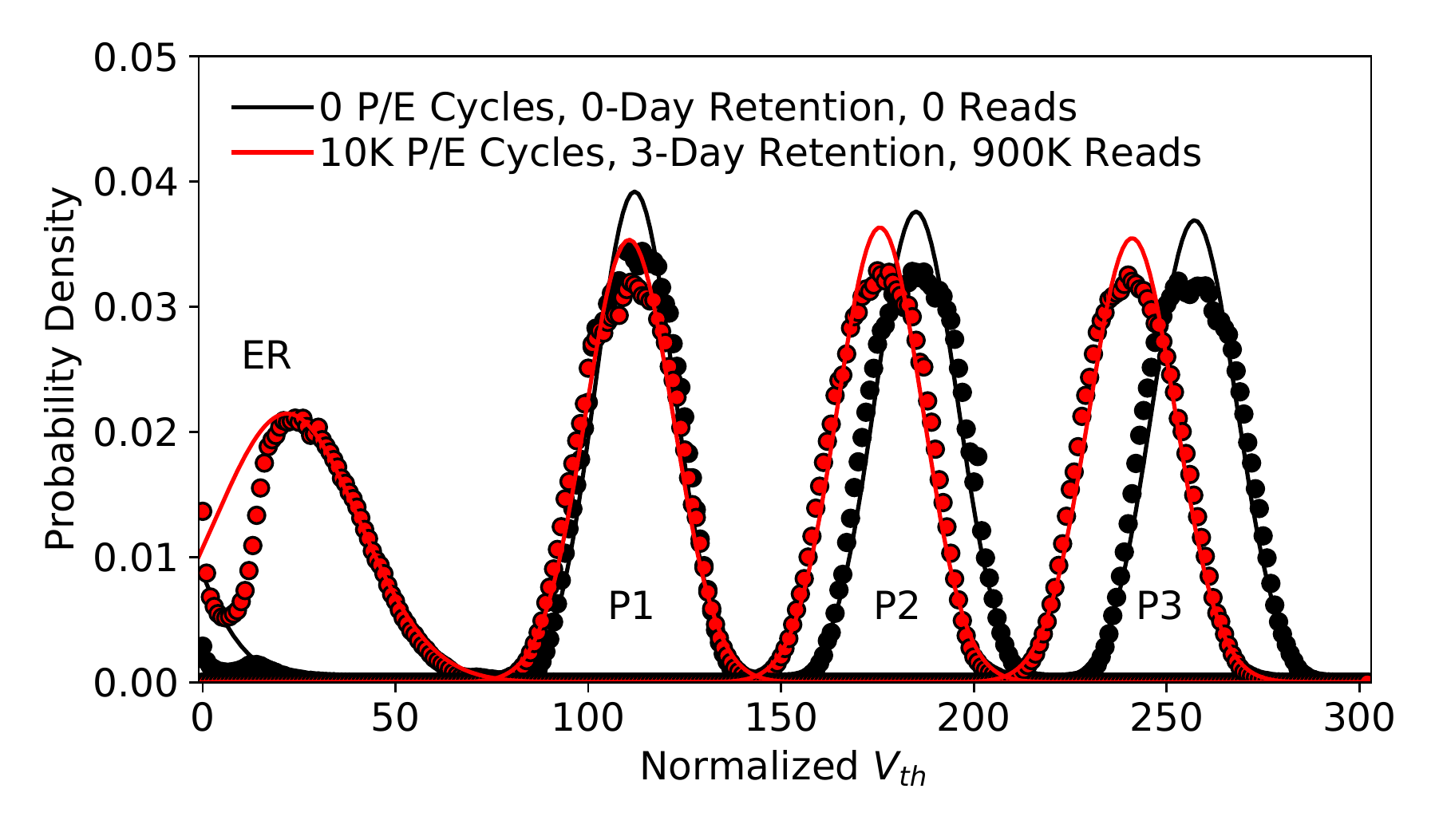}
\caption{3D NAND threshold voltage distribution before (black) and after (red)
the data is subject to a high number of errors \chVI{(due to P/E cycling,
retention loss, and read disturb)}.}
\label{fig:distribution-shape}
\end{figure}

We show two examples in Figure~\ref{fig:distribution-shape} to
visualize how well this simple Gaussian model captures the change in
the measured threshold voltage distribution.
Figure~\ref{fig:distribution-shape} shows the measured and modeled
distributions under two conditions: (1)~\chVII{after} 0~P/E cycles,
0-day retention \chVII{time}~\chVI{\cite{cai.hpca15}}, and 0~read disturbs (i.e., the data contains few
errors); and (2)~\chVII{after} 10K~P/E cycles, 3-day
retention \chVII{time}~\chVI{\cite{cai.hpca15}}, and 900K~read
disturbs (i.e., the data contains a high number of errors). \chVII{Dotted
points} plot
the \chVI{measured threshold voltage \chVII{distributions}} from \chX{the real} 3D NAND
\chVII{memory chips}. \chV{Note that we are
unable to show the ER state distribution when \sg{the P/E cycle count} is low (\chVI{i.e.,} the
black dots), because the erase operation cleanly resets the threshold voltage
to a negative value that is lower than the observable voltage range \chVII
{under a low P/E cycle count}.}
We use a solid
line to show a fitted Gaussian distribution for each state. The
\chVII{Kullback-Leibler divergence \chXI{error values}~\cite{luo.jsac16, parnell.globecom14}
of the fitted \chVII{Gaussian} distributions 
are 0.034 and 0.23.}\footnote{\chVII{A KL-divergence error of $x$ means that
the model loses $x$ natural \chVIII{units} of information (i.e., nats) due to modeling
error.}} We observe, from this figure,
that after the chip \chVIII{is
used}, the threshold voltage distribution \chVI{shifts due to P/E cycling,
retention loss, and read disturb}, reducing the
error margins between neighboring states, \chVI{and} leading to more raw bit errors in the
data. Thus, \chVII{depicting and understanding} how \chVII{threshold voltage}
distributions are affected by various \chVI{factors helps}
us understand how raw bit errors occur and \chVI{thus devise mechanisms to}
mitigate \chVI{various} errors more effectively.

In the following sections, we directly show the mean and the
standard deviation of the \chVII{\emph{fitted}} threshold voltage distributions
instead of the distribution itself, \chVI{to simplify the presentation of our}
results.

\sph{\textbf{Limitations.} In our experiments, we randomly sampled 27 flash
\chVI{blocks throughout} our characterizations. Note that each sampled flash
block consists of tens of millions of flash cells. Thus, we believe that our
observations are representative of the general behavior that takes place in
the model of 3D NAND chips that we tested. While adding
more data samples (i.e., flash blocks to test) can add to the statistical
strength of our results,
we do not believe that this would change the \emph{general qualitative
findings} that we make and the \emph{models} that we develop in this work.
This is because the new error characteristics we observe are caused by the
underlying architecture of 3D NAND flash memory (see
Section~\ref{sec:background:3d}).}

\sph{Note that we do not characterize \emph{chip-to-chip}
process variation, as an accurate study of such variation requires a large-scale
study of a large number (e.g., hundreds) of 3D NAND flash memory chips, which
we do not have access to. Hence, we leave such a large-scale study for future
work.}


\subsection{Layer-to-Layer Process Variation}
\label{sec:variation}

Process variation refers to the variation in the attributes of flash cells
when they are fabricated (see Section~\ref{sec:background:errors}). Due to
process variation, some flash cells can have a higher RBER than others, making
these cells the limiting factor of overall flash memory reliability.
In 3D NAND flash memory, process variation can occur along all three axes of
the memory (see Figure~\ref{fig:organization}). Among
the three axes, we expect the variation along the z-axis (i.e.,
layer-to-layer variation) to be the most significant,
due to the new challenge of stacking multiple flash cells across layers.
\chI{Prior work has shown that current circuit etching technologies are unable
to produce identical 3D NAND cells when punching through multiple stacked
layers, leading to significant variation in the error characteristics of flash
cells that reside in different layers~\cite{wang.tecs17, hung.jssc15}.}

To characterize layer-to-layer process variation errors within a
flash block, we first wear out the block by
programming random data to each page in the block until the block
\chX{endures} 10K P/E cycles. Then, we compare the collective
characteristics of the flash cells in
one layer with those in another layer. We repeat this experiment for
flash blocks on multiple chips to verify all of our findings.

\begin{figure}[h]
\centering
\includegraphics[trim=0 10 0 10,clip,width=\figscale\linewidth]
{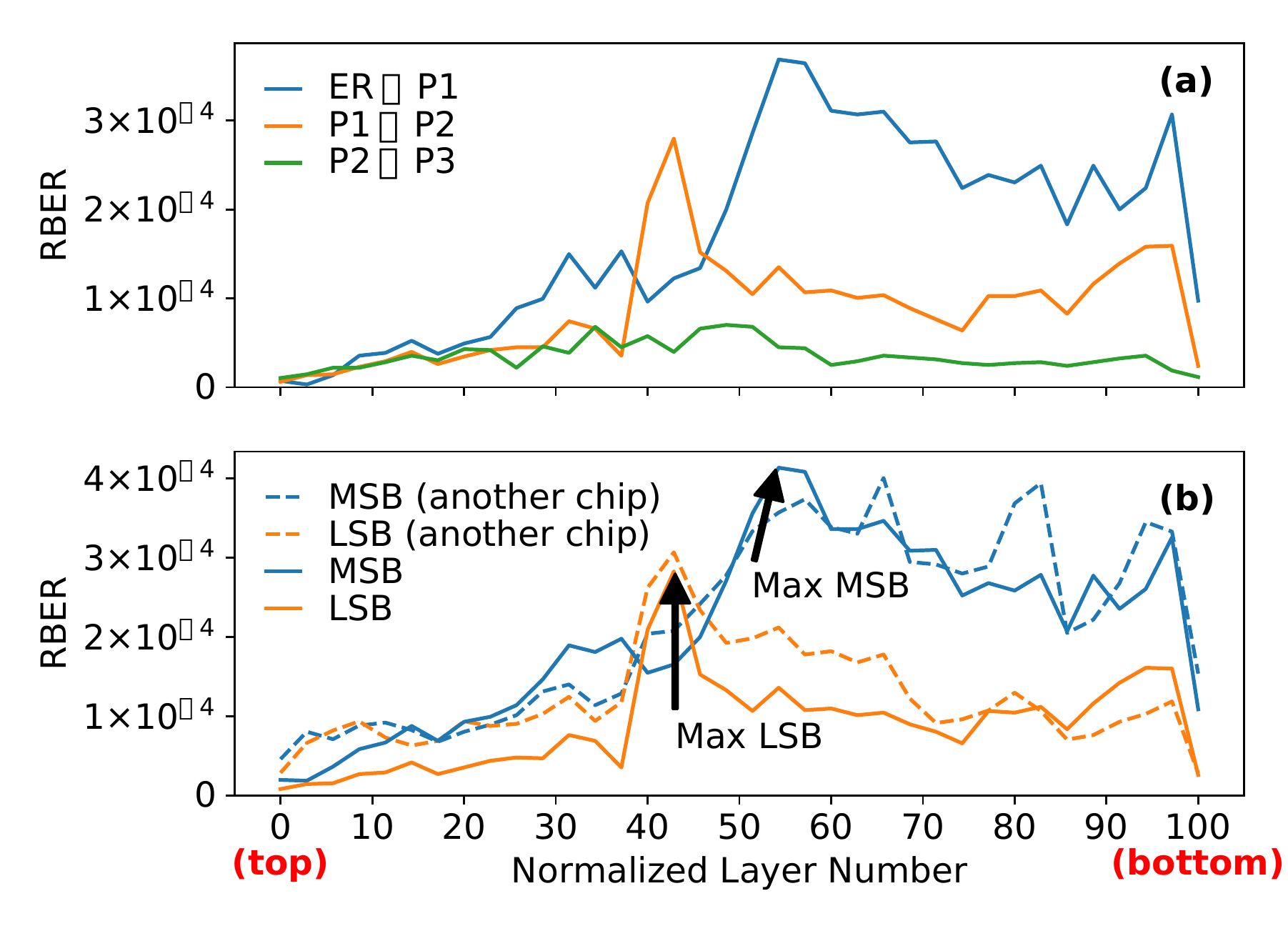}
\caption{\chIX{Variation of RBER across layers.}}
\label{fig:variation-wlopterr}
\end{figure}

\textbf{Observations.} 
Figure~\ref{fig:variation-wlopterr} shows the RBER
variation along the z-axis (i.e., across layers) for a flash block \chX{that has endured} 10K
P/E cycles. \sph{\chV{The chips we use for characterization have between 30 and 40 layers. We
normalize the number of layers from 0 (the top-most
layer) to 100 (the bottom-most layer) by multiplying the actual layer number
with a constant, \sg{to maintain the anonymity of the chip vendors}.}}
\chVII{Figure~\ref{fig:variation-wlopterr}a}
breaks down the errors
according to the \chX{originally-programmed state and the} current state of each cell;
\chVI{Figure~\ref{fig:variation-wlopterr}b} breaks down the
errors into MSB and LSB page errors.
\chVI{In Figure~\ref{fig:variation-wlopterr}b}, the solid curve and
the dotted curve show the results for two blocks that were randomly selected
from two different flash chips. 
\sph{We make five observations from Figure~\ref{fig:variation-wlopterr}.
First, ER~$\leftrightarrow$~P1 and P1~$\leftrightarrow$~P2 errors vary
significantly across layers, while P2~$\leftrightarrow$~P3 errors remain
similar across layers.
\chVII{The variation in ER~$\leftrightarrow$~P1 errors is mainly caused by the
large variation in mean threshold voltage of \chVIII{the} ER state across layers; the
variation in P1~$\leftrightarrow$~P2 is caused by the variation in the
threshold voltage distribution width of the P1 state across layers
(Section~\ref{sec:appendix:variation}).}
Second, both the MSB and LSB error rates vary significantly across layers.
\chVI{\chVII{We call this
phenomenon} \emph{layer-to-layer process variation}.}
For example, MSB page on normalized layer~55 in the middle (i.e., \emph{Max
MSB}) has an RBER 21$\times$
\chVIII{that of} normalized layer~0.
Third, MSB error rates are much higher than LSB error rates in a majority of
the layers, \chVI{on average by 2.4$\times$}. \chVI{\chVII{We call this phenomenon} \emph{MSB--LSB RBER variation}. MSB error rates are
usually higher than LSB error rates because reading \chVII{an} MSB page requires two
read reference voltages ($V_a$ and $V_c$), whereas reading \chVII{an} LSB page requires
only one ($V_b$).}
Fourth, the top half of the layers have lower error rates than the bottom half.
\chVII{This is likely caused by the variation in the flash cell size \chVIII{across layers}.}
\chVII{Fifth}, the RBER
variation we observe is consistent across two \chX{randomly-selected} blocks from
two different chips. \chVII{This indicates \chVIII{that layer}-to-layer process
variation and MSB--LSB RBER variation are \chVIII{consistent characteristics}
of 3D NAND flash memory.}}

\begin{figure}[h]
\centering
\includegraphics[trim=0 10 0 10,clip,width=\figscale\linewidth]
{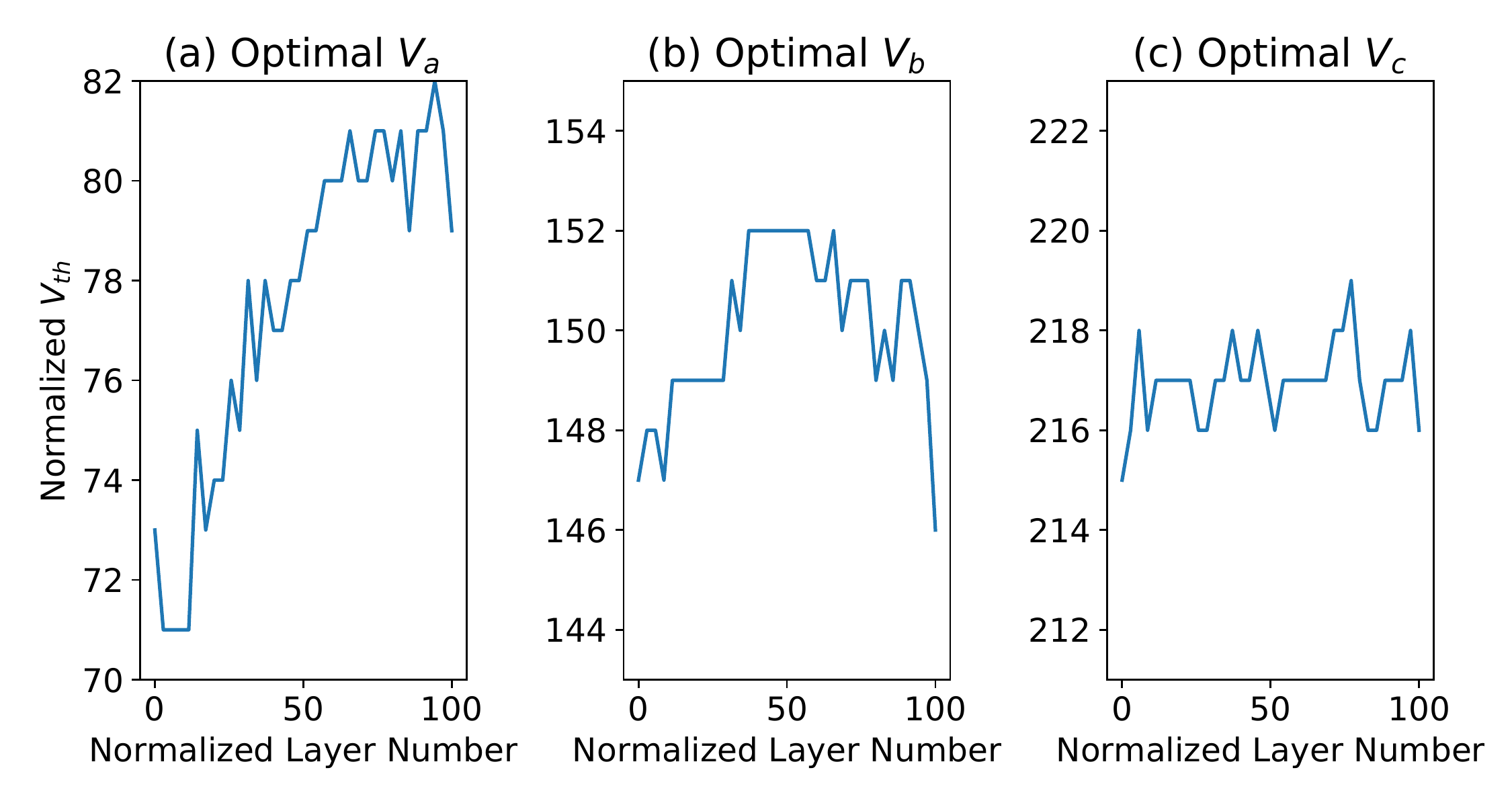}
\caption{\chVII{Variation of optimal} read reference voltage across layers.}
\label{fig:variation-wloptvrefs}
\end{figure}

Figure~\ref{fig:variation-wloptvrefs} shows how the optimal read reference
voltages vary across layers. Three subfigures show the optimal read reference
voltages for $V_a$, $V_b$, and $V_c$.  We make two observations \chXI{from
Figure~\ref{fig:variation-wloptvrefs}}.  First, the optimal voltages for $V_a$ and $V_b$ vary significantly
across layers, but the optimal \chX{voltage for} $V_c$ does not change by much. \chVI{This is
because process variation mainly affects the threshold voltage distributions of
the ER and P1 states, whereas the threshold voltage distributions of the
P2 and P3 states, \chVII{which are more accurately controlled by ISPP (see
Section~\ref{sec:background}),} are similar across layers. We discuss this
further in Appendix~\ref{sec:appendix:variation}.}
\sph{Second, \chVI{the optimal read reference voltages for $V_a$ and $V_b$ 
 are lower for cells in the top half of the layers than for cells in the bottom half.}} \chVI{This is because
process variation \chVII{significantly affects} the threshold voltage of the
ER and P1 states (see Appendix~\ref{sec:appendix:variation}).}

\textbf{Insights.} We show that the \chVII{\chVIII{phenomena} of}
layer-to-layer process variation \chVIII{and MSB-LSB RBER variation},
which \chVIII{are} unique to 3D
NAND flash memory, \chX{are} significant. We refer to
Appendix~\ref{sec:appendix:variation} for \chVI{a comparison
between layer-to-layer process variation and}
bitline-to-bitline process variation. In the future, as 3D NAND flash devices
scale along the z-axis, more layers
will be stacked vertically along each bitline. This will \chVI{likely} further
exacerbate
the effect of layer-to-layer process variation, making it even more
important to study and mitigate \chVII{its negative effects}.


\subsection{Early Retention Loss}
\label{sec:retention}

Retention errors are flash \chVI{memory} errors that accumulate after data has been
programmed to the flash cells~\chVI{\cite{cai.iccd12, cai.hpca15, cai.itj13}}
(see Section~\ref{sec:background:errors}).
\chI{Because 3D NAND flash memory typically uses a different cell design
(i.e., the charge trap cell described in Section~\ref{sec:background:3d})
than planar NAND flash memory (which uses floating-gate cells), it has
drastically different retention error characteristics.
The charge trap flash cells used in 3D NAND flash memory
suffer from \chVI{\emph{early retention loss}}, i.e., fast charge loss within a few seconds.
This phenomenon \chVI{has been} observed by prior works
using circuit-level characterization~\cite{chen.iedm10, choi.svlsi16}.
However, due to limitations of the \chVI{circuit-level characterization}
methodology used by these prior works, openly-available characterizations of
early retention loss in 3D charge trap NAND flash devices document retention
loss behavior for up to only 5~minutes after the data is written (i.e., for
a maximum \emph{retention time} of 5~minutes).  This limited window is insufficient for
understanding early retention loss under real workloads, which typically have
much longer \chVI{\emph{retention time} \chVII{requirements}}~\cite{luo.msst15}, \chVI{i.e., the length of time 
that has elapsed since programming \chVII{until the data is accessed again}}.}

\chI{Our goal is to experimentally characterize early retention loss in 3D NAND
flash memory for a large range of retention times (e.g., from several minutes
to several weeks).}
\sph{First, we randomly select 11 flash blocks within each chip and write pseudo-random data to each
page within the block to wear the blocks out.  We wear \chX{out} each block to a different
\chX{P/E cycle count}, so that we have error data for every 1K P/E cycles
between 0 and 10K P/E cycles.\footnote{For all
experiments throughout the paper, we consistently assume a 0.5-second
\emph{dwell time}, which is the length of time between consecutive
program/erase operations~\cite{luo.hpca18}.}
Then, we program pseudo-random data to each flash block, and wait for
up to 24~days under room temperature. To characterize retention loss, we measure
the RBER and the threshold voltage distribution at
nine different retention times, ranging from 7~minutes to 24~days. To
minimize the impact of other errors, and to allow us to include very low retention
times, we characterize only the
first 72~flash pages within each block. We believe that the observations we
make on these flash cells are representative of the entire chip, and we can 
generalize the observations to \chVI{a majority of} 3D NAND \chX{flash memory} cells.}
\chX{We analyze the threshold voltage distribution} in
Appendix~\ref{sec:appendix:retention}.

\textbf{Observations.} Figure~\ref{fig:retention-3d-vs-2d} shows the comparison
between the retention error rate of 3D NAND and planar NAND flash memory at
10,000 P/E cycles \chVI{\chX{using} both \chX{a} logarithmic time scale \chVII{on the x-axis}
(Figure~\ref{fig:retention-3d-vs-2d}a) and \chX{a} linear time scale \chVII{on the x-axis}
(Figure~\ref{fig:retention-3d-vs-2d}b) for different retention times after
programming}. \sph{To \chVII{make} this comparison,
we perform the same experiment as above for planar NAND flash memory chips.
\chVI{Due to limitations of the available data, we} extend \chVII{our data to
the same retention time range} using \chV{a linear model \sg{that was
proposed by} prior work~\cite{mielke.irps08, luo.hpca18}: $\log(RBER) = A \cdot
\log(t) + B$, where $t$ is the retention time, and $A$ and $B$ are parameters
of the linear model.}} The dotted portions of the lines represent the RBER 
\sg{that is predicted} by the linear model.

\chVI{We make two observations from this figure. First, in
Figure~\ref{fig:retention-3d-vs-2d}a, we} observe that the retention error rate
changes much more slowly for planar NAND \chX{flash memory} than for 3D NAND flash memory.
Although the 3D NAND flash \chX{memory} chip has lower RBER than the planar NAND flash \chX{memory} chip
shortly after programming, the RBER becomes higher on the 3D NAND flash \chX{memory} chip
after \chVII{$7\times 10^3$ seconds ($\sim$2 hours)} of
retention time. This means that
3D NAND flash memory is more susceptible to \chVII{the} retention loss
phenomenon than
planar NAND flash memory. Second, in Figure~\ref{fig:retention-3d-vs-2d}b,
we observe that the RBER of 3D NAND flash memory quickly increases by an order
of magnitude in \chVI{$10^4$ seconds ($\sim$3~hours)}, and by another order of
magnitude in \chVI{$10^6$ seconds ($\sim$11~days)}. However, we do \chVII{\emph{not}}
observe a large difference in retention loss between low and high retention
times \chVI{for \chVII{\emph{planar}} NAND flash memory \chVII{(also
\chVIII{shown} by prior works~\cite{mielke.irps08, cai.hpca15})}.}
This shows that the retention loss
is \emph{steep} when retention time is \emph{low}, but the retention loss
flattens out when the retention time is high. This is a result of the early
retention \chVI{loss phenomenon} in 3D NAND flash memory.

\chX{Early} retention loss
can be caused by two possible reasons. First, the tunnel oxide layer is thinner
in 3D NAND \chVI{flash memory} \chVII{than in planar NAND flash
memory}~\cite{zhang.acsnano14, samsung.whitepaper14}.
Since a 3D charge trap cell uses an insulator to store charge, which is immune
to the short circuiting caused by stress-induced leakage current
(SILC)~\chVI{\cite{naruke.iedm98, degraeve.ted04}}, the tunnel
oxide layer in 3D NAND flash memory is designed to be thinner to improve
\chVI{programming} speed~\cite{park.jssc15}. \chVI{This causes charge to leak
\chVII{very fast soon} after programming.} \chVIII{Second, 
\chX{cells connected on the same bitline
share the same \chX{charge trap} layer. \chX{As a \chXI{result, charge} that is}
programmed to a flash cell quickly leaks to 
\chXII{adjacent cells that are on the same bitline}
due to \chIX{\emph{electron diffusion} through the shared \chX{charge trap}
layer}~\cite{choi.svlsi16}, which we discuss further
in Section~\ref{sec:retention:interference}.}

\begin{figure}[h]
\centering
\includegraphics[trim=0 0 0 0,clip,width=.7\linewidth]{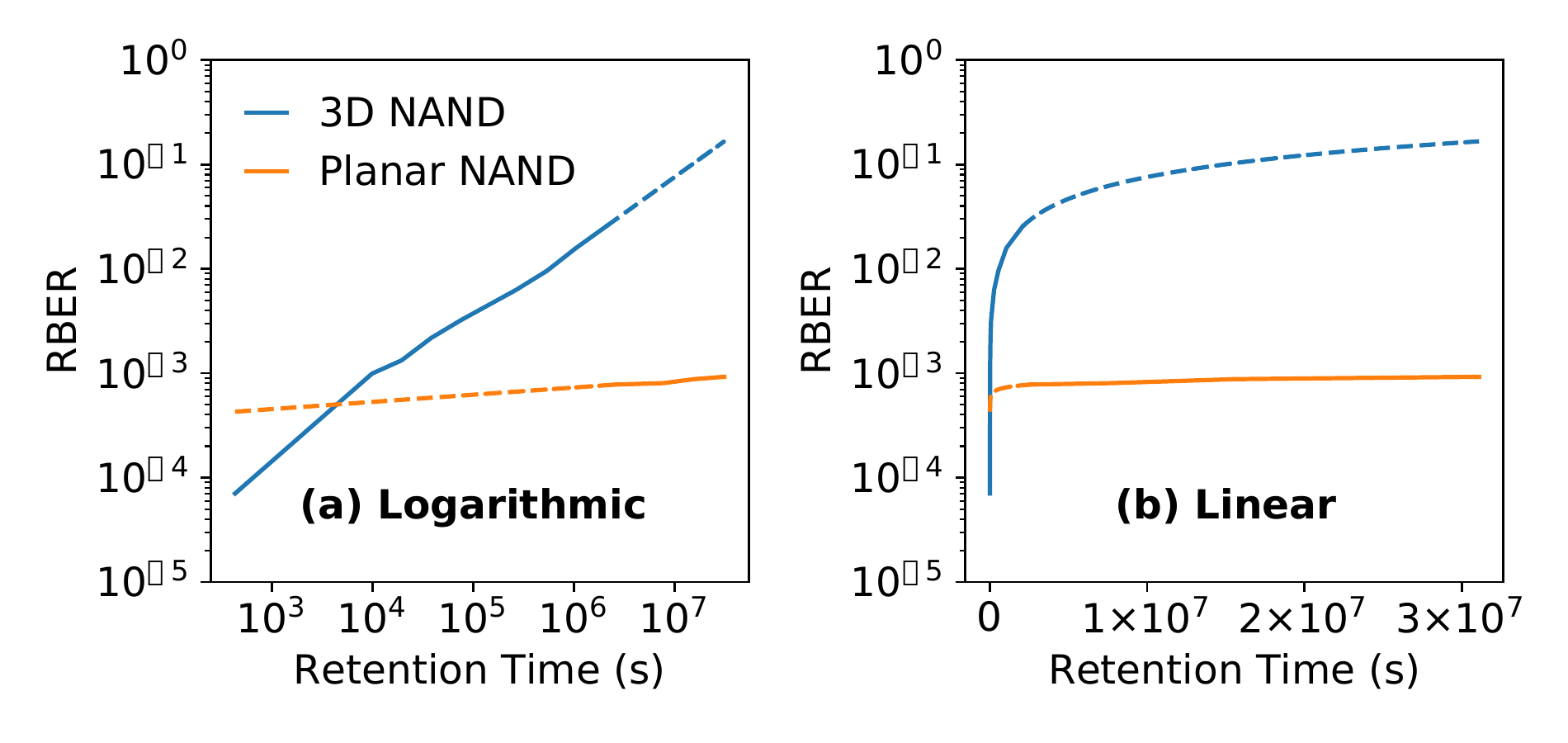}
\caption{Retention error rate comparison between 3D NAND and planar NAND flash
memory \chVI{at 10K P/E cycles}. \chVII{Dotted portions of lines represent
the RBER predicted by the linear model \chVIII{proposed by prior
work~\cite{mielke.irps08, luo.hpca18}}. 
\chX{We show the retention time on the x-axis using both (a)~a \emph{logarithmic}
time scale and (b)~a \emph{linear} time scale.}}}
\label{fig:retention-3d-vs-2d}
\end{figure}

Figure~\ref{fig:retention-optvrefs} plots how the optimal read reference
voltage changes with retention time. The three subfigures show the optimal
voltages for $V_a$, $V_b$, and $V_c$. We make three observations from this
figure.  First, the relation between the optimal read reference voltages
\chVI{of} $V_b$ or $V_c$ and the retention time can be modeled
as~\cite{mielke.irps08, luo.hpca18}: $V = A \cdot \log(t) + B$, similar to the
logarithm of RBER \chVII{(which we \chX{discuss} above)}. Second, the
optimal read reference voltages for $V_b$ and $V_c$ decrease significantly as
retention time increases, whereas $V_a$ remains relatively constant. Third,
\chVI{due to the early retention loss phenomenon,}
the optimal read reference voltages \chVI{for} $V_b$ and $V_c$ change rapidly when the
retention time is low \chVII{(e.g., $V_c$ changes by 5 voltage steps within
the first 3 hours)}, but
they change slowly when the retention time is high \chVII{(e.g., $V_c$ changes
by another 5 voltage steps after 11 days)}.

\begin{figure}[h]
\centering
\includegraphics[trim=0 0 0 0,clip,width=\figscale\linewidth]
{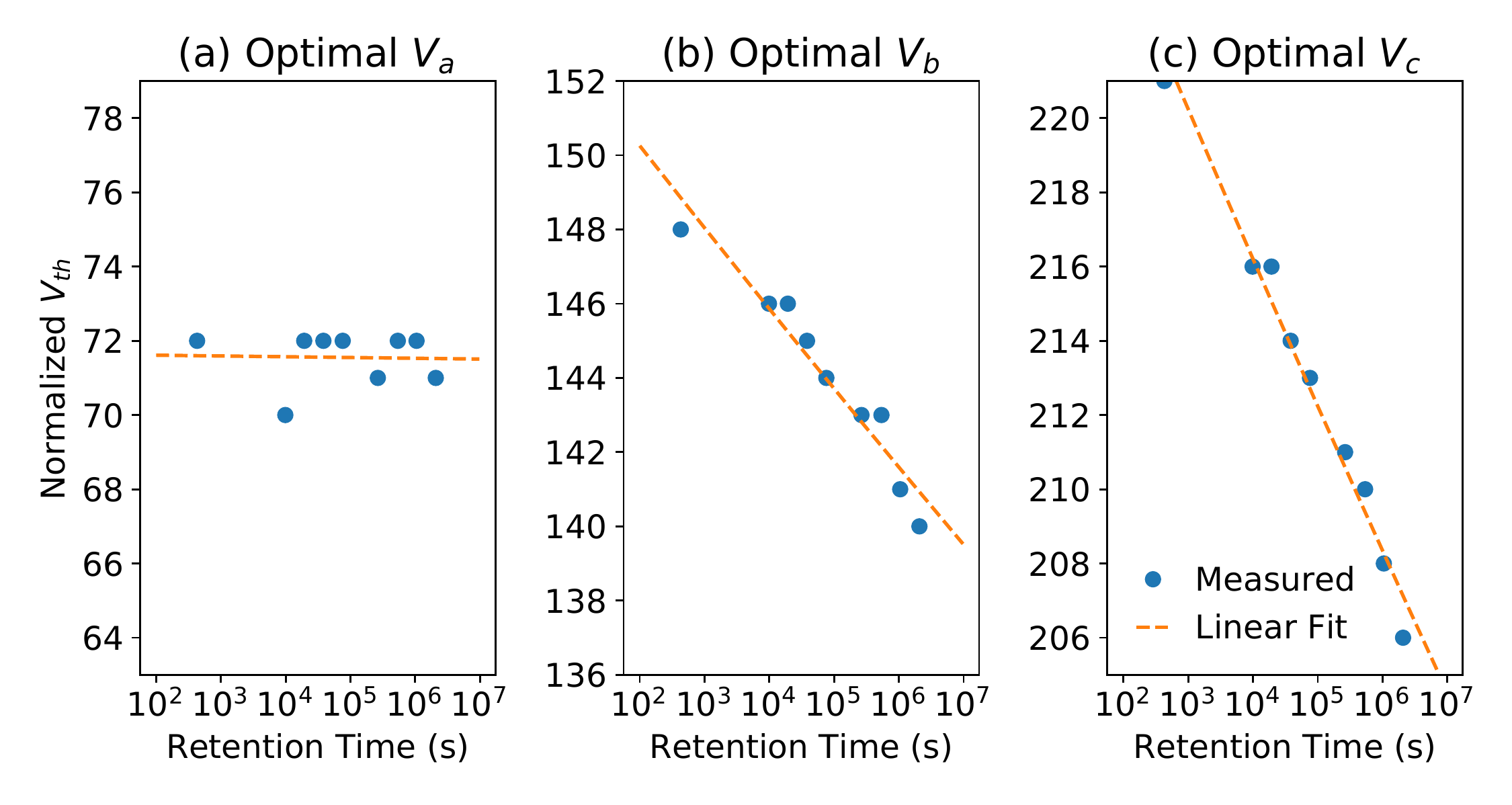}
\caption{Optimal read reference voltages \chVIII{for different} retention
times.
\chVII{Note that \chX{the x-axis uses a} logarithmic time scale.}}
\label{fig:retention-optvrefs}
\end{figure}

\textbf{Insights.} We compare the errors caused by retention loss in 3D NAND \chX{flash memory} to
that in planar NAND \chX{flash memory}, using our results in Figure~\ref{fig:retention-3d-vs-2d}
\chX{and} the results reported in prior work~\cite{mielke.irps08,
cai.hpca15, cai.iccd12}. We find two major differences in 3D NAND \chVI{flash
memory, which we summarize below. More}
results and insights are in Appendix~\ref{sec:appendix:retention}.
First, \chVI{3D NAND flash memory is more susceptible to retention errors than planar
NAND flash memory, and} its error \chVI{rate} increases much faster when the
retention time is low than when the retention time is high. \chVI{This is a result of
the early retention \chVI{loss} phenomenon in 3D NAND flash memory, which is
due to the use of a different flash cell design and thus is likely \chVII{to}
remain in future generations of 3D NAND flash memory.}
Second, the optimal read reference \chVIII{voltages} for $V_b$ \chVII{and $V_c$} in 3D
NAND flash memory
\chX{change} \chVIII{significantly} with retention time. However, in planar NAND
flash memory, the optimal
voltage \chVI{for} $V_b$ does \chX{\emph{not}} change by much~\cite{cai.hpca15}, \chVIII
{indicating that retention loss is a more pressing phenomenon in 3D NAND
flash memory}. This makes
adjusting the optimal read reference voltages even more important for 3D NAND
flash memory than for planar NAND flash memory. \chVI{We conclude that it is
necessary to develop novel mechanisms to mitigate \chVII{the} early retention
loss \chVII{phenomenon} in 3D NAND flash memory.}

\subsection{Retention Interference}
\label{sec:retention:interference}

Retention interference is the phenomenon that the speed of retention loss for a
cell depends on the threshold voltage of a
\emph{\chX{vertically}-\chXI{adjacent neighbor} cell} \chVI{whose charge trap layer
is directly connected to the victim cell along the bitline}. Retention
interference is unique to 3D NAND flash memory, as cells along the
\chVI{\emph{same}} bitline in 3D NAND flash memory share the same charge trap
layer. If two neighboring cells \chVI{have} different threshold voltages \chVII{over time},
charge can leak away from the cell with a higher threshold voltage to the cell
with a lower threshold voltage~\cite{choi.svlsi16}.
\chXI{Figure~\ref{fig:retention-interference-cell} shows an example of this
phenomenon, where charge leaks from the top cell (which is in a higher-voltage
state) to the bottom cell (which is in a lower-voltage state) through
the \chXII{shared} charge trap layer.  \chXII{This} charge leakage reduces the threshold voltage of the top cell
while increasing the threshold voltage of the bottom cell.}

\begin{figure}[h]
\centering
\includegraphics[trim=0 225 635 0,clip,width=.35\linewidth]
{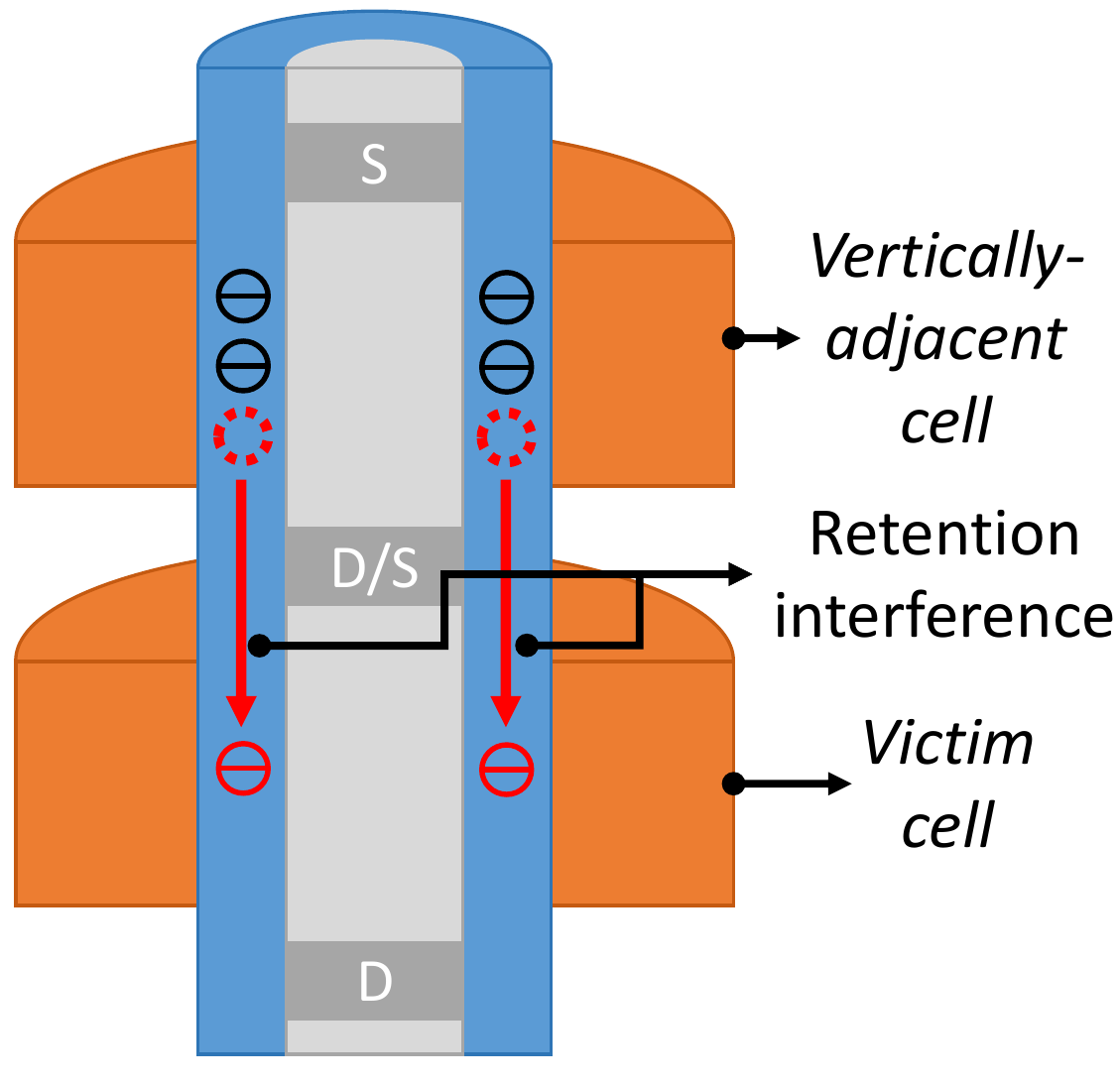}
\caption{\chXII{Retention interference phenomenon: a vertically-adjacent cell
leaks charge into a victim cell.}}
\label{fig:retention-interference-cell}
\end{figure}

We use the same data used for retention loss in Section~\ref{sec:retention} to
observe the \chX{effects} of retention interference. To eliminate any noise due to
program interference, we use \chVI{only} the neighboring cells \chVI{that are
programmed \emph{before} the victim cells} to establish the \chVI{retention}
interference correlation, as these cells \chVI{do \chVII{\emph{not}}
induce program interference on the victim cells}. We also ignore victim cells that
are in the ER state, as they are significantly affected by program interference
\chVI{even though they are programmed after their neighbors~\cite{cai.hpca17}}.
\chVI{Once} program interference \chVI{is eliminated},
the cells should experience a similar threshold voltage shift \chVII{due to
retention loss} \emph{except for}
the effects of retention interference. To find the retention interference, we
first group all \chX{of} the victim cells based on their threshold voltage states and
the states of \chX{their} neighboring cells. Then, we compare the amount by which the
threshold voltages shift over a 24-day retention time, for each group, to
observe how the cells are \chVI{affected} by \chVI{the retention interference
\chVII{caused by}} neighboring cells.

\textbf{Observations.} 
Figure~\ref{fig:retention-interference} shows the average threshold voltage
shift over \chVI{a} 24-day retention time, broken down by the state of the
victim cell (V) and the state of the neighboring cell (N). Each bar represents
a different (V, N) pair. Different shades represent the different states of the
neighboring cell, as labeled in the legend. Every 4 bars are grouped
by the state of the victim cell, as labeled on the \chVI{y-axis}. \chVI{The
length of each bar represents the amount of threshold voltage shift \chVII
{over the 24-day retention time.}}
From Figure~\ref{fig:retention-interference}, we observe that the
threshold voltage shift over retention time is lower when the neighboring
cell is in a higher-voltage state (e.g., the P3~state).

\begin{figure}[h]
\centering
\includegraphics[trim=0 285 0 0,clip,width=\figscale\linewidth]
{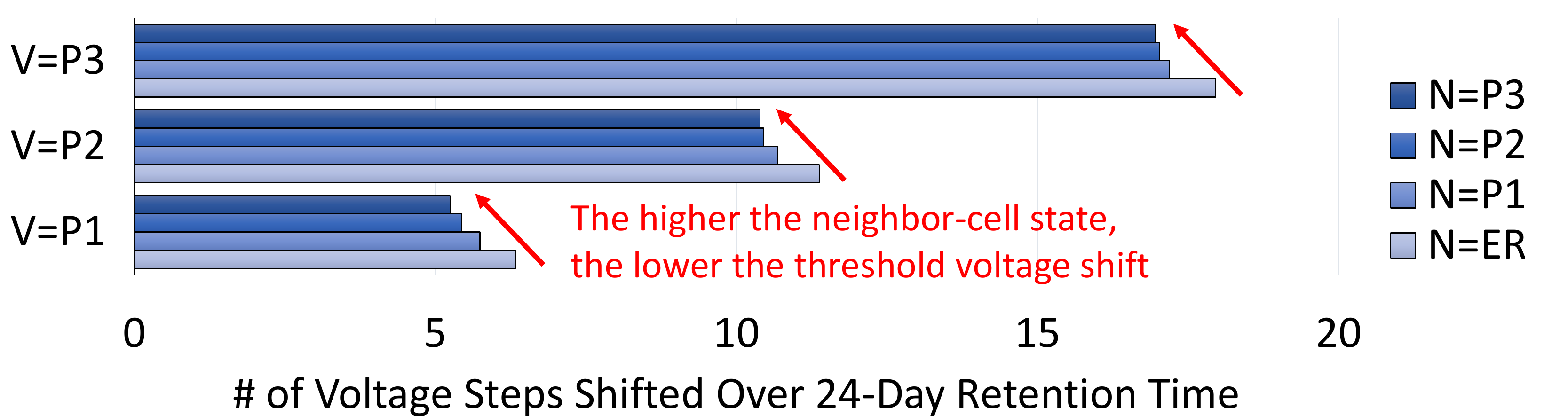}
\caption{Retention interference \chVI{phenomenon observed} at 10K P/E cycles.}
\label{fig:retention-interference}
\end{figure}

\textbf{Insights.} We are the first to \chVI{quantify the retention
interference phenomenon} in 3D NAND flash memory. Our observation from
Figure~\ref{fig:retention-interference} shows that the amount of retention
loss for a flash cell is correlated with its neighboring cell's state. We
expect retention interference to become stronger as we shrink the manufacturing
process technology \chVII{node} in future 3D NAND flash memory devices. This
is because the
distance between neighboring cells will decrease, and fewer electrons will be
stored within each flash cell, increasing the susceptibility of \chVII{a cell} to
interference from neighboring cells.


\subsection{\chVI{Other Error Characteristics}}
\label{sec:summary}

In addition to the three \chVII{\emph{new}} error sources we find in 3D NAND flash memory, we
also \chI{characterize the behavior} of other \chVII{\emph{known}} error sources in 3D NAND
flash memory \chI{and compare them to their behavior in} planar NAND flash
memory. \chI{We present a high-level summary of our findings for these errors
here, and provide} detailed results and analyses \chVI{for them} in
Appendix~\ref{sec:appendix}:

\begin{itemize}[topsep=0pt,partopsep=0pt,noitemsep,leftmargin=10pt]

\item Unlike in planar NAND \chVI{flash memory}, we do \emph{not} \chX{find any evidence of}
\emph{program errors}~\chVI{\cite{luo.jsac16, parnell.globecom14,
cai.hpca17}} in 3D NAND \chVI{flash memory}
\chI{(Section~\ref{sec:programerror})}.

\item \sph{P/E cycling error in 3D NAND flash memory follows a linear trend, which
is similar to that in planar NAND flash memory using an older
manufacturing process technology \chX{node} (e.g.,
\SIrange{20}{24}{\nano\meter})~\cite{cai.date13}. However, in
sub-\SI{20}{\nano\meter} planar NAND flash
memory, P/E cycling error exhibits a \emph{power law}
trend~\cite{parnell.globecom14, luo.jsac16} \chI{(Appendix~\ref{sec:pecycle})}.}

\item 3D NAND flash memory experiences 40\% \emph{less program interference} than
\SIrange{20}{24}{\nano\meter} planar NAND flash
memory
\chI{(Appendix~\ref{sec:interference})}.

\item 3D NAND flash memory experiences 96.7\% \emph{weaker read disturb} than
\SIrange{20}{24}{\nano\meter}
planar NAND flash memory.
The impact of read disturb is low enough
in 3D NAND flash memory that it does \emph{not} require significant error
mitigation \chI{(Appendix~\ref{sec:read:disturb})}.

\end{itemize}
\chI{Note that these differences are
mainly due to the larger manufacturing process technology \chX{nodes} currently used in 3D
NAND flash memory, and thus are not the focus of this paper.
In comparison, the new error characteristics that we focus on (layer-to-layer
process variation, early retention loss, \chX{and} retention interference) are 
caused by \chVII{the architectural and circuit-level} changes introduced
in 3D NAND flash memory.}



\begin{sidewaystable}[htbp]
\vspace{400pt}
\centering
\small
\setlength{\tabcolsep}{5pt}
\begin{tabular}{C{2.4cm}C{5cm}C{4.4cm}C{6cm}}
\toprule
\multirow{2}{\linewidth}{\centering \textbf{Attribute}} & 
\multirow{2}{\linewidth}{\centering \textbf{Observation in 3D NAND}} & 
\textbf{Cause of Difference} & 
\multirow{2}{\linewidth}{\centering \textbf{Future Trend}} \\

& & \textbf{\chVII{in 3D vs. Planar}} & \\ \midrule

\emph{Process Variation} (Section~\ref{sec:variation}, Appendix~\ref{sec:appendix:variation},
\ref{sec:bitline-variation})
  & Layer-to-layer process variation \newline is significant
  & Vertical stacking of flash cells
  & Process variation will increase \newline as we stack more cells vertically \\ \midrule

\multirow{2}{\linewidth}[-4pt]{\centering \emph{Retention Loss} (\chVII{Sections}~\ref{sec:retention},
\ref{sec:retention:interference}, Appendix~\ref{sec:appendix:retention})}
  & Early retention loss
  &\chX{Charge trap} cell
  & Early retention loss will continue \newline if \chX{charge trap} cell is used \\ \cmidrule{2-4}
  & Retention interference
  & Vertical stacking of flash cells
  & Retention interference will increase \newline when smaller process technology \chVI{node} is used \\ \midrule

\emph{P/E Cycling} (Appendix~\ref{sec:pecycle})
  & Distribution parameters change \chX{with} \newline P/E \chVI{\chX{cycle count} following a} linear trend
  instead of \chVI{a} power-law trend
  & Larger manufacturing process technology \chVI{node}
  & P/E cycle trend will go back to power-law trend when smaller process technology \chVI{node} is used \\ \midrule

  \multirow{2}{\linewidth}[-2pt]{\centering \emph{Program Interference} (Appendix~\ref{sec:interference})}
  & Wordline-to-wordline interference along \chVI{the} z-axis
  & Vertical stacking of flash cells
  & Will \chVI{continue to} \chVII{exist} in 3D NAND \\ \cmidrule{2-4}
  & 40\% lower program \chVI{interference} 
  & Larger manufacturing process technology \chVI{node}
  & Program \chVI{interference will} increase \newline when smaller process technology \chVI{node} is used \\ \midrule

  \emph{$V_{th}$ Distribution} (\chVII{Section}~\ref{sec:methodology})
  & ER and P1 states have \newline no programming errors
  & Use of one-shot programming instead of two-step programming
  & Programming errors may \chVI{start occurring} \newline if two-step programming is used \\ \midrule

  \emph{Read Disturb} (Appendix~\ref{sec:read:disturb})
  & 96.7\% smaller read disturb effect 
  & Larger manufacturing process technology \chVI{node}
  & Read disturb effect will increase \newline when smaller process technology \chVI{node} is used \\

\bottomrule
\end{tabular}
\vspace{5pt}
\caption{Summary \chVI{of error} characteristics of 3D NAND and planar NAND flash memory.}
\label{tbl:summary}
\end{sidewaystable}

\subsection{Summary}
\label{sec:summary:summary}

We summarize the key differences between 3D NAND and planar NAND flash memory, in terms of error
characteristics and the expected trends for future 3D NAND \chX{flash memory} devices, in
Table~\ref{tbl:summary}. The first column of this table lists \chX{each} attribute \chX{that} we 
study. The second column shows the key difference in the observation that we
find in 3D NAND flash memory \chVII{versus planar NAND flash memory,} \chX
{for each attribute that we study}. The
third column shows the fundamental cause of each
difference. The last column \chVII{describes} the expected trend of this
difference in
future 3D NAND flash \chX{memory} devices. \chI{We provide the necessary characterizations
and models that help us quantitatively understand these differences in
\chVII{Appendix~\ref{sec:pecycle}, \ref{sec:interference},
\ref{sec:appendix:retention}, 
\ref{sec:read:variation}, \ref{sec:read:disturb}, \chX{and} \ref{sec:appendix:variation}}.}


\section{3D NAND Flash Memory Error Models}
\label{sec:model}

In \chX{the} previous sections, we have established a basic understanding of the
similarities and differences between 3D NAND and planar NAND flash memory in terms of
\chVII{error characteristics and} reliability. In this section, we quantify these differences by developing
analytical models of the process variation (Section~\ref{sec:model:variation})
and retention loss (Section~\ref{sec:model:retention}) \chVIII{phenomena} in 3D NAND flash memory. 
\chVII{These models are useful for at least two major purposes. First,} the
insights \chVIII{obtained from using} these models \chVII{can} motivate \chVII{and enable} us to develop new error mitigation
mechanisms for 3D NAND flash memory. \chVII{Second,} the retention model and the model parameters 
are also useful for comparing the reliability of newer or older
generations of \chX{planar} NAND flash memory with our tested 3D NAND flash \chX{memory} chips.
\sph{\chX{We} focus on developing these models using our existing
characterization data \chVI{from real 3D NAND flash \chVII{memory} chips
\chVII{(some of which was presented in Section~\ref{sec:errors})}}.
In Section~\ref{sec:mitigation}, we discuss \chVII{(1)~}how to
efficiently \chVII{\emph{learn}} the \sg{models} for each chip \emph{online}
\chVIII{within the SSD controller}
by performing the characterization and model fitting \chVIII{online},
and \chVII{(2)~}how to use the online \sg{models to
develop mechanisms that improve the lifetime of 3D NAND flash memory}.}

\subsection{\chVI{RBER} Variation Model}
\label{sec:model:variation}

Since the layer-to-layer variation in 3D NAND \chX{flash memory} causes variation in RBER within a
flash block, 
\chX{it is no longer sufficient to use a single RBER value to represent the
reliability of \emph{all} pages in that block.}
Instead, we model the
variation in per-page RBER within a flash block as a gamma distribution
(i.e., $gamma(x, a, s) = \frac{x^{a-1} e^{-\frac{x}{s}}}
{\Gamma(a) s^a}$). \chV{In this model, $x$ is the RBER;
$a$ is the shape parameter, which \chX{controls} how the RBER distribution is skewed; and
$s$ is the scale parameter, which \chX{controls the width of the RBER distribution}.}

Figure~\ref{fig:rber-distribution} \chV{shows} the probability density for per-page
RBER within a block \chVIII{that has \chX{endured}} 10K P/E cycles. The bars show the measured per-page
RBERs categorized into 50 bins, and the \chX{blue and orange} curves are the fitted gamma
distributions whose parameters are shown on the legend. 
\chX{The blue bars and curve represent the measured and fitted RBER distributions
when the pages are read using the \emph{variation-agnostic $V_{opt}$}.
To find the variation-agnostic $V_{opt}$, we use techniques designed for planar NAND flash
memory to learn a single optimal read reference voltage ($V_{opt}$) for each flash block,
such that the chosen voltage minimizes the overall RBER \emph{across the entire 
block}~\cite{papandreou.glsvlsi14, luo.jsac16}.
The orange bars and curve represent the measured and fitted RBER distributions
when the pages are read using the \emph{variation-aware $V_{opt}$}, \chXI{on a
per-page basis}.
To find the variation-aware $V_{opt}$, we use techniques that are described in
Section~\ref{sec:mitigation:variation} to efficiently learn an optimal read reference
voltage \emph{for each page in the block}, such that we minimize the \chXI{\emph{per-page}}
RBER.}

\begin{figure}[h]
\centering
\includegraphics[trim=0 10 0 10,clip,width=.85\linewidth]
{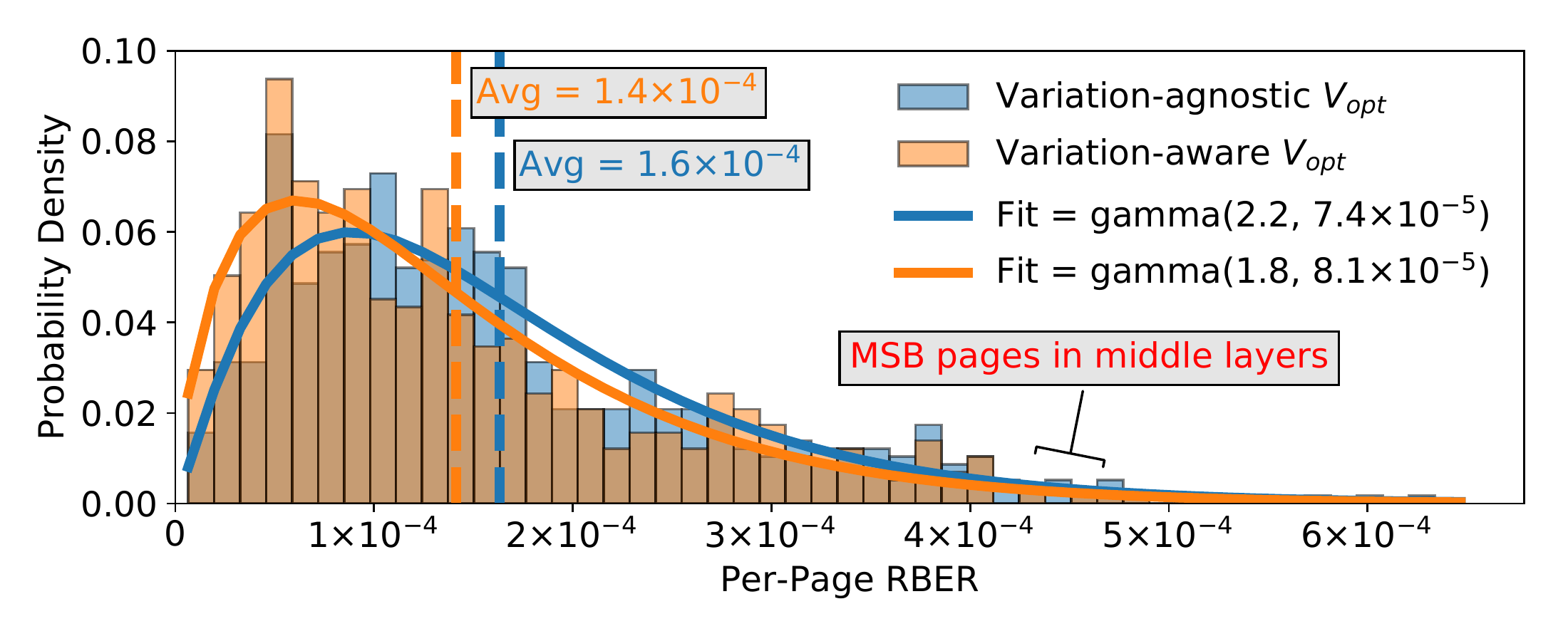}
\caption{RBER distribution \chVII{across pages} within a flash block.}
\label{fig:rber-distribution}
\end{figure}

We make \chVI{three} observations from \chX{the}
figure. \chVI{First, the gamma distribution fits well with the \chX{measured} probability
density function of RBER variation across layers: the Kullback-Leibler
divergence \chXI{error value}~\cite{kullback.math51} between the \chX
{measured} and fitted
distributions is \chVII{only} $0.09$.} Second, the average RBER
\chVII{reduces} from $1.6 \times 10^{-4}$ to
$1.4 \times 10^{-4}$ \chVII{when we \chVIII{use} the} variation-aware $V_{opt}$.
Third, \chVII{some} flash pages have a much higher RBER than the average RBER 
(e.g., $>4 \times 10^{-4}$) \chVI{even when \chVIII{we \chX{use} the}
variation-aware $V_{opt}$}.
\chVI{This large gap between the worst-case RBER and the average RBER is caused
by both layer-to-layer process variation and MSB--LSB RBER variation
(see Figure~\ref{fig:variation-wlopterr} in Section~\ref{sec:variation}).}
\chVI{The pages that have the highest RBER} are MSB pages \chVII{that} reside
in the middle layers.
\chVI{This observation indicates that there is potential to significantly improve
reliability by \chVIII{minimizing} the RBER \chX{variation across flash pages}
(for which we describe a
mechanism in Section~\ref{sec:mitigation:raid}).}

\subsection{Retention Loss Model}
\label{sec:model:retention}

\chVI{We} construct a model \chVI{to describe the early retention
loss phenomenon and its impact on RBER ($\log(RBER)$) and threshold voltage 
($V$)} in 3D NAND \chVI{flash memory}, as a function of retention time ($t$) and
\chX{the} P/E cycle \chVI{count} ($PEC$):
$\log(RBER) = A \cdot \log(t) + B$; $V = A \cdot
\log(t) + B$. \chVII{\chX{For both equations,} $A = \alpha \cdot PEC + \beta$ and
$B = \gamma \cdot PEC + \delta$}, \chX{where 
$\alpha$, $\beta$, $\gamma$, and $\delta$ are constants that
change depending on which variable we are solving for.}
\sph{We use ordinary least squares \chXI{method} implemented
in Statsmodel~\cite{seabold.statsmodels10} to fit the model
\chVI{to \chX{our} real characterization data \chX{described} in Section~\ref{sec:retention}.}
\chVI{Recall that this data is}
collected from 72 flash pages \sg{belonging to} 11 randomly-selected flash blocks.}
Following the \chVI{experimental} observations in Section~\ref{sec:retention}
and \chVI{in} prior work~\cite{mielke.irps08, luo.hpca18},
we break down our model into two \chX{parts}.
The first \chX{part ($A$)} models the retention loss at a certain P/E cycle \chVI{count} as a
logarithmic function of retention time. The second \chX{part ($B$)} models how
\chVI{the P/E cycle count changes} the parameters of retention loss.

\sph{\chV{Table~\ref{tbl:retention-model} shows all of the parameters we
use to model the RBER and the threshold voltage as a function of the 
\chVI{retention time ($t$) and the P/E cycle count ($PEC$)}. 
In this table, the first column
shows the modeled variable for each row.
\chVII{The second to fifth columns show the parameters (i.e., $\alpha$,
$\beta$, $\gamma$, and $\delta$) fitted to our model. Note that the \chX{model for the}
optimal $V_a$ does not have
$\alpha$ and $\beta$ parameters because \chXI{$V_a$} is insensitive to
retention time.}
The last column shows the \chVII{adjusted coefficient of determination 
(\chXII{\emph{adjusted $R^2$})} of our model.} \chVII{We find that \chXII{our model achieves
high adjusted $R^2$ values for} all variables except for
$\sigma_{ER}$ and $V_a$,
meaning that our model explains $>$89\% of the variation in the
characterized data. The adjusted $R^2$ values are relatively small for
$\sigma_{ER}$ and $V_a$ because these two variables do not change much \chX{with the
retention time or the P/E cycle count}.}
\chVI{\chVII{We conclude that our} model is
accurate and \chX{easy to \chXI{compute}} \chVIII{(as it can be \chX{computed using simple} linear regression)}.
\chX{Thus, our model} is suitable to use 
\chX{online in the SSD controller} (for which we will describe a mechanism in
Section~\ref{sec:mitigation:retention}).}}}

\begin{table}[h]
\centering
\begin{tabular}{ccccccc}
\toprule
\multicolumn{2}{c}{\multirow{3}{*}{\textbf{Variable}}}
	& \multicolumn{4}{c}{\textbf{Model Parameters for:}}
	& \multirow{3}{*}{\textbf{Adjusted $R^2$}} \\
 & & \multicolumn{4}{c}{$Variable = (\alpha \cdot PEC + \beta) \cdot \log(t) +
 \gamma \cdot PEC + \delta$} & \\ \cmidrule{3-6}
 & & $\alpha$ & $\beta$ & $\gamma$ & $\delta$ & \\
\midrule
MSB RBER & $\log(RBER_{MSB})$ &  $5.49\times10^{-6}$ &  0.16 & $1.33\times10^{-4}$ & -13.11 & 97.17\% \\
LSB RBER & $\log(RBER_{LSB})$ &  $7.92\times10^{-6}$ &  0.25 & $3.28\times10^{-5}$ & -12.72 & 90.05\% \\
ER Mean & $\mu_{ER}$ &  $1.01\times10^{-4}$ &  0.74 & $1.52\times10^{-3}$ & -27.27 & 96.86\% \\
P1 Mean & $\mu_{P1}$ & -$1.94\times10^{-5}$ & -0.40 & $3.51\times10^{-4}$ &  114.47 & 95.88\% \\
P2 Mean & $\mu_{P2}$ & -$4.71\times10^{-5}$ & -0.70 & $3.23\times10^{-4}$ &  189.58 & 98.50\% \\
P3 Mean & $\mu_{P3}$ & -$7.37\times10^{-5}$ & -1.20 & $5.75\times10^{-4}$ &  264.85 & 98.29\% \\
ER Stdev & $\sigma_{ER}$ &  $1.20\times10^{-5}$ & -0.10 & $1.63\times10^{-6}$ &  17.01 & 56.33\% \\
P1 Stdev & $\sigma_{P1}$ & -$1.34\times10^{-6}$ &  $9.83\times10^{-3}$ & $7.55\times10^{-5}$ &  10.20 & 93.20\% \\
P2 Stdev & $\sigma_{P2}$ & -$2.12\times10^{-6}$ &  $9.85\times10^{-3}$ & $6.69\times10^{-5}$ &  10.65 & 89.02\% \\
P3 Stdev & $\sigma_{P3}$ &  $2.87\times10^{-6}$ &  $1.40\times10^{-2}$ & $3.30\times10^{-5}$ &  10.83 & 93.00\% \\
Optimal $V_a$ & $V_a$ & --- & --- & $1.20\times10^{-3}$ &  60.52 & 71.20\% \\
Optimal $V_b$ & $V_b$ & -$3.72\times10^{-5}$ & -0.57 & $4.20\times10^{-4}$ &  150.56 & 94.27\% \\
Optimal $V_c$ & $V_c$ & -$6.51\times10^{-5}$ & -1.06 & $4.81\times10^{-4}$ &  227.24 & 97.72\% \\
\bottomrule
\end{tabular}
\vspace{5pt}
\caption{\sph{\chVI{Retention loss model for 3D NAND flash memory and its model parameters}.
\emph{PEC} is P/E cycle lifetime, \emph{t} is retention time.}}
\label{tbl:retention-model}
\end{table}


\section{3D NAND Error Mitigation Techniques}
\label{sec:mitigation}

Motivated by our new findings in Section~\ref{sec:errors},
we aim to \chI{design new techniques that} mitigate
the three unique error effects \chVI{(i.e., layer-to-layer process variation, early
retention loss, and retention interference)} in 3D NAND flash memory. We propose
four error mitigation mechanisms.  To mitigate layer-to-layer process
variation, we propose LaVAR and LI-RAID\@. LaVAR learns \chX{our new} \chXI{RBER} variation
model \chX{(see Section~\ref{sec:model:variation})} online in the SSD
controller, and uses this model to predict and apply \chVI{an} optimal read reference
voltage that is fine-tuned \chXI{to} each layer (Section~\ref{sec:mitigation:variation}).
LI-RAID is a new RAID scheme that reduces
the RBER variation induced by layer-to-layer process variation in 3D NAND flash memory
(Section~\ref{sec:mitigation:raid}). To
mitigate retention loss in 3D NAND flash memory, we propose ReMAR, a new technique that
tracks the retention time information within the SSD controller and uses \chX{our new}
retention \chVI{loss} model \chX{(see Section~\ref{sec:model:retention})}
to predict and apply the optimal read reference voltage \chX{that is} fine-tuned \chXI{to} the
retention time of the data (Section~\ref{sec:mitigation:retention}). To mitigate
retention interference, we propose ReNAC, which is adapted from \chVI{neighbor-cell
assisted correction (NAC)~\cite{cai.sigmetrics14},} an existing
technique originally designed to reduce program interference in planar NAND
\chVI{flash memory,} to
also account for retention interference in 3D NAND flash memory
\chI{(Section~\ref{sec:mitigation:nac})}.

\subsection{LaVAR\@: Layer Variation Aware Reading}
\label{sec:mitigation:variation}

In planar NAND \sg{flash memory}, existing techniques assume that \chX{the RBER} is
the \chVII{\emph{same}} across all pages within a flash \chVI{memory} block,
\chX{and, thus, a single $V_{opt}$ value can be used for all pages in the block}~\cite{cai.hpca15,
papandreou.glsvlsi14}.
\chX{This approach is called} \chVI{\emph{variation-agnostic $V_{opt}$}}.
However, as our results in Section~\ref{sec:variation}
show, this assumption no longer \chVII{holds} in 3D NAND \chVI{flash memory} \chVII{due to}
layer-to-layer process variation, \chX{as each page in a block resides in a different layer}. 
\chVI{\chVII{We} aim to improve flash memory
lifetime by mitigating layer-to-layer process variation and reducing \chX{the} RBER. 
The key idea is to identify how much the read reference voltage
must be offset by for \chVII{\emph{each}} layer in a flash chip, to account for the
layer-to-layer process variation, \chVII{instead of using a single read
reference voltage for the \emph{entire} block} \chVIII{irrespective of layers}.
When the SSD controller performs a read request, it \chX{accounts for 
\chXI{(1)~\chXII{per-block variation in RBER,} by predicting a variation-agnostic $V_{opt}$
based on the P/E cycle count of the flash block;
and (2)}~layer-to-layer variation, by adding} the layer-specific 
offset to the variation-agnostic $V_{opt}$ for \chX{the target} block.  \chX{This generates} a
\emph{variation-aware $V_{opt}$} that the controller uses as the
read reference voltage.}

\textbf{Mechanism.} \chVIII{We devise a new
mechanism called \chX{\emph{Layer Variation Aware Reading} (LaVAR)},
which (1)~learns the voltage offsets for each layer and records them in
per-chip tables in the \chX{SSD} controller, and (2)~uses the variation-aware
$V_{opt}$ during a read operation
by reading the appropriate voltage offset for the request from the
\chX{per-chip table that corresponds to the layer} of the request.}
\chVI{LaVAR constructs}
\chV{a model of the optimal read reference voltage ($V_{opt}$)
variation across different layers. Since there are only a limited number of layers,
this model can be represented as a table (i.e., it is a non-parametric model)
of the offset between the $V_{opt}$ \chVII{for} each layer 
(\emph{variation-aware
$V_{opt}$}) and the overall $V_{opt}$ for the entire flash block
(\emph{variation-agnostic $V_{opt}$}). Any previously-proposed model for
$V_{opt}$\chVII{~\cite{luo.jsac16, papandreou.glsvlsi14, cai.hpca15}} can \sg{be used to calculate the} variation-agnostic
$V_{opt}$.}
Since the \chVI{layer-to-layer} process variation is similar across blocks and
is consistent across P/E \chX{cycle counts}, the $V_{opt}$ variation model can be learned 
\emph{offline} for each chip
through an extensive characterization of a single flash block. To do this, the
SSD controller randomly picks a flash block and records the difference between
the variation-aware $V_{opt}$ and the variation-agnostic $V_{opt}$. 
\sph{\chVI{LaVAR uses the existing read-retry functionality in modern NAND flash
\chX{memory} chips (see Section~\ref{sec:methodology}) to find the 
variation-aware $V_{opt}$ online.}} The
controller then computes and stores the average $V_{opt}$ offset for each
layer in a lookup table \chVIII{stored for each chip}. Note that $V_c$
variation does not need to be modeled,
since $V_c$ is unaffected by layer-to-layer process variation \chVI{(see
Figure~\ref{fig:variation-wloptvrefs} \chVII{in} Section~\ref{sec:variation})}.

\chVI{When performing a read operation,} the SSD controller
simply looks up the $V_{opt}$ \chVI{offset} 
\chVI{that corresponds to the layer \chVIII{and the chip} that contains the
data being read,}
and \chVI{adds} the offset to the per-block $V_{opt}$ predicted by existing
techniques~\cite{luo.jsac16, papandreou.glsvlsi14, cai.hpca15}. By \chVII{using}
variation-aware $V_{opt}$, LaVAR \chVI{enables the use of} a more accurate
$V_{opt}$ for 3D NAND \chVI{flash memory} than existing techniques, and \chVII
{thus} reduces the RBER (see Figure~\ref{fig:rber-distribution} in
Section~\ref{sec:model:variation}).

\textbf{Overhead.} \sph{\chV{\chVI{LaVAR} can be implemented fully in the SSD
controller firmware, and, thus, does not require any modification to the
hardware.} Assuming that the 3D NAND \chX{flash memory chip} has $N$~layers and that it takes
\sg{1~Byte to store each} $V_{opt}$ \chVI{offset for each layer}, the memory
overhead of \chVI{storing} the lookup table \chVI{for $V_a$ and $V_b$} in the
SSD controller is $2N$~Bytes. The latency overhead of each read operation is
negligible as LaVAR requires \chVII{only} a table lookup and an addition to obtain
variation-aware $V_{opt}$, which take less than \SI{100}{\nano\second}. Since the lookup table
is shared across \chVIII{\emph{all} blocks in a chip}, it needs to be learned
\chVII{only} once, and \chX{it} can be
\chVI{constructed} gradually in the background. \chVI{Thus,} the performance
overhead \chVI{of LaVAR} is negligible.}

\textbf{Evaluation.} Figure~\ref{fig:vopt-variation} compares the RBER \chVIII
{obtained} by \chVIII{using} LaVAR (\emph{variation-aware
$V_{opt}$})\chVII{~\cite{luo.jsac16, papandreou.glsvlsi14, cai.hpca15}} to
that \chVIII{obtained} by \chVIII{using} an existing read
reference voltage tuning technique (\emph{variation-agnostic $V_{opt}$})
designed for planar NAND \chX{flash memory}.
We evaluate the average RBER \chVIII{obtained} by each
mechanism by simulating \chX{read operations} using our characterization data in
Section~\ref{sec:variation}. \chXI{Averaged} \chVI{across all P/E cycle counts},
LaVAR reduces \chX{the} RBER by 43.3\%.
The benefit comes from tuning the read reference voltage towards the
\chVI{variation-aware $V_{opt}$} by an offset learned by our model. \chVI{The
RBER} reduction becomes smaller as \chX{the} P/E cycle \chVI{count} increases, because the
overall RBER increases exponentially as the \chX{NAND} flash memory wears out, decreasing
the fraction of process variation errors. \sph{\chV{While the flash lifetime
improvements \chVII{produced} by LaVAR might \chVIII{seem} small (as we show
in Section~\ref{sec:mitigation:all}), 
(1)~they are achieved with negligible overhead, and
(2)~the RBER reduction enabled by LaVAR throughout
the flash memory lifetime reduces the average flash read latency~\cite{cai.hpca15}.
As the number of layers within
a 3D NAND flash memory \chX{chip} grows (e.g., \chX{vendors}
are already bringing chips with 96~layers to the market~\cite{anandtech.web17}),
we expect that layer-to-layer process variation will increase,
which in turn will increase the magnitude of the lifetime benefits
provided by LaVAR.}}

\begin{figure}[h]
\centering
\includegraphics[trim=0 0 0 0,clip,width=.65\linewidth]
{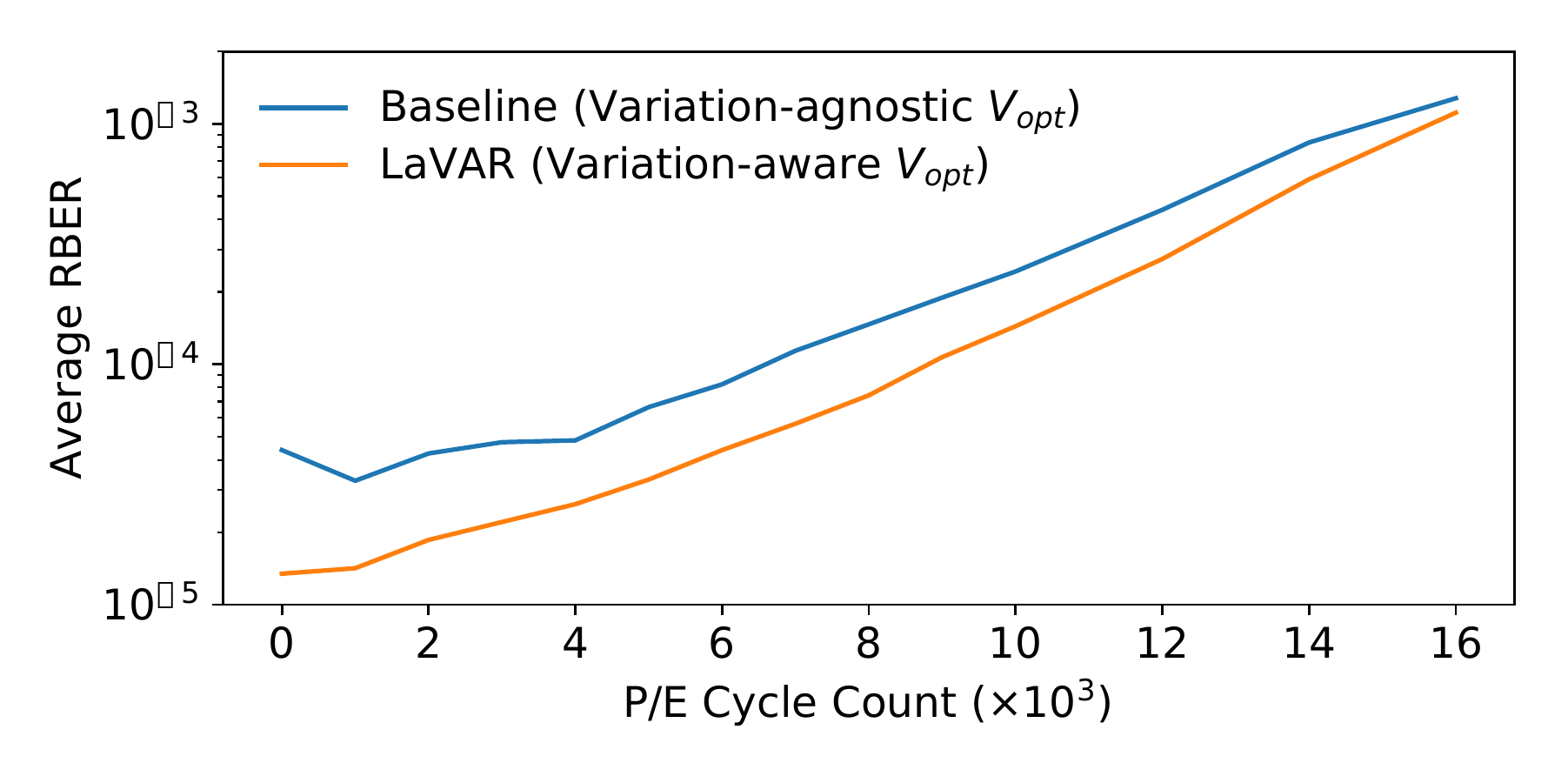}
\vspace{-3mm}
\caption{RBER reduction using LaVAR.}
\label{fig:vopt-variation}
\end{figure}

\subsection{LI-RAID\@: Layer-Interleaved RAID}
\label{sec:mitigation:raid}

As we \chVI{observe} in Section~\ref{sec:model:variation}, even
after applying the variation-aware $V_{opt}$, \chVI{the per-page RBER is
distributed over a wide range according to a \chVIII{fitted} gamma
distribution}
due to
layer-to-layer process variation and MSB--LSB RBER variation
(see Figure~\ref{fig:variation-wlopterr} in Section~\ref{sec:variation}). In
enterprise SSDs, \chVI{in addition to ECC, the Redundant Array of Independent Disks
(RAID)~\cite{patterson.sigmod88, balakrishnan.tos10} error recovery
technique} is \chVII{used} across \chVIII{\emph{multiple}} flash chips to tolerate
chip-to-chip process variation \chVIII{in error rates}. \chVI{RAID in modern
SSDs typically combines one flash page from
each flash chip into a logical unit called a \emph{RAID group}, and uses one of the
pages to store the parity information for the entire group.}
However, state-of-the-art RAID schemes do \chVI{\emph{not}} consider
layer-to-layer process variation and MSB--LSB RBER variation.  These schemes
group MSB \chI{or} LSB pages in the \chVI{\emph{same}} layer together \chI{in
a RAID group}. As a
result, the reliability of the SSD is limited by the RBER of the weakest 
(i.e., the least reliable) RAID group that contains the MSB or LSB pages from
the least reliable layer across all chips. We devise a new RAID scheme called
\chX{\emph{Layer-Interleaved RAID} (LI-RAID)}, \chVI{which} eliminates these low-reliability RAID
groups by equalizing the RBER among different RAID groups. \chVI{LI-RAID makes
use of two key ideas: (1)~group flash pages in less reliable layers
with pages in more reliable layers, and (2)~group MSB pages with LSB pages.}

\textbf{Mechanism.}
Instead of
grouping pages in the same layer together \chVI{in the same RAID group}, we
select pages from different chips
and different layers and group them together, such that \chVI{the low-reliability
pages (either due to layer-to-layer process variation or MSB--LSB RBER
variation) are \emph{distributed} to different} RAID groups. Thus,
the new groups formed by LI-RAID have a more evenly-distributed \chVII{RBER}
than the groups formed using
traditional \chVI{layer-unaware} RAID schemes. We assume, without loss of
generality, that there
are $m$ chips in the SSD, and each RAID
group contains $m$ pages, \chI{one from each chip}. We also assume that each block
contains $n$ wordlines, and that the layer numbers of each wordline are in ascending
order  (e.g., the wordline in layer~$i$ has a lower wordline number than its
neighboring wordline in layer~$i+1$). Thus, LI-RAID groups together
the MSB page of wordline~$0$, the LSB page of wordline~$\frac{n}{m}$, the MSB
page of wordline~$2\cdot\frac{n}{m}$, \chVI{the LSB page} \ldots, \chVII
{the MSB page of wordline~$(m-2)\cdot \frac{n}{m}$}, the \chVI{LSB} page
of wordline~$(m-1)\cdot \frac{n}{m}$.
\chI{Figure~\ref{fig:li-raid} shows an example LI-RAID
layout on an SSD with 4 chips and with 4 wordlines within each flash block.
Flash pages in the same RAID group are highlighted in the same color.}
In this way, LI-RAID \chVI{distributes} the less reliable
pages within each chip across different RAID groups, \chVIII{thereby avoiding}
the \chVIII{formation of significantly less reliable} RAID groups that \chVI
{bottleneck} SSD reliability.

\begin{figure}[h]
\centering

\begin{tabular}{ccccccc}
\toprule
\textbf{Wordline \#} & \textbf{Layer \#} & \textbf{Page} & \textbf{Chip 0} & \textbf{Chip 1} & \textbf{Chip 2} & \textbf{Chip 3} \\
\midrule
0 & 0 & MSB & \cellcolor{olive!25}Group 0 & Blank & \cellcolor{red!25}Group 4 & \cellcolor{green!25}Group 3 \\
0 & 0 & LSB & \cellcolor{blue!25}Group 1 & Blank & \cellcolor{cyan!25}Group 5 & \cellcolor{gray!25}Group 2 \\
1 & 1 & MSB & \cellcolor{gray!25}Group 2 & \cellcolor{blue!25}Group 1 & Blank & \cellcolor{cyan!25}Group 5 \\
1 & 1 & LSB & \cellcolor{green!25}Group 3 & \cellcolor{olive!25}Group 0 & Blank & \cellcolor{red!25}Group 4 \\
2 & 2 & MSB & \cellcolor{red!25}Group 4 & \cellcolor{green!25}Group 3 & \cellcolor{olive!25}Group 0 & Blank \\
2 & 2 & LSB & \cellcolor{cyan!25}Group 5 & \cellcolor{gray!25}Group 2 & \cellcolor{blue!25}Group 1 & Blank \\
3 & 3 & MSB & Blank & \cellcolor{cyan!25}Group 5 & \cellcolor{gray!25}Group 2 & \cellcolor{blue!25}Group 1 \\
3 & 3 & LSB & Blank & \cellcolor{red!25}Group 4 & \cellcolor{green!25}Group 3 & \cellcolor{olive!25}Group 0 \\
\bottomrule
\end{tabular}

\caption{LI-RAID layout example for an SSD with 4 chips and with 4 wordlines
in each flash block.}
\label{fig:li-raid}
\end{figure}

\sph{\chI{Note that, \chVI{since \chVII{the} order of RAID group number is different
in each flash chip}, the LI-RAID layout may potentially violate the
program sequence recommended by flash vendors,
where wordlines within each flash block
must be programmed \emph{in order} to minimize harmful program
interference~\cite{cai.sigmetrics14, cai.iccd13,
park.dac16, cai.procieee17}.
For example, in Chip~2 in Figure~\ref{fig:li-raid},
\chVIII{Wordline~3 (\chX{Groups 2 and 3}) is programmed \emph{after}
Wordline~2 (\chX{Groups 0 and 1}).
In Chip~2, we leave \chX{Wordline~1} \emph{blank} 
(marked as``Blank'' \chX{in Figure~\ref{fig:li-raid}}).  \chX{Otherwise, 
Wordline~1 would} cause
program interference to \chX{the data in Wordline~2,
which already experiences program interference when Wordline~3
is programmed, significantly increasing the error rate of
Wordline~1~\cite{cai.iccd13, cai.sigmetrics14} (see 
Appendix~\ref{sec:interference}).}}
\chX{By laying out the data in the proposed manner}, 
LI-RAID provides the same reliability guarantee as
the recommended program sequence, by guaranteeing that any data stored in a
flash page experiences program interference from \chVII{\emph{at most}} one
neighboring wordline.}}

\chI{\textbf{Overhead.} \sph{\chV{The grouping of flash pages by LI-RAID is
implemented entirely in the SSD controller firmware. This requires the
firmware to be aware of the physical-page-to-layer \chVI{mapping.}}
The \chX{flash pages} left blank in LI-RAID incur a small
additional storage overhead compared to a conventional RAID scheme. Only
\emph{one} wordline \chX{(\chX{i.e., two pages in MLC NAND flash memory)} 
within a flash block is left blank, \chX{to mitigate the
impact of program interference on Groups 0 and 1}. 
\chX{Without this blank wordline, the data in Groups 0 and 1 would be the
only data to experience program interference twice: once when Groups 2 and 3
are programmed, and once when the last two groups are programmed.}
In modern NAND flash
memory, each flash block typically contains \chXI{at least 256} flash pages. Thus,
the additional storage overhead \chX{for the blank pages} is less than \chVII{0.8\%}.
LI-RAID does not incur additional computational overhead because it computes
parity in the same way as a conventional RAID \chVI{scheme}, and only
\chVI{\emph{reorganizes}} the RAID
groups differently. Because we do \chVI{\emph{not}} change the data layout across flash
blocks, the flash translation layer (FTL) and the garbage collection (GC)
\chX{algorithms} remain the same as in a conventional RAID scheme.}}

\textbf{Evaluation.} Figure~\ref{fig:99p-rber} plots \chVI{the \emph{worst-case
RBER} (i.e., the highest per-page RBER within a flash block)} when \chVIII{we
use} different error mitigation techniques at 10,000
P/E cycles. \chVI{Recall that the per-page RBER within a flash block follows
a gamma distribution (see Figure~\ref{fig:rber-distribution} in
Section~\ref{sec:model:variation}).}
Thus, several \chVI{least-reliable} flash pages within a block may become
unusable (i.e., their RBER exceeds the ECC correction capability)
before the \emph{overall} RBER of the flash chip
exceeds the ECC correction capability. We use the worst-case RBER to represent the reliability
of these \chVI{least-reliable} flash pages. In this figure, the baseline
\chVIII{uses} the
per-block variation-agnostic optimal read reference voltage (i.e.,
variation-agnostic $V_{opt}$), achieving a worst-case RBER of $4.8\cdot10^{-4}$. When
we \chVIII{use} the variation-aware $V_{opt}$ proposed in
Section~\ref{sec:mitigation:variation}, the worst-case RBER is reduced by 9.6\% over the
baseline, to $4.3\cdot10^{-4}$. LI-RAID reduces the
worst-case RBER by 66.9\% over the baseline, to only $1.6\cdot10^{-4}$. Thus, by
grouping flash pages on less reliable layers with pages on more reliable
layers, and by grouping MSB pages with LSB pages, LI-RAID
reduces the probability of unusable pages within a block, \chVI{thereby
\chVIII{reducing}} the
number of retired flash blocks due to ECC failures.

\begin{figure}[h]
\centering
\includegraphics[trim=0 390 295 0,clip,width=\figscale\linewidth]
{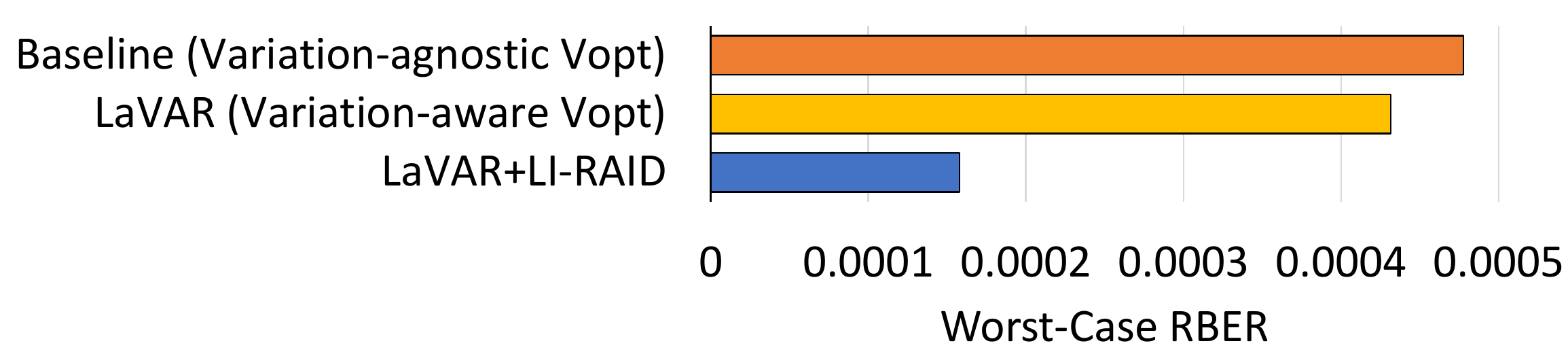}
\caption{\chVII{Effect of LaVAR and LI-RAID on} \chVI{worst-case} RBER at
10,000 P/E cycles.}
\label{fig:99p-rber}
\end{figure}

\sph{\chVI{Note that LaVAR and LI-RAID do
\chVI{\emph{not}} rely on whether the RBER variation is consistent across all chips. LaVAR
learns a different lookup table for each chip.
So, even if there is some chip-to-chip process variation that is present,
our models are effective at dynamically capturing the behavior
of \emph{any} \chVIII{NAND flash memory chips. Conventional RAID tolerates
only chip-to-chip process variation. LI-RAID improves
flash reliability over conventional RAID by eliminating the strong correlation
between RBER and layer number, which we show in
Figure~\ref{fig:variation-wlopterr}. We conclude that both LaVAR and LI-RAID
are effective \chX{at reducing the impact of layer-to-layer variation
on the RBER}.}}}

\subsection{ReMAR\@: Retention Model Aware Reading}
\label{sec:mitigation:retention}

As we show in Section~\ref{sec:retention}, due to early
retention loss, retention errors increase much faster \chVI{after programming
a page} in 3D NAND \chVI{flash memory} than \chX{they do} in
planar NAND \chVI{flash memory}. Thus, mitigating retention errors has become
more important in
3D NAND than in planar NAND \chX{flash memory}, as \chX{the errors have a} greater impact on SSD
reliability.
However, as we show in our model in Section~\ref{sec:model:retention}, \chVIII
{the RBER impact of} early
retention loss is proportional to the logarithm of \chVIII{retention} time.
This means that \chX{a large}
majority of the retention errors and threshold voltage shifts happen \chX{\emph{shortly}}
after programming. As a result, traditional retention \chVI{error} mitigation
techniques
developed for planar NAND \chX{flash memory, which are optimized for
\chXI{much} larger retention times,} may become less effective on 3D NAND
\chVIII{flash
memory}. For example,
\chVI{Flash Correct-and-Refresh (FCR)}~\chVII{\cite{cai.iccd12, cai.itj13}},
a mechanism that remaps all data
periodically, allows planar NAND to tolerate $50\times$ more P/E cycles with
\chVII{a}
3-day refresh period. \sg{However,} \chVI{according to our evaluations, the
P/E cycle lifetime improvement of FCR} reduces to \chVI{\emph{only}}
$2.7\times$ for 3D NAND \chX{flash memory}
due to \chVI{the} early retention loss \chVI{phenomenon}. This motivates us to
explore new ways to mitigate
retention errors in 3D NAND \chVI{flash memory}.

\textbf{Mechanism.} 
\sg{We propose a new mechanism called \chX{\emph{Retention Model Aware Reading} (ReMAR)},
whose key idea} is to accurately track the retention
time of the data and apply the optimal read reference voltage predicted by our
model in Section~\ref{sec:model:retention}.
First, ReMAR \chVI{constructs} the same linear models proposed in
Section~\ref{sec:model:retention} \emph{online} to accurately predict the
optimal $V_a$, $V_b$, and $V_c$. Similar to the distribution parameter model
used in Section~\ref{sec:model:retention}, we model the optimal $V_b$ and $V_c$
as: \chVIII{$V = (\alpha \cdot PEC + \beta) \cdot \log(t) + \gamma \cdot PEC
+ \delta$}. We model the
optimal $V_a$ as: \chVIII{$V_a = \gamma \cdot PEC + \delta$}, since $V_a$ is
\chVI{\emph{not}} affected by
retention time \chX{(as we show empirically in Section~\ref{sec:retention})}. 
To construct this model \emph{online}, the
controller randomly selects a \chVI{flash block and} records the
optimal read reference voltage of the block \chX{(which the controller learns} by sweeping the read
reference voltages\chVII{,} \chVI{as done} in prior work~\cite{cai.hpca15}), \chVI
{along with the block's P/E cycle count} ($PEC$) and retention time ($t$). Over
time,
these data samples \chVI{would} cover a range of P/E cycle \chVI{counts} and
retention times.\footnote{The SSD controller can also perform \chVI
{additional} characterization if \chVI{a}
certain data range is missing.} Note that as the P/E cycle count of the SSD
increases, the accuracy of the model increases, because more data samples
are collected. Once this online model is constructed, it is used in the
controller to predict the optimal read reference voltage \chVI{to be used for}
each read
operation. To do this, the SSD controller stores the P/E cycle \chX{count} and
the program time of each block as metadata. During each read operation, the
controller computes the retention time for each read by subtracting the program
time from the read time. Using the recorded P/E cycle \chVI{count} and the
computed
retention time of the data, ReMAR applies the online model to predict $V_a$,
$V_b$, and $V_c$. By accurately predicting and applying the optimal read
reference voltages, ReMAR \chVII{increases} the accuracy of read
operations and
thereby \chVI{decreases} the raw bit error rate.

\textbf{Overhead.} \sph{\chV{Like LaVAR, \chVI{ReMAR is}
implemented fully in the SSD controller firmware, and \chVI{does}
\emph{not} require any modifications to the hardware.}
\chVI{Assuming that the flash block size is \SI{5}{\mega\byte}, and \chX{that} \chVII{ReMAR}
stores the program time in the UNIX Epoch time format~\cite{matthew.book08},
which takes up \SI{4}{\byte},
the} memory and storage overhead of \chVII{ReMAR} is 800KB for a 1TB SSD.
The performance overhead of each read operation is small, \chVI{as ReMAR needs \chX{only}
a few dozen CPU cycles (on the order of \SI{100}{\nano\second} in total)} in the
\chX{SSD} controller to compute $V_{opt}$, which is negligible
compared to flash read latency (on the order of \SI{10}{\micro\second}). The performance
overhead of learning the model can be hidden by \chVI{\chX{(1)~}performing \chX{learning
in} the background and \chX{(2)~}deprioritizing the requests issued for
characterization} \chVIII{purposes}.}

\sph{The controller uses the UNIX Epoch time format~\cite{matthew.book08}
for program and read times, such that
the recorded time is valid after reboot. To do this, the controller needs
a real-time clock to keep track of the current time. Without a power source on
the SSD, the controller needs a special command to synchronize the current
time with the host when it boots up. The program time of each block is \chVI{stored}
in the memory of the controller, along with other metadata that already
exists such as the logical address map and the P/E cycle \chX{count} of each block.}

\textbf{Evaluation.} Figure~\ref{fig:vopt-retention} compares the RBER
achieved by ReMAR to \chX{that of} the state-of-the-art
read reference voltage tuning technique~\chVIII{\cite{luo.jsac16}}
\chVI{designed for planar NAND \chX{flash memory}
(\emph{Baseline})}. The results are based on the
characterization data in Section~\ref{sec:retention}.
We assume that the average retention time of the
data is 24~days. \chVI{The \emph{Baseline} technique is unaware of the 
retention time. Thus, \emph{Baseline}
uses a \emph{retention-agnostic} $V_{opt}$ based on only the P/E cycle
count of the flash page. ReMAR uses a \emph{retention-aware} $V_{opt}$ based
on both the P/E cycle count and the retention time of the flash page.}
On average \chVI{across all P/E cycle counts}, ReMAR reduces \chVI{the} RBER by 51.9\%.
As \chVI{the} P/E cycle \chVI{count} increases, the benefit of ReMAR \chVI{(i.e., the
RBER improvement of ReMAR over \emph{Baseline}) also increases}. \chVII{We
conclude that,} by accurately tracking retention time, and by
using our \chVI{retention loss} model, ReMAR \chVII{accurately adapts the
read reference voltage} to the threshold
voltage \chVII{shifts} \chVIII{that occur} due to retention loss, and \chVI
{hence} \chVII{it effectively reduces the} RBER\@.

\begin{figure}[h]
\centering
\includegraphics[trim=0 10 0 0,clip,width=.65\linewidth]
{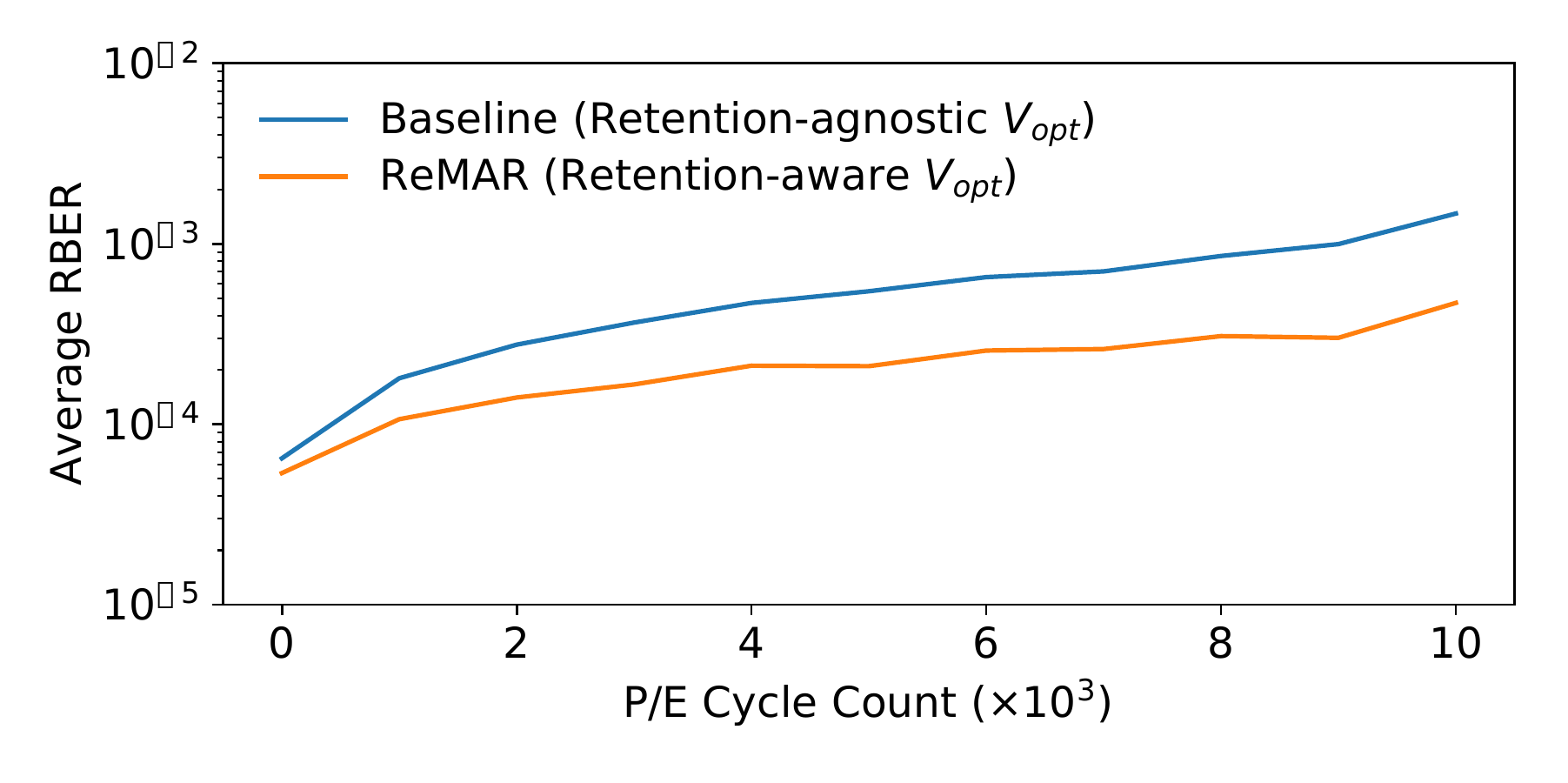}
\caption{RBER reduction using ReMAR.}
\label{fig:vopt-retention}
\end{figure}

\subsection{ReNAC\@: Retention Interference Aware Neighbor-Cell Assisted Correction}
\label{sec:mitigation:nac}

As we observe in Section~\ref{sec:retention:interference}, due
to retention interference, the amount of threshold voltage shift of a
victim cell during a certain amount of retention time is affected by
the value stored in \chXII{a \chVI{\chX{vertically-\chXI{adjacent neighbor}}
cell}}. This \chVII{phenomenon}
presents a similar
data dependency as that induced by program interference, where the amount
of \chVII{the} threshold voltage shift of a victim cell during programming
operation also depends on the value stored in the \chX{directly-neighboring
cells}~\cite{cai.iccd13, cai.sigmetrics14}.
To mitigate program interference errors, \chV{prior
work proposes neighbor-cell assisted correction
(NAC)~\cite{cai.sigmetrics14}. \chVI{The goal of NAC is to reduce \chVII{the}
raw bit
error rate by reading each cell at the read reference voltage
optimized for the amount of program interference induced by its \chVIII{\chX{directly-}neighboring
cells}.} \sg{\chVI{To achieve this goal,} after error correction fails on a
flash page, NAC \chVI{reads the data stored in the \chVII{\emph{neighboring}} wordline and}
re-reads the \chVI{failed} page using a set of
read reference voltage values that \chVII{are adjusted based on} the data
\chVII{values} stored in \chX{the directly-}neighboring \chVII{cells}~\cite{cai.sigmetrics14}.}} However,
this mechanism does \chVI{\emph{not}}
account for
\chXI{\emph{retention interference}} \chX{induced by the neighboring cells}, which is new in 3D
NAND flash memory. \chVII{We} adapt NAC
for 3D NAND flash memory to account for the new retention interference
\chX{phenomenon, and call this adapted mechanism
\emph{Retention Interference Aware Neighbor-Cell Assisted Correction}
(ReNAC)}.

\textbf{Mechanism.} The key idea of ReNAC is to use the data stored in 
\chX{a vertically-adjacent neighbor cell} to predict the amount of retention interference on
a victim cell. Using similar techniques from
Section~\ref{sec:model:retention}, ReNAC first develops an online model of
retention interference as a function of the retention time and the neighbor
cell's state. The SSD controller obtains the retention time of each block using a
mechanism \chVII{similar} to ReMAR, and computes and applies the
neighbor-cell-dependent read offset at that retention time from the model.
\sph{\chV{\chVI{For ReNAC, we} are \chVIII{currently} unable to show
any meaningful improvements in flash lifetime for the current generation of 3D
NAND flash memory, because retention interference shifts the threshold
voltage by \chVII{only} less than two \chX{voltage} steps (Figure~\ref{fig:retention-interference}),
which is much smaller than the voltage \chX{changes} due to process variation
(Figure~\ref{fig:variation-wloptvrefs}) and early retention loss 
(Figure~\ref{fig:retention-optvrefs}). However, we expect that retention
interference will increase in future 3D NAND flash memory devices due to \chX{decreasing} cell
sizes and \chX{decreasing} distances between neighboring cells
\chVII{(Table~\ref{tbl:summary}), which, in turn, will likely increase} the
benefit of \chVIII{using} ReNAC.} \chXI{We also expect ReNAC to have a
relatively larger benefit in 3D NAND flash memory chips that use
\emph{triple-level cell} (TLC) or \emph{quadruple-level cell} (QLC)
\chXII{technologies.  A TLC or QLC NAND flash memory chip stores more bits in a cell than 
an MLC NAND flash memory chip}, by splitting up the same voltage range into a greater
number of states (eight for TLC and sixteen for QLC).  \chXII{Doing so} reduces the
voltage margin between neighboring threshold voltage distributions.
\chXII{Therefore}, shifting the read reference voltage by two voltage
steps may affect more cells \chXII{in TLC} and QLC 3D NAND flash memory than in
MLC 3D NAND flash memory, and, thus, \chXII{ReNAC can reduce a greater number of} 
raw bit errors in future TLC or QLC NAND flash memory.}
We leave a quantitative evaluation of ReNAC
\chVIII{on future \chX{3D} NAND flash memory chips to} future work.}

\subsection{\chVI{Putting It All Together: Effect on System Reliability \chXI
{and Performance}}}
\label{sec:mitigation:all}

\chI{The \chVIII{mechanisms we} propose in this section can be combined together to
achieve significant reductions in average and worst-case RBER\@.}
\chVI{For a consumer-class 3D NAND flash memory device, these reductions
improve \emph{flash memory lifetime}, i.e., the device can tolerate more
P/E cycles before failing. For \chVII{an} enterprise-class device which is expected to
be replaced after a fixed amount of time, these reductions improve the
sustainable workload write intensity \chVIII{or} reduce the ECC storage
overhead.} \chVI{We \chVIII{evaluate these} potential effects of our mechanisms on
storage system \chVIII{reliability and performance}.}

\textbf{Flash Lifetime \chXI{(or Performance)} Improvement.}
\chI{In Figure~\ref{fig:this-work-rber}, we compare and contrast the
reliability \chVII{(i.e., the RBER)} of \chV{five} example SSDs:
(1)~\emph{Baseline}, an SSD that uses a
fixed, default read reference voltage and \chX{employs} a conventional RAID scheme;
(2)~\emph{State-of-the-art}, an SSD that uses the optimal read reference voltage
predicted by existing mechanisms designed for planar
NAND \chVI{flash memory}~\cite{luo.jsac16, parnell.globecom14, cai.hpca15,
papandreou.glsvlsi14}
and \chX{employs} a conventional RAID scheme; \sph{\chV{(3)~\emph{LaVAR}, an SSD that uses
the optimal read reference voltage for each layer predicted by LaVAR in
addition to \emph{State-of-the-art}; (4)~\emph{LaVAR+LI-RAID}, an SSD that
\chX{uses} the LI-RAID scheme in addition to \emph{LaVAR};} and (5)~\emph{This
Work} \chXI{(LaVAR + LI-RAID + ReMAR)}, an SSD
that uses the optimal read reference voltage predicted by LaVAR and ReMAR, and
\chX{also employs} the
LI-RAID scheme.} In this figure, we plot the \chVII{\emph{worst-case RBER}}
\chVI{(i.e., the highest per-page RBER within a flash block)} instead of the
average RBER, because the worst-case RBER \chX{\emph{limits}} \chVIII{the} flash memory
lifetime. Because
RBER increases \sg{with} P/E cycle count, if the worst-case RAID group has a high enough
\chVIII{worst-case} RBER, NAND flash memory can no longer guarantee reliable
operation.}

\begin{figure}[h]
\centering
\includegraphics[trim=0 290 420 0,clip,width=\figscale\linewidth]
{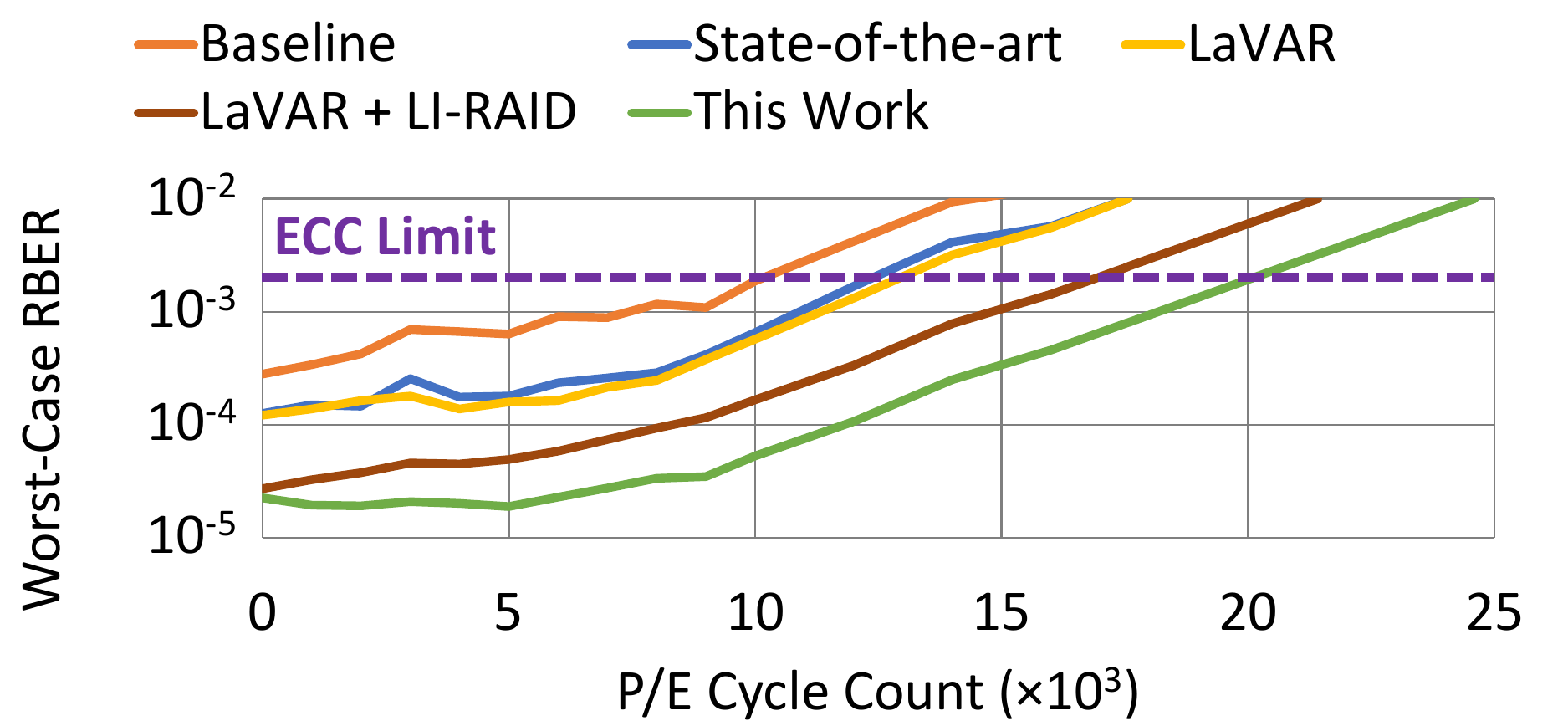}
\caption{\sph{\chVII{Effect of} \chVI{LaVAR, LI-RAID, and ReMAR} \chVII{on
worst-case RBER experienced by any flash block}.}}
\label{fig:this-work-rber}
\end{figure}

\chI{Assuming that the ECC deployed on the SSD can correct \chX{errors up to 
an RBER of $3\cdot10^{-3}$~\cite{cai.hpca15, cai.procieee17}} 
(i.e., the \chXI{\emph{ECC limit}}, \chVI{shown as} \chX{a} purple \chX{dashed} line in
Figure~\ref{fig:this-work-rber}), we can calculate the lifetime of \chVIII
{each SSD we evaluate}.\footnote{\chV{Note that we are \emph{unable} to 
\emph{directly}
measure the flash lifetime improvements on real devices, because manufacturers
do \emph{not} provide us with the ability to modify the SSD firmware
directly, which prevents us from
evaluating our techniques on the real devices themselves.
Unfortunately, we also do \emph{not} have the \chVI{resources} to measure the
lifetime of a large number of real flash chips by emulating the behavior of our mechanisms,
as this would require many additional months to years of effort.}
\chV{Instead, \sg{we follow the precedent of prior work to} evaluate the flash memory
lifetime based on real \chVI{RBER characterization data we obtain from the
testing of real flash memory devices}.}} In our
evaluations, the flash memory lifetime ends when the \chVI{worst-case RBER
exceeds} the ECC limit. We find that \emph{State-of-the-art}, \chV{\emph{LaVAR},
\emph{LaVAR+LI-RAID}, and \emph{This Work} \chVII{improve} flash memory lifetime by
23.8\%, 25.3\%, 57.2\%, and 85.0\%, respectively,} over the
\emph{Baseline}. When the SSD is \chX{used} in a server, which has a fixed device
lifetime, the server has to throttle the write frequency to a certain 
\emph{drive writes per day} (DWPD) to ensure \chVIII{that} the SSD can operate
\chVIII{reliably}
during the fixed lifetime. In this case, \chX{our combined mechanisms (\emph{This Work})
increase the maximum write frequency (i.e., the maximum DWPD) of 
the SSDs in a server by 85.0\%.}
\chVII{Thus, our mechanisms either \chVIII{improve}
lifetime or \chVIII{improve} performance under a fixed lifetime.}

\chVI{\textbf{ECC Storage Overhead Reduction.}}
\chI{In modern SSDs, the storage overhead for error \chVI{correction
increases} in each generation to better tolerate the degraded flash
reliability due to aggressive scaling. For example, to tolerate an RBER of up to
$3\cdot10^{-3}$ for the \emph{Baseline} SSD at the end of its lifetime, a
modern BCH code~\cite{hocquenghem.chiffres59} requires 12.8\% storage overhead for
the redundant ECC bits~\cite{deal.whitepaper09} \chVI{(i.e., \emph{ECC
redundancy})}. \chVI{By deploying all of our
proposed error mitigation techniques in an enterprise-class SSD, the
RBER at the end of the fixed flash memory lifetime is significantly
lower compared to \emph{Baseline}. Thus, we can redesign the ECC deployed in
the SSD to tolerate only up to the reduced RBER, which requires fewer \chVIII
{ECC} bits
\chX{and, thus,} lower ECC \chVI{redundancy} than \chVII{the ECC required for the}
\emph{Baseline}.}
Assuming all five \chX{of the evaluated} SSDs achieve the same lifetime, and
the same reliability (i.e., uncorrectable error rate) at the end of their lifetime,
\chVIII{\emph{State-of-the-art}, \emph{LaVAR}, \emph{LaVAR+LI-RAID}, and 
\emph{This Work} reduce ECC redundancy by 42.2\%, 45.3\%, 68.8\%, and 78.9\%,
respectively, over \emph{Baseline}.}
We leave the evaluation of the performance improvements due to a weaker ECC
requirement~\cite{lee.isscc12, chen.tit81} for future work.}

\sg{We conclude that by combining LaVAR, LI-RAID, and \chVI{ReMAR,}
we can \chX{(1)~}achieve significant improvements in the lifetime of
3D NAND flash memory, \chVIII{\chX{(2)~}enable \chX{higher} write intensity in workloads within
a given lifetime requirement,} \chVII{or \chVIII{(3)}~keep the lifetime constant but
greatly reduce the storage cost of reliability in 3D NAND flash memory}.}


\section{Related Work}
\label{sec:related}

To our knowledge, this paper is the first in open literature
to (1)~show the \chVII{differences} between the error characteristics of 3D
NAND \sg{flash memory} and that of planar NAND flash memory through extensive
characterization using real \chVI{3D NAND flash memory} chips, (2)~develop models of layer-to-layer
process variation and early retention loss for 3D NAND flash memory, 
and \sph{(3)}~propose \chVI{and show the benefits of} four new
mechanisms based on the new error characteristics of 3D NAND flash memory.
Due to the importance of NAND flash memory reliability \chVI{in storage systems}, 
there is a large body of related work. We \chVI{treat} this related
work \chVI{in} \chI{five} \chVI{different} categories.

\textbf{3D NAND \chVI{Flash Memory} Error Characterization.} \chVI{Two recent}
works \chVI{compare} the retention loss \chVI{phenomenon} between 3D NAND and
planar NAND \sg{flash memory}~\cite{mizoguchi.imw17, luo.hpca18} through real device
characterization, and \chVI{report} findings \chVI{similar} \sg{to our work}
regarding \chVI{the} early retention loss \chVI{phenomenon}.
\chVI{Two other} recent \chVI{works use} \sg{a methodology similar to ours} to characterize
3D NAND devices based on \chVII{\emph{different}} 3D NAND \chVI{flash memory cell}
technologies (\chVI{i.e.,} 3D floating-gate cell
and 3D vertical gate cell)~\cite{xiong.sigmetrics17, xiong.tos18, hung.jssc15}, which are
\chVII{\emph{less common}} than the 3D charge trap NAND \sg{flash memory \chVI{cell
technology} that we test} in this paper.
\sph{Other recent works~\chVI{\cite{park.jssc15, wang.tecs17, choi.svlsi16,
grossi.bookchapter16, park.iedm12} report several differences
of 3D NAND flash memory from planar NAND flash memory}.
These differences include (1)~smaller
program variation at high P/E cycle \sg{counts}~\cite{park.jssc15}, (2)~smaller program
interference~\cite{park.jssc15}, (3)~layer-to-layer process
variation~\cite{wang.tecs17}, (4)~early retention
loss~\cite{choi.svlsi16, grossi.bookchapter16, park.iedm12}, and (5)~retention
interference~\cite{choi.svlsi16}.
\chV{While prior works have reported 
on the existence of these errors, \emph{none} of them provide a comprehensive characterization
of \emph{all} of the different errors \chVII{using} the \emph{same} \chVII{chips}.
Only one of these prior works~\cite{choi.svlsi16} provides a detailed analysis
based on
\chI{\emph{circuit-level}} \chVI{measurements and characterizations}, and does
so only for early retention loss and
retention interference. Other works provide only a high-level \chVI{\emph{summary}} of
real device characterization~\cite{park.jssc15} or do \chVI{\emph{not}} provide any real device characterization
results at all~~\cite{wang.tecs17, grossi.bookchapter16, park.iedm12}. 
Our work performs an extensive detailed analysis of \emph{all} known sources of error in 
3D NAND flash memory chips, which allows us to understand
the relative impact of each error source on the same chip.
We \chVI{report \chVII{the first set of} extensive results on} three error characteristics \chVI{that
are new in 3D NAND flash memory}: layer-to-layer
process variation, early retention loss, and retention interference.}}

\textbf{Planar NAND \chVI{Flash Memory} Error Characterization.}
\chVI{A large body of prior work studies} all types of \chVI{error sources on}
planar NAND \chVI{flash memory},
including P/E cycling errors~\cite{cai.procieee17, cai.date13,
parnell.globecom14, luo.jsac16}, programming errors~\cite{cai.hpca17,
parnell.globecom14, luo.jsac16},
cell-to-cell program interference errors~\cite{cai.iccd13, cai.sigmetrics14},
retention errors~\chVII{\cite{cai.procieee17, cai.iccd12, cai.hpca15, fukami.di17}}, and read disturb
errors~\cite{cai.procieee17, cai.dsn15}. These works characterize how \chX{the} raw bit error
rate and threshold voltage change \chVII{due to} various types of
\chVI{error \chX{sources}}. \chVI{A detailed survey of such prior works on
planar NAND flash memory can be found in \chX{our \chXI{recent} survey articles}~\cite{cai.procieee17, cai.book18}.}
Our \chVII{paper} \chVI{experimentally studies} all \chX{of} these
\chVI{error mechanisms} in the new 3D NAND \chVI{flash memory context}, and
\chVI{compares} 3D NAND \chVI{flash memory} error characteristics with results in
these \chVI{prior works to \chVII{show}} the
differences between 3D NAND and planar NAND \chVI{flash memory}. Prior work
\chVII{demonstrates} \chVI{the} early retention loss \chVI{phenomenon} in planar NAND \chVI{flash
memory} based on charge trap
transistors~\cite{chen.iedm10}, which is similar to, \chVII{but not as severe as,} \chVI{the} early retention
loss \chVI{phenomenon} in 3D NAND \chVI{flash memory}. \chVII{We} \chVI
{investigate} retention interference and process variation \chVIII{related} \chX{errors, in
addition to these other} error types \chVIII{discovered before in planar NAND flash memory}.

\textbf{Planar NAND Error Modeling and Mitigation.}
Based on characterization results, prior work \chVI{proposes} models for planar
NAND \chVI{flash memory} threshold voltage distribution, and models for \chVI
{estimating the effect of P/E cycling} on the \chVI{threshold voltage}
distribution~\cite{parnell.globecom14, luo.jsac16, cai.date13}. Our work uses a
simpler \chVII{threshold voltage} distribution model, \chVI{since more complex models are designed to
handle programming errors in planar NAND flash memory that do not exist in the
3D NAND flash \chVII{memory chips} that we \chX{test}.} \chVII{We develop} a unified model
of retention loss and wearout for \chX{the} RBER, threshold voltage distribution,
\chVI{and $V_{opt}$} in 3D NAND \chVI{flash memory}. There is a large body of prior
work that proposes mechanisms to mitigate planar NAND \chVI{flash memory}
errors~\cite{cai.book18, cai.procieee17, cai.hpca17, cai.hpca15, cai.iccd12,
pan.hpca12, cai.dsn15, cai.iccd13, cai.sigmetrics14, cai.itj13, luo.jsac16,
ha.apsys13, ha.tcad16, jeong.fast14, luo.msst15, wilson.mastots14, li.fast15,
huang.fast17, zhang.fast16, jimenez.fast14, pan.fast11}. \chVI{In
Section~\ref{sec:mitigation},} we
have \chVI{already} compared \chVI{our mechanisms} to several of \chVI{these
techniques \chVII{that} are state-of-the-art}, and \chVI{have
shown} that \chVII{prior techniques developed for planar NAND flash memory}
are less effective in 3D NAND \chVI{flash memory} than our
techniques due to the new
error characteristics \chVIII{of 3D NAND flash memory}.

\chI{\textbf{3D NAND \chVI{Flash Memory} Error Mitigation.}
Prior work \chVI{proposes} circuit-level and system-level techniques to tolerate
layer-to-layer process variation in 3D NAND \chVI{flash memory}. \sg{Two
recent works} propose to use
different read reference voltages for different layers~\cite{hung.jssc15,
ye.fms17}, which is similar to the LaVAR technique \sg{that we propose} in
Section~\ref{sec:mitigation:variation}. \sph{\chVI{Unlike} our work, these
\chVI{prior} works do not
(1)~design a detailed mechanism like LaVAR to learn and use the $V_{opt}$ in a
lookup table, or (2)~evaluate \chVI{their techniques} using real
characterization data.} Wang et al.\ propose to
apply different read reference \chVII{voltages for} less-reliable pages storing
critical metadata~\cite{wang.tecs17}. As we have shown in
Section~\ref{sec:mitigation:variation}, while these prior techniques improve
average RBER, they do \chVI{\emph{not}} significantly reduce worst-case RBER,
which limits the flash memory lifetime. In this work, we \chVI{propose} a
series of mitigation techniques that not only significantly reduce the
\chVI{average and }worst-case RBER but also tolerate other new error characteristics
we find in
3D NAND \chX{flash memory}, such as early retention loss and retention interference.}

\textbf{Large-Scale SSD Error Characterization.}
Prior work \chVI{performs} large-scale \chVI{studies of errors found in} flash
memories deployed in data centers~\cite{meza.sigmetrics15, schroeder.fast16,
narayanan.systor16}.
Since \chVI{the operating system is unaware of the raw bit errors in the NAND
flash memory} \chX{devices}, these studies can \chVI{only} \chVII{use} drive-level
statistics
provided by the \chX{SSD} controller, such as overall RBER and uncorrectable error rate, 
average P/E \chX{cycle count}, and \chX{a coarse} estimation of
retention time and read disturb counts. \chVI{In contrast, in our studies, we
have complete} access to the
physical location, P/E cycle \chVI{count}, retention time, \chVI{and} read disturb count of
each read/write
operation, and thus can provide deeper insights \chVI{and controlled
experimental results compared to} large-scale studies, \chVI{which have to
be correlational in nature}.

\chVI{\textbf{DRAM Error Characterization.} Like a flash memory cell, a DRAM cell
stores charge to represent a piece of data. Hence, DRAM has
many error characteristics \chX{that are} similar to NAND flash memory. For example, charge
leaks from \chX{a} DRAM cell over time, at a speed much faster than 
\chX{that for} NAND flash memory 
(i.e., on the order of \emph{milliseconds} to \emph{seconds} \chX{in DRAM}~\chVII{\cite{R104,
liu.isca12}}),
leading to \emph{data retention errors}. This phenomenon in DRAM is analogous to the
retention loss phenomenon in NAND flash memory (see
Section~\ref{sec:retention} and Appendix~\ref{sec:appendix:retention}), and its
effect has been
studied through extensive experimental characterization of DRAM
chips~\chX{\cite{R126, R127, R149, R206, R125, R119, R104, R147, kim.hpca18,
jung.dt17, hamamoto.ted98, R157}}. Similar
to the retention interference phenomenon found in 3D NAND flash memory (see
Section~\ref{sec:retention:interference}), DRAM
exhibits data-dependent retention behavior,
or \emph{data pattern dependence} (DPD)~\cite{R104}, where the retention time
of a DRAM cell is dependent on the values written to \emph{nearby}
DRAM cells~\cite{R104, R126, R127, R147, R149, R206}. Conceptually similar to
the read disturb errors found in NAND flash memory (see
Appendix~\ref{sec:read:disturb}), commodity DRAM
chips that are sold and used in the field today exhibit read
disturb errors~\cite{R116}, also called \emph{RowHammer}-induced
errors~\cite{R131}. \chVII{These errors are affected by \emph{process
variation}, which we \chVIII{comprehensively} examine} in 3D NAND flash memory
(see Section~\ref{sec:variation} and Appendix~\ref{sec:appendix:variation}).
\emph{Process variation} in DRAM \chVII{is shown to also affect access
latency, retention time, and power consumption}\chX{~\cite{R126, R127, R149, 
R206, R125, ghose.sigmetrics18, R119, R128, R120, R178, R104, R147, kim.hpca18,
jung.patmos16, mathew.date18, jung.dt17, R157, hamamoto.ted98, R152,
R174, R175, liu.isca12}}.}


\section{Conclusion}
\label{sec:conclusion}

\chI{We develop \chVI{a new} understanding of three new error characteristics in
3D NAND flash memory through rigorous experimental characterization of real,
state-of-the-art 3D NAND flash memory chips: layer-to-layer process
variation, early retention loss, and retention interference. We analyze and
show that these new error characteristics are fundamentally caused by changes
introduced in the 3D NAND flash memory architecture \chX{compared to the
planar NAND flash memory architecture}.}
To handle these three new error characteristics in 3D NAND \chVI{flash memory}, we
develop new analytical models for layer-to-layer process variation and early
retention loss in 3D NAND flash memory. \chVI{Our models can accurately
predict/estimate the optimal read reference voltage and the raw bit error
rate based on the retention time and the layer number of \chVII{each flash memory} page. We}
propose four new \chX{error mitigation} techniques that
utilize our \chVI{new} models to improve the reliability of 3D NAND flash
memory. 
\chI{Our evaluations show that our newly-proposed techniques successfully
mitigate the new error patterns that we discover in 3D NAND flash memory.}
We hope that the \chVI{rigorous and comprehensive}
error characterization and \chX{analyses} performed in this work \chX{motivate} future
\chVI{rigorous}
studies on \chX{3D NAND \chVII{flash memory} reliability}, and that \chX{they inspire} new error mitigation
mechanisms that cater to \chVII{the} \chX{new} \chVI{error \chX{characteristics} found in} 3D NAND flash
memory.

\section*{Acknowledgments}
We thank our shepherd, Benny Van Houdt, the anonymous reviewers, and SAFARI
members for their feedback.
This work is partially supported by
\chII{grants from Huawei and Seagate, and gifts from \chXI{Huawei,} Intel, \chVIII{Microsoft,} and Samsung}.

{
\bibliographystyle{IEEEtranS}
\bibliography{refs}
}
\newpage

\appendix

\section{Appendix}
\label{sec:appendix}


\subsection{Write-Induced Errors}
\label{sec:write}

We analyze how each type of write-induced error affects the
RBER and the threshold voltage distribution of 3D NAND flash memory.

\subsubsection{Program Errors}
\label{sec:programerror}

\chX{Program errors occur when the data is incorrectly written to the
NAND flash memory~\cite{park.jssc08, cai.hpca17, cai.procieee17,
cai.book18}.  Such errors are introduced when multiple programming operations
are required to write data to a single cell.  For example, in many MLC NAND
flash memory devices, \emph{two-step programming}~\cite{park.jssc08,
cai.hpca17} \chXI{is employed. Two-step programming} uses two separate \emph{partial programming} steps to write data
to an MLC NAND flash cell.  In the first step, the flash controller writes
only the LSB to the cell, setting the cell to a temporary voltage state.
In the second step, the controller writes the MSB to the cell, but in order
to perform this write, the controller must first determine the current
voltage state of the cell.  This requires reading the partially-programmed data
from the cell, during which an error may occur.  This error \chXI{causes the
controller to} incorrectly \chXI{set} the final voltage state of the cell during the
second programming step, and, thus, is called a program error.
Prior work~\cite{cai.hpca17} \chXI{shows} that program errors occur in
state-of-the-art planar MLC NAND flash memory.}

\chIX{\chX{Current generations of 3D NAND flash memory use} \emph{one-shot
programming}~\cite{park.jssc08, cai.hpca17, cai.procieee17, cai.book18}, which
programs \chX{\chXI{\emph{both}} the LSB and MSB of a cell at the \chXI{\emph{same time}}.
As a result, current 3D NAND flash memory devices do \chXI{\emph{not} experience} program
errors.
Our measurements in Figure~\ref{fig:distribution-shape} confirm the lack of
program errors \chXI{in 3D NAND flash memory}.  In an MLC NAND flash memory
that has program errors,
the threshold voltage distributions of the ER and P1 states have secondary
peaks near the P2 and P3 states, respectively~\cite{cai.hpca17}.  This is
because program errors affect only the LSB, since only the LSB is being read
during the second programming step.  Since there is no second peak in
Figure~\ref{fig:distribution-shape}, there are no program errors.}}

\chIX{\chX{Program errors may appear in future 3D NAND flash memory devices.
In planar NAND flash memory, two-step programming was introduced when
planar MLC NAND flash memory transitioned to the \SI{40}{\nano\meter} 
manufacturing process technology \chXI{node,} in order to reduce the number of
program interference errors~\cite{park.jssc08}.  A similar transition may
occur in the future to continue scaling the density of 3D NAND flash
memory, especially as it becomes increasingly difficult to add more layers
into a 3D NAND flash memory chip.}
Thus, we conclude that today's 3D NAND flash memories
do \chX{\emph{not}} have program errors, but program errors may appear in future
generations.}

\subsubsection{Program/Erase Cycling Errors}
\label{sec:pecycle}

A P/E cycling \chVIII{error occurs} \chVI{because of
the natural variation of the threshold voltage of cells in each
state~\cite{mielke.irps08, cai.date13} due to the inaccuracy of each program
and erase operation} (see Section~\ref{sec:background:errors}). \chVI{Such
inaccuracy during program and erase operations increases as \chX{the} 
P/E cycle count increases.}
To study the impact of P/E cycling errors, we randomly select
a flash block within each 3D NAND chip, and wear out the block by
programming random data to each page in the block until the block reaches 
16K P/E cycles.
Using the methodology described in Section~\ref{sec:methodology}, 
we obtain the overall RBER and the threshold voltage of each cell at various
P/E cycle counts.\footnote{Due to limitations with our \chVI{experimental testing}
platform, each data
point at a particular P/E cycle count has a retention time of 50 minutes.}

\textbf{Observations.}
Figure~\ref{fig:pec-mean-var} shows how the
mean and standard deviation \chVI{of the threshold voltage distribution
\chX{of each state}} change
\chX{as a function of the P/E cycle count, when we 
\chXI{fit our voltage measurements for each state to a Gaussian model}}.
\chVI{Each subfigure in the top row represents the mean for a
different state; each subfigure in the bottom row represents the standard
deviation for a different state. The blue dots shows the measured data; each
orange line shows a linear trend fitted to the measured data. The x-axis
shows the P/E cycle count; 
\chX{the y-axis shows the mean 
(Figures~\ref{fig:pec-mean-var}a--\ref{fig:pec-mean-var}d) or the
standard deviation 
(Figures~\ref{fig:pec-mean-var}e--\ref{fig:pec-mean-var}h) of the threshold
voltage distribution of each state, in voltage steps.}}
We make four observations from Figure~\ref{fig:pec-mean-var}. 
First, the mean and standard deviation of all states \chX{increase linearly
as the P/E cycle count increases.  We fit a line using linear regression, 
shown as an orange dotted line in each subfigure.}%
\footnote{For the ER state, a linear fit has a
5.9\% higher root mean square error than a power-law
fit. However, we choose the linear fit due to its simplicity.} 
\chVI{Second, 
\chX{the threshold voltage distributions of the ER and P1 states}
shift to higher voltages, while the distributions of the \chX{P2 and
P3 states shift to lower voltages, causing the distributions to
move closer to the middle of the threshold voltage range}.}
Third, the threshold voltage distributions \chX{of} all four states become wider
\chVI{(i.e., the standard deviation increases)} as the P/E cycle count increases.
\chVI{\chX{Since the distributions shift towards the middle of the threshold
voltage range} and become wider as \chX{the} P/E
cycle count increases, the distributions become closer to each other,
\chX{which increases} the raw bit error rate.
Fourth, the magnitude of the \chX{threshold voltage shift and the
widening of the distributions} is much larger for the ER~state than
it is for the other three states (i.e., P1, P2, P3). \chVI{Therefore,
ER$\leftrightarrow$P1 errors \chX{(i.e., an error that shifts a cell
that is originally programmed in the ER state to the P1 state, or vice versa)} 
increase faster than other errors \chX{with the P/E cycle count.}}}

\begin{figure}[h]
\centering
\includegraphics[trim=10 10 10 10,clip,width=\linewidth]
{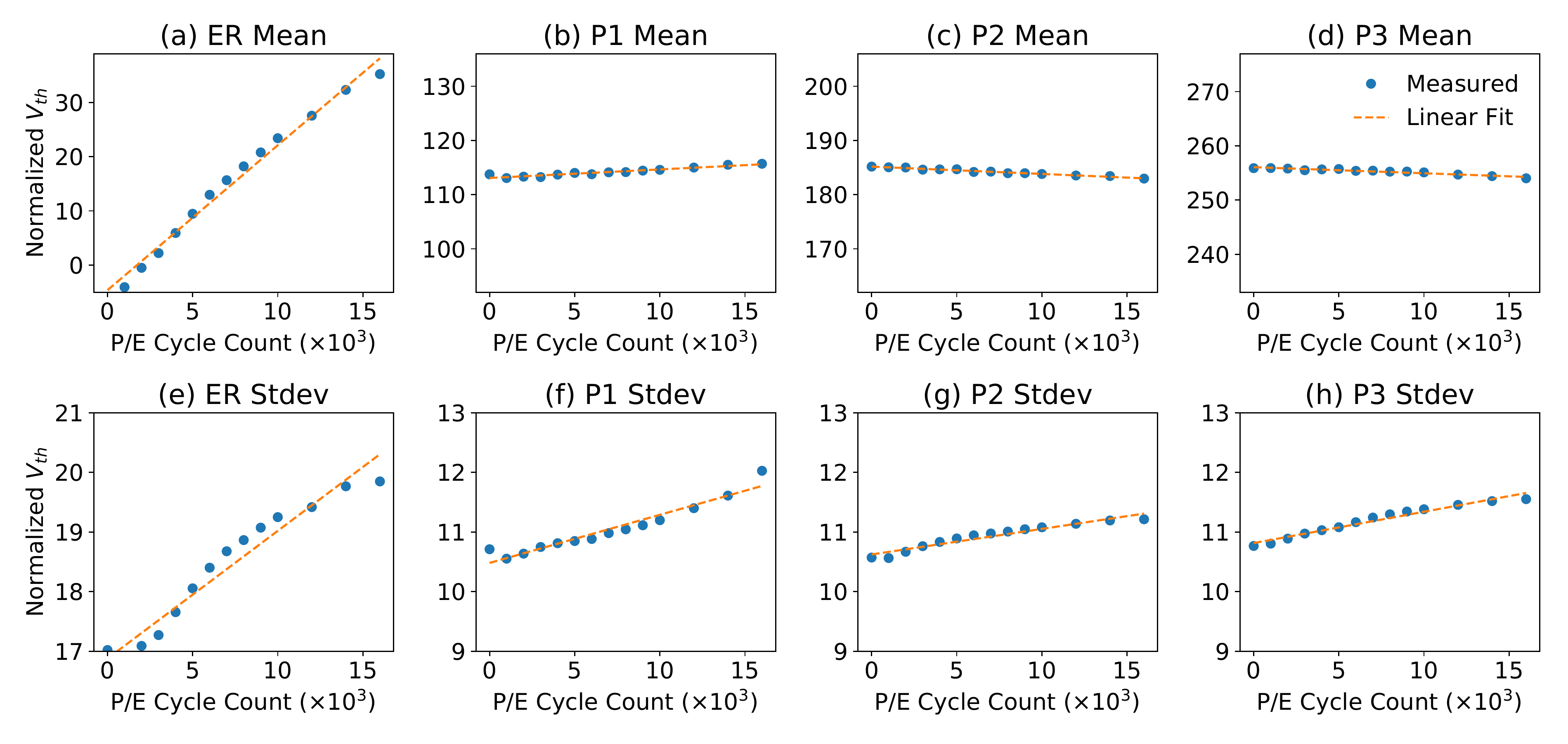}
\caption{\chVI{Mean and standard deviation of \chIX{our Gaussian} threshold
voltage} distribution \chIX{model of each state}, \chVI{versus} P/E cycle count.}
\label{fig:pec-mean-var}
\end{figure}

Figure~\ref{fig:pec-opterr} shows how the RBER increases as the
P/E cycle count increases.  The top graph breaks down the errors into which \chX{bit}
(i.e., LSB or MSB) they occur in.  The bottom graph
breaks down the errors based on how the error changed the cell state due to a
shift in the cell threshold voltage.  If
the error \chXI{caused} \emph{either} the LSB or MSB (but not both) \chXI{to
be read incorrectly}, we refer to \chX{that error} as a single-bit error
\chX{(ER~$\leftrightarrow$~P1, P1~$\leftrightarrow$~P2, and P2~$\leftrightarrow$~P3
in the graph)}.
If \chX{\emph{both}} the LSB and MSB are \chXI{read incorrectly} as a result
of the \chXII{error}, we refer to that
\chX{error} as a multi-bit error.
We make four observations from Figure~\ref{fig:pec-opterr}.
First, both LSB and MSB errors increase as the P/E cycle count increases,
following an exponential trend. 
Second, ER~$\leftrightarrow$~P1 errors increase at \chX{a} much faster rate as the P/E
cycle count increases,
\chVI{compared} to the other types of cell state changes, and
ER~$\leftrightarrow$~P1 errors become the
dominant MSB error type when the P/E cycle count reaches \chVI{8K P/E cycles (6K
is the cross-over point)}. \chVI{This is because the electrons trapped in the
cell during wearout \chX{prevent the cell from being set to very low threshold
voltages}.
\chX{As a result, the threshold voltage distribution of the ER~state \chXI{shifts
and widens} more than the distributions of the other states, as we see in
Figure~\ref{fig:pec-mean-var}.}}
Third, multi-bit errors are \chXI{less common}, \chVI{but} they occur 
as early as \chXII{at} 1K P/E cycles.
\chVI{\chX{Only} a large difference between the target and actual
threshold voltage can lead to a multi-bit error, which is unlikely to happen.}
Fourth, \chX{MSBs} have a $2.1\times$ higher error rate than \chX{LSBs}, on
average across all P/E cycle counts. 
\chX{This is because the flash controller must use two read reference voltages
to read a cell's MSB, but needs \chXII{\emph{only one}} read reference \chXI{voltage} to read
a cell's LSB.}

\begin{figure}[h]
\centering
\includegraphics[trim=10 10 10 10,clip,width=\figscale\linewidth]
{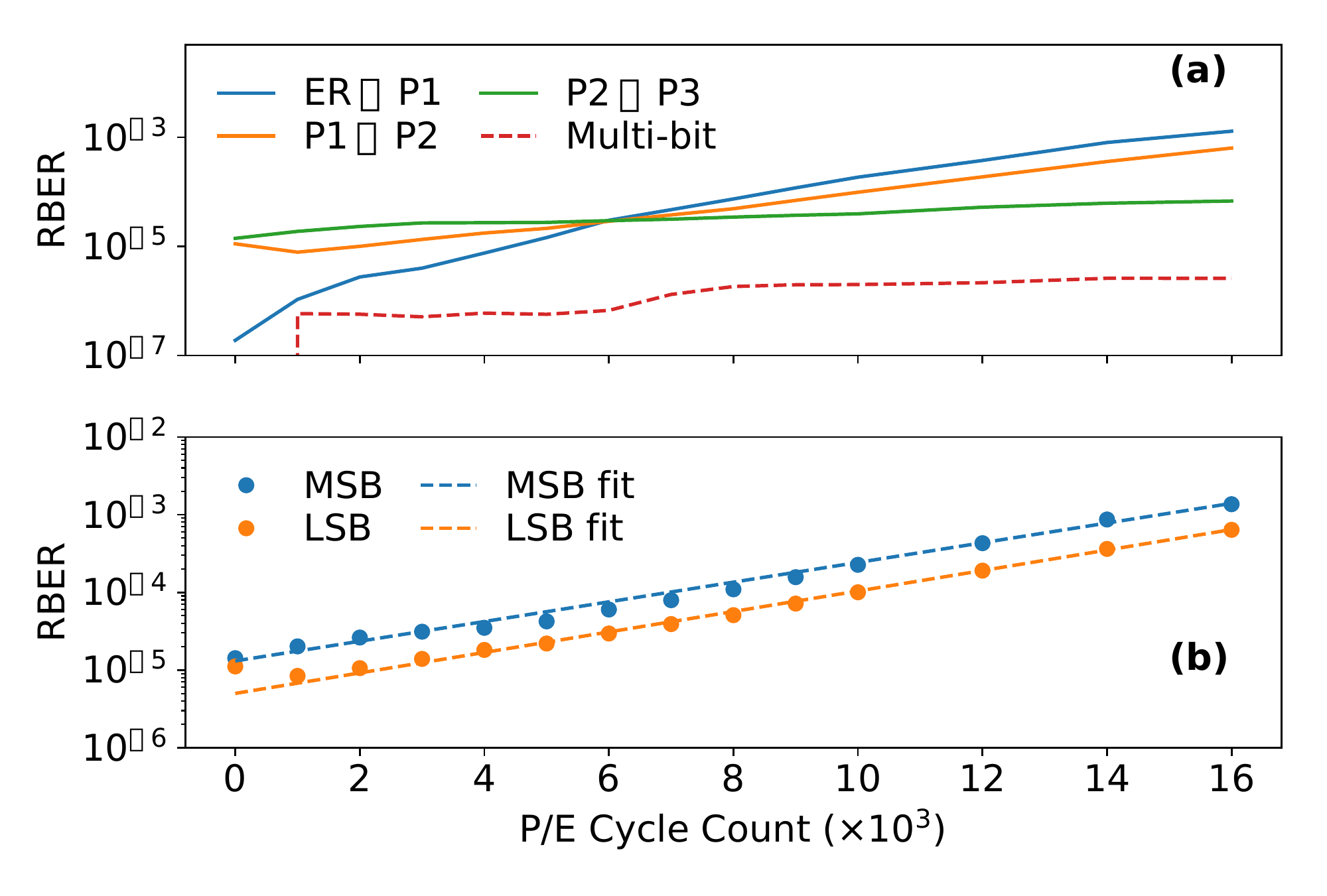}
\caption{\chX{RBER due to} \chVI{P/E cycling} errors vs. P/E \chX{cycle count}.}
\label{fig:pec-opterr}
\end{figure}

Figure~\ref{fig:pec-optvrefs} shows how the optimal read reference voltages
change as the P/E cycle count increases. This figure contains three
subfigures, each of which shows the
optimal voltage for $V_a$, $V_b$, and $V_c$ (see Figure~\ref{fig:mlc}). 
We make two observations from this figure.
First, the optimal voltage for $V_a$ increases rapidly as the P/E cycle count
increases: after 16K P/E cycles, the voltage goes up by more than 20 voltage 
steps.
Second, the optimal \chX{voltages} for 
$V_b$ and $V_c$ remain almost constant as the P/E cycle count increases:
neither voltage changes by more than 4~voltage steps after 16K P/E cycles,
\chVI{as expected from the lack of change in P1, P2, \chX{and P3 distribution}
means shown in Figure~\ref{fig:pec-mean-var}.}

\begin{figure}[h]
\centering
\includegraphics[trim=10 10 10 10,clip,width=.8\linewidth]
{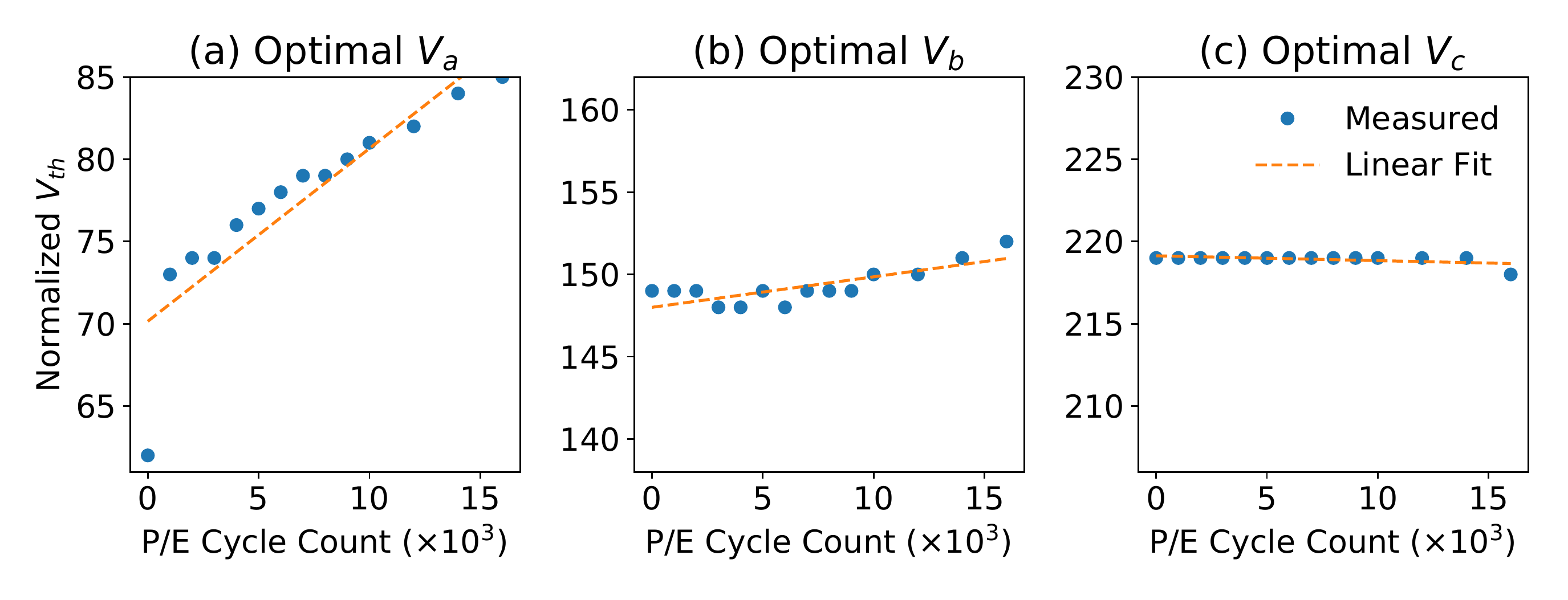}
\caption{Optimal read reference voltages vs. P/E cycle \chX{count}.}
\label{fig:pec-optvrefs}
\end{figure}

\textbf{Insights.} To compare the error characteristics of 3D NAND \chX{flash memory} to that of
planar NAND flash memory, we \chX{take the} equivalent observations on planar NAND
\chX{flash memory} reported by prior
works~\cite{cai.date13, parnell.globecom14,
luo.jsac16}, and compare them to our findings for 3D NAND \chVI{flash \chX{memory,}
which we just described}.
We find two key differences.
First, for 3D NAND \chX{flash memory}, the threshold voltage distributions for the P2~state and
the P3~state shift to \chVI{\emph{lower}} voltages as the P/E cycle count increases.  In
contrast, for planar NAND flash memory, the distributions of both states shift
to \emph{higher} voltages~\cite{cai.date13, parnell.globecom14, luo.jsac16}.
\chX{One possible source of this change is the \chXI{increased impact of early retention loss
with P/E cycle count,}}
which lowers the threshold voltage of cells in \chX{higher-voltage} states (i.e.,
P2 and P3)~\cite{choi.svlsi16}. 
Second, for 3D NAND \chX{flash memory}, the \emph{change} in the mean \chVI{threshold voltage} of
each state distribution
exhibits a linear \chX{increase}.  However, in sub-\SI{20}{\nano\meter} planar NAND flash memory,
the change in the mean \chVI{threshold voltage} exhibits a \chX{\emph{power-law}-based
increase with P/E cycle count}~\cite{parnell.globecom14, luo.jsac16}.
In sub-\SI{20}{\nano\meter}
planar NAND flash memory, the mean \chVI{threshold voltage} of each state distribution increases more rapidly
at lower P/E cycle counts than in higher P/E cycle counts, resulting in \chX{the power-law-based}
behavior.  However, we note that planar NAND flash memory using an older manufacturing
process technology (e.g., \SIrange{20}{24}{\nano\meter}) exhibits a linear \chX{increase
with P/E cycle count} for the
distribution mean~\cite{cai.date13}, just as we \chVI{observe} for 3D NAND
\chVI{flash memory}.
Thus, \chVI{there is evidence that} when the manufacturing process technology
scales below
a certain size, the change in the distribution mean transitions from linear
behavior to power-law\chX{-based} behavior \chX{with respect to P/E cycle count}.  
As a result, when future 3D NAND \chX{flash \chXI{memory scales}} down
to a sub-\SI{20}{\nano\meter} manufacturing process technology \chX{node}, we \chVI
{might} expect that it too will exhibit power-law behavior \chXI{for the change in the
distribution mean}.
\chIX{We conclude that the differences we observe between the P/E cycling
effect in 3D NAND flash memory and planar NAND flash memory are mainly caused
by \chXI{the use of a significantly} different manufacturing process technology \chX{node}.}

\subsubsection{Program Interference}
\label{sec:interference}

When a cell (which we call the \emph{aggressor cell}) is being programmed, 
cell-to-cell program interference can cause
the threshold voltage of nearby flash cells (which we call \emph{victim cells})
to increase unintentionally~\cite{cai.sigmetrics14, cai.iccd13}
(see Section~\ref{sec:background:errors}).
In 3D NAND \chVI{flash memory}, there are two types of program interference that can occur.
The first, \emph{wordline-to-wordline program interference}, affects
victim cells along the z-axis \chVI{of} the cell \chX{that is} programmed (see Figure~\ref{fig:organization}).
These victim cells are physically next to the cell \chX{that is} programmed, and
belong to the same bitline (and thus the same flash block).
The second, \emph{bitline-to-bitline program interference}, affects
victim cells along the x-axis or y-axis \chVI{of} the cell \chX{that is} programmed.  
Bitline-to-bitline program interference can affect victim cells in the same
wordline (i.e., cells on the y-axis), or it can affect victim cells that
belong to other flash blocks (i.e., cells on the x-axis).

To quantitatively analyze the effect of program interference \chVI{on
cell threshold voltage and raw bit error rate}, we use the
same experimental data that we have for \chVI{P/E cycling} errors
(see Section~\ref{sec:pecycle}).  
\chX{A correlation exists between the amount by which program interference 
changes the threshold voltage of a victim cell ($\Delta V_{victim}$) and the 
threshold voltage change of the aggressor cell ($\Delta V_{aggressor}$)~\cite{cai.iccd13}.
As a result of this \emph{interference correlation}, the threshold voltage 
of a victim cell is \emph{dependent} on the threshold voltage}
of the aggressor cell. The strength of this correlation can be
quantified as $\frac{\Delta V_{victim}}{\Delta V_{aggressor}}$, which is a
property of the NAND device and is largely dependent on the distance between the 
cells~\cite{lee.iedl02}. \chVI{After programming \chX{randomly-generated} data to
the victim cells and the aggressor cells, we} estimate $\Delta V_{aggressor}$
by calculating the threshold voltage difference between the aggressor cell's
\chVI{threshold voltage in its} final state and \chVI{that in} the ER~state.
We estimate
$\Delta V_{victim}$ by calculating the difference between the
victim cell's threshold voltage with and without program
interference.\footnote{\chXI{The cell threshold voltage \chXII{\emph{without}} program
interference is obtained by reading the cell \chXII{\emph{before}} the next wordline is 
programmed.}}

\textbf{Observations.} 
Figure~\ref{fig:interference-correlation} shows the
interference correlation 
for wordline-to-wordline interference and bitline-to-bitline
interference on a victim cell, for aggressor cells of varying
distance from the victim cell. \chVI{For example, the victim cell
in BL~M, WL~N has \chX{an interference correlation of} 2.7\%
with the \emph{next wordline} aggressor cell in BL~M,
WL~N+1, which means
that, if the threshold voltage of the aggressor cell increases by $\Delta V$,
the threshold voltage of the victim cell \chX{increases by $0.027
\Delta V$} due to wordline-to-wordline program interference.}
We make two observations from this figure. 
\chVI{First, the interference correlation of the \emph{next wordline}
aggressor cell (i.e., 2.7\%) is over an order of magnitude higher than 
\chX{that of} any
other aggressor \chX{cell, of which the maximum
interference correlation is only 0.080\% (the \chXII{\emph{previous wordline}}
aggressor cell in \chXI{BL~M, WL~N-1}).} 
Thus, the program interference to the victim cell,} is dominated by
wordline-to-wordline interference from the \chXI{\emph{next}} wordline.
Second, all of the other types of interference have \chX{much smaller
interference correlation values}.

\begin{figure}[h]
\centering
\includegraphics[trim=0 180 370 0,clip,width=.8\linewidth]
{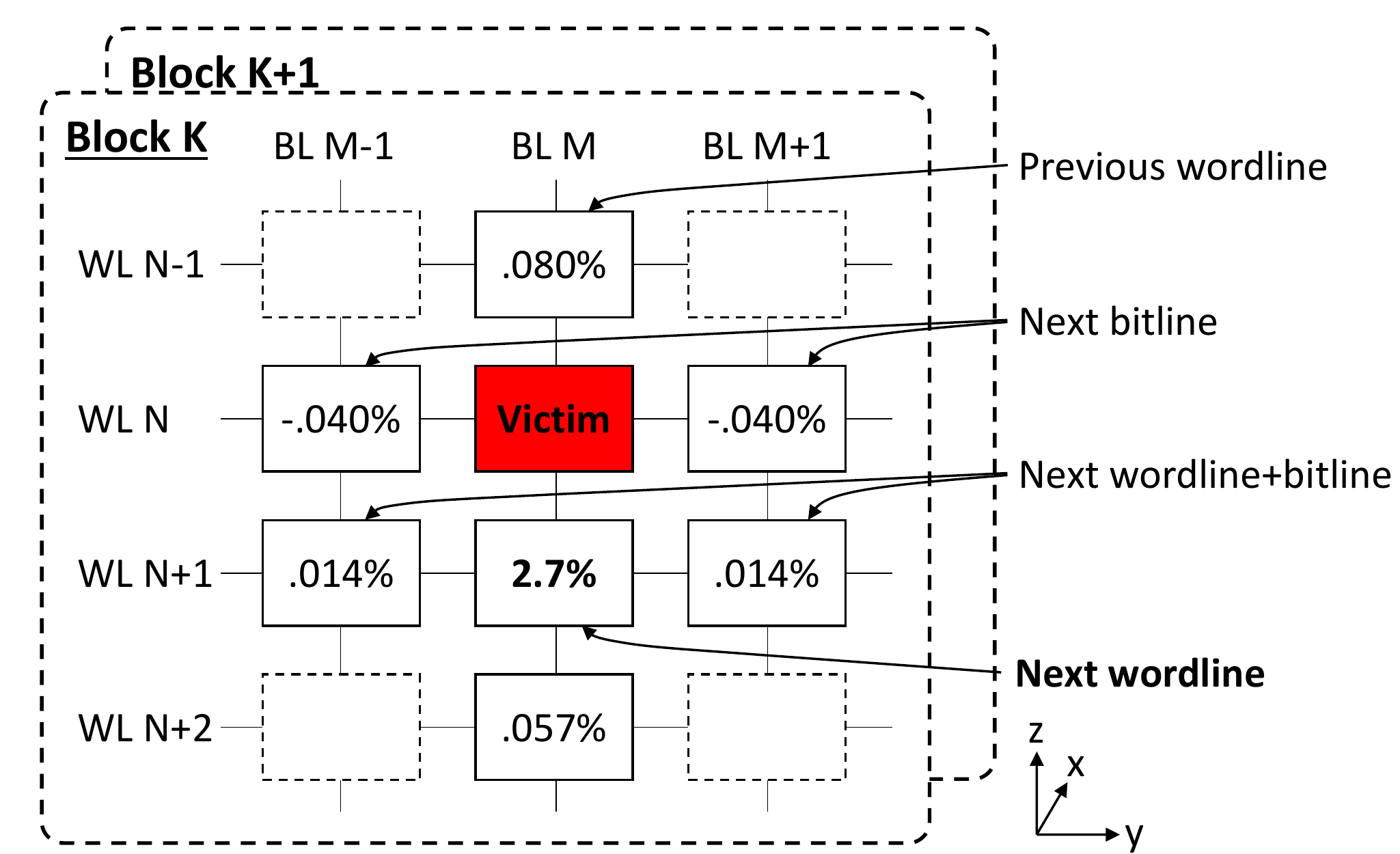}
\caption{Interference correlation for a victim cell, as a result of programming \chX{aggressor cells of varying 
distances from the victim cell}.}
\label{fig:interference-correlation}
\end{figure}

Figure~\ref{fig:interference-trend} \chX{shows how} much the threshold voltage of a victim
cell shifts \chVI{($\Delta V_{victim}$)} when a neighboring aggressor cell is
programmed to the P3~state, which
generates the largest possible program interference.
Each curve represents a certain program interference type \chVI{
(i.e., Next WL or Prev WL)}
and a certain state of the victim cell \chVI{(V)}. \chVI{The curves \chX{that} have
a significant amount of threshold voltage shift (e.g., $>$6 voltage steps) due
to program interference are shown in Figure~\ref{fig:interference-trend}(a);
the curves \chX{that} have \chX{a} small amount of threshold voltage shift are shown in
Figure~\ref{fig:interference-trend}(b).}
We make three observations from Figure~\ref{fig:interference-trend}.
First, the effect of program interference decreases as the 
P/E cycle count increases \chVI{(along the x-axis, from left to right).
\chX{As we discuss in Section~\ref{sec:pecycle}, electrons trapped in a
flash cell due to wearout prevent the cell from returning to the lowest 
threshold voltage values during an erase operation.  As a result, as the
P/E cycle count increases, the \chXI{mean} threshold voltage \chXI{of 
the ER~state increases.  This causes $\Delta V_{aggressor}$ to decrease
as the P/E cycle count increases, because the starting voltage of the 
aggressor cell increases but its target voltage after programming remains 
the same.  As we discuss above, the interference correlation (i.e., the ratio
between $\Delta V_{aggressor}$ and $\Delta V_{victim}$) is largely a function 
of the distance between flash cells.  Thus, since $\Delta V_{aggressor}$
decreases, $\Delta V_{victim}$ also decreases with the P/E cycle count.}}
Second, the \chX{amount of} program interference induced by an aggressor cell in the next
wordline decreases
when the victim cell is in a higher-voltage state \chVI{(Next WL curves in
Figure~\ref{fig:interference-trend}a, from top to bottom). \chX{This} is likely
because the voltage difference between the aggressor cell and the victim cell
is lower when the victim cell is in a higher-voltage state, reducing the
the threshold voltage shift due to program interference.}
Third, the program interference induced by an aggressor cell in the previous
wordline \chVI{(Prev WL curves in Figure~\ref{fig:interference-trend})}
affects the threshold voltage
distribution of only the ER~state for a victim cell, but it has little effect
on the distributions of the other three states (i.e., P1, P2, P3).
\chX{This is a result of how programming takes place in NAND flash memory.
A program operation can only \emph{increase} the voltage of a cell due to
circuit-level limitations.  When the aggressor cell in the previous wordline is
programmed, the victim cell is already in the ER~state, and the victim cell's
voltage increases due to program interference.  Some time later, the victim cell
is programmed.  If the target state of the victim cell is P1, P2, or P3, the
programming operation needs to further increase the voltage of the cell,
and any effects of program interference from the aggressor cell in the
previous wordline are eliminated.
If, however, the target state of the victim cell is ER, the programming
operation does not change the victim cell's voltage, and the effects of
program interference from the aggressor cell in the previous wordline remain.}

\begin{figure}[h]
\centering
\includegraphics[trim=0 10 0 10,clip,width=.7\linewidth]
{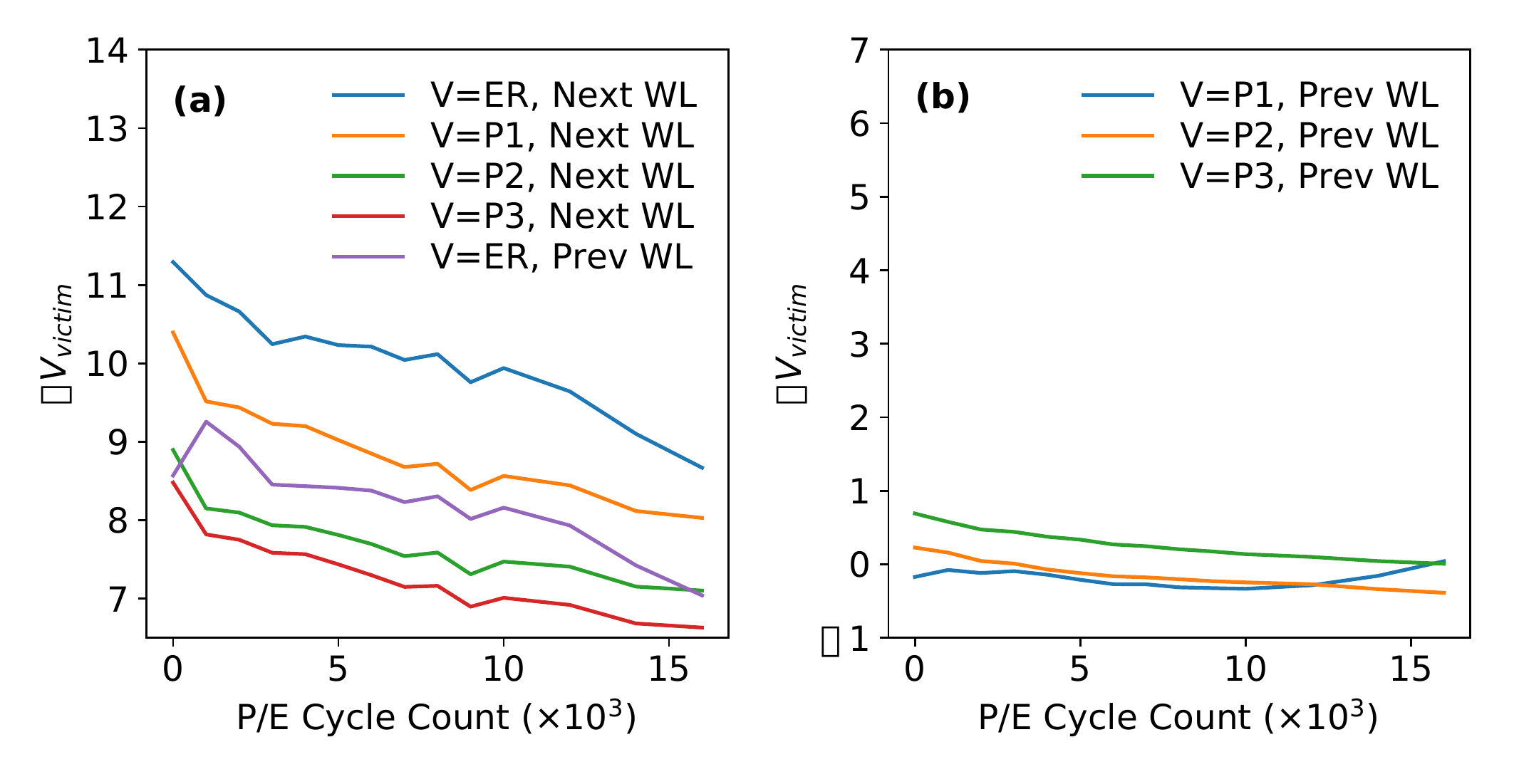}
\caption{\chVI{Amount of threshold voltage shift due to program interference}
vs. P/E cycle \chVI{count}.}
\label{fig:interference-trend}
\end{figure}

\textbf{Insights.} We compare the program interference in 3D NAND \chX{flash memory}
to the program interference observed in
planar NAND \chVI{flash memory}, as reported in prior work~\cite{cai.iccd13,
cai.sigmetrics14}. We
find one major \chVI{difference.} 
\chX{The maximum interference correlation of program interference from a 
directly-adjacent cell is 40\% lower in 3D NAND flash memory (2.7\%) than in
state-of-the-art (\SIrange{20}{24}{\nano\meter}) planar NAND flash memory
(4.5\%~\cite{cai.iccd13}).
This is corroborated by findings in prior work~\cite{park.jssc15}, which
shows that 3D NAND flash memory has 84\% lower program interference than
\SIrange{15}{19}{\nano\meter} planar NAND flash memory.
The lower interference correlation in 3D NAND flash memory is due to the
larger manufacturing process technology node
\chXI{(\SIrange{30}{50}{\nano\meter} for the chips we test)} 
that it uses compared to state-of-the-art
planar NAND flash memory.  \chXI{The} amount of interference correlation
between neighboring cells is a function of the distance between the cells~\cite{lee.iedl02}.
In a larger manufacturing process technology node, the flash cells are farther 
away from each other, causing the interference correlation to decrease.
We note that when future 3D NAND flash \chXI{memory chips} use smaller manufacturing
process technology nodes, the impact of programming interference will increase,
similar to what happened in planar NAND flash memory.}

Note that we are the first to compare how the threshold voltage shift caused
by program interference changes with the P/E cycle count. As we discuss in our
first observation for Figure~\ref{fig:interference-trend},  the program
interference \chXI{effect decreases} as the P/E cycle count increases
\chX{because the increasing effects of wearout reduce the value of
$\Delta V_{aggressor}$ during programming}.
\chIX{We conclude that the 40\% reduction in the program interference effect
we observe in 3D NAND flash memory compared to planar NAND flash memory is
mainly caused by the difference in manufacturing process technology.}


\subsection{Early Retention Loss}
\label{sec:appendix:retention}

In this \chX{section}, we present the results and analysis of retention loss in
3D NAND \chX{flash memory} in addition to the key findings in Section~\ref{sec:retention}. We use
the same methodology as described in Section~\ref{sec:retention}.

\textbf{Observations.}
\chX{Figure~\ref{fig:retention-mean-var} shows} how the
mean and the standard deviation of \chVI{the threshold voltage}
distribution \chVI{change} with retention time.
Each subfigure \chVI{in the top row \chX{shows} the mean for a different
state; each subfigure in the bottom row \chX{show} the standard deviation for
a different state. The blue dots \chX{show} the measured data; each orange line
shows a linear \chX{trend line} fitted to the measured data. \chVI{The x-axis shows the
retention time in log scale; the y-axis shows the mean or standard deviation
value in voltage steps.}}
We make five observations from \chVI{this figure}.
\chVI{First}, the threshold voltage distribution shifts \chX{more} when the
retention time is low. \chVI{This is the early retention \chXI{loss} phenomenon, \chX{which
occurs because charge that is trapped near the surface of the charge
trap layer is detrapped soon} after programming.}
\chVI{Second}, \chX{as the retention time increases,} 
the \chX{voltage values of cells in the P1, P2, and P3 states decrease, while
the voltage values of cells in the ER state increase.  This is} because the
cells in \chXI{the} ER state have \emph{negative} threshold \chX{voltages, and hence 
they} \emph{gain} charge over retention time.
\chVI{Third}, the \chX{threshold voltage distributions of the ER and P3 states 
shift faster than the distributions of the P1 and P2 states as the retention time
increases.  This is because the}
ER and P3 states have larger voltage \chX{differences} \chXI{from} the ground than the other
states.}
\chVI{Fourth}, retention loss has little effect on the width of the threshold
voltage distribution (i.e., standard deviation values change by less than 1
voltage step \chX{after 24 days}).  \chX{This is because the effects of retention loss
(i.e., charge leakage) impact cells at a similar rate, causing all of the cells
within the threshold voltage distribution to lose a similar amount of voltage.}
\chVI{Fifth}, the correlation between any distribution parameter ($V$) and the
retention time ($t$) can be modeled as a linear function (\chX{shown by} the dotted lines in
\chXI{Figure~\ref{fig:retention-mean-var}}): $V = A \cdot \log(t) + B$. 
\chX{$A$ and $B$ are constants that change based on which parameter $V$
is modeling (i.e., the threshold voltage distribution mean or standard deviation).}
\chVI{Prior
work shows that planar NAND flash memory has a similar trend
for retention loss, \chX{even though it uses} a different flash cell design. We have
already compared \chX{and evaluated the differences} between 3D NAND and planar NAND flash memory
in retention loss speed in Section~\ref{sec:retention}, and \chX{provided} more
detail about the linear \chX{function \chXI{that models the}
threshold voltage distribution parameters} in Section~\ref{sec:model:retention}.}

\begin{figure}[h]
\centering
\includegraphics[trim=10 10 10 10,clip,width=\linewidth]
{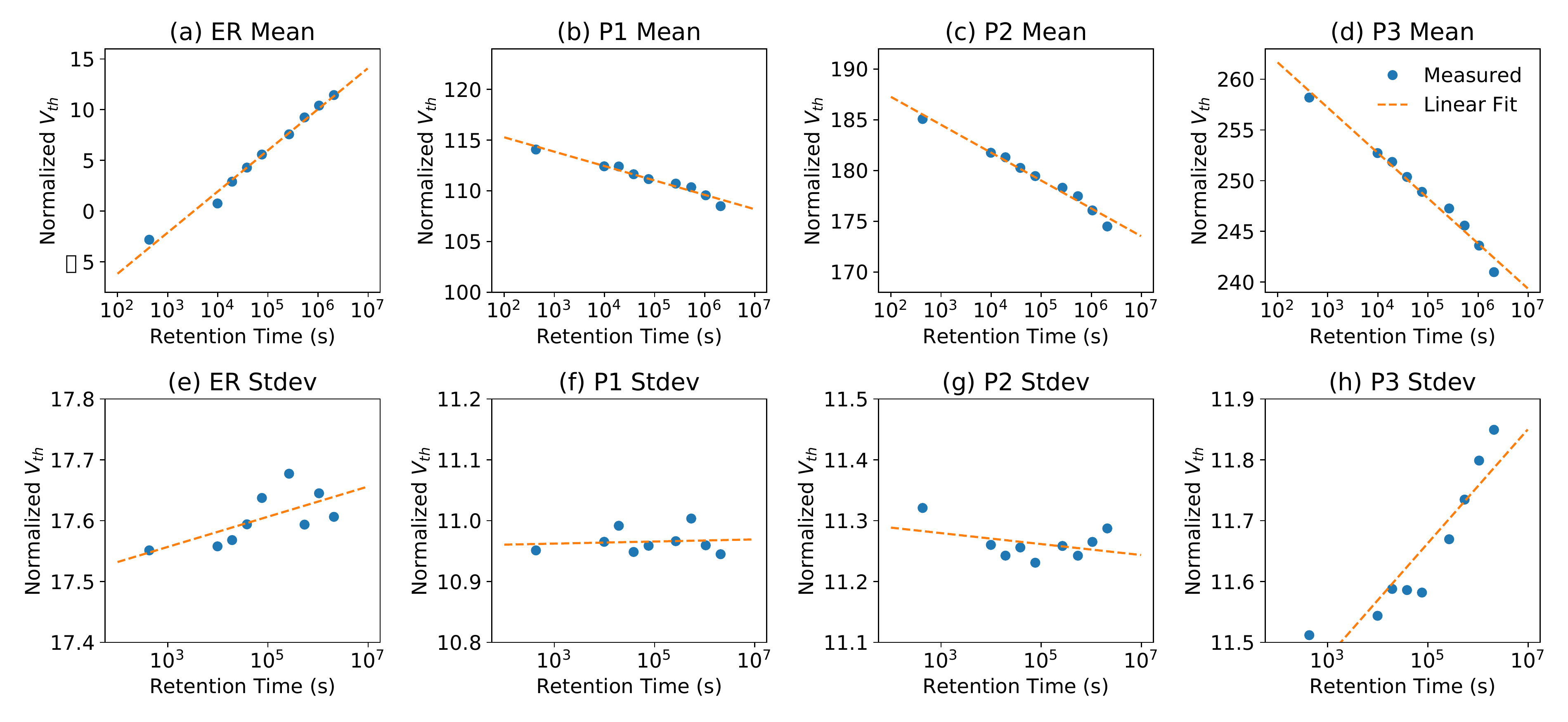}
\caption{Mean \chVI{and standard deviation of \chIX{our Gaussian} threshold voltage} distribution
\chIX{model of each state}, \chVI{versus retention time}.}
\label{fig:retention-mean-var}
\end{figure}

Figure~\ref{fig:retention-opterr} \chVI{shows how} \chXI{the} RBER
increases with retention time for a block \chX{that has endured} 10K P/E cycles. The top \chVI
{graph} breaks down the errors according to the change in cell state as a
result of the errors; the bottom graph breaks down
the errors into MSB and LSB page errors.
We make \chVI{two} observations from Figure~\ref{fig:retention-opterr}, \chVI
{in addition to our observations in Section~\ref{sec:retention}}.
\chVI{First}, retention errors are dominated by P2~$\leftrightarrow$~P3
errors, \chVI{because \chX{the threshold voltage distribution of the} P3 state not 
only shifts \chX{more} but also widens \chX{more with}
retention time than \chX{the distributions of} the other states (see Figure~\ref{fig:retention-mean-var}). 
Although \chX{the distribution of the} ER
state also shifts \chX{significantly}, there are fewer ER~$\leftrightarrow$~P1 errors to
begin with.}
\chVI{Second}, the MSB error rate increases
faster than the LSB error rate \chX{as the retention time increases. This is}
\chVI{because \chX{as the distributions of both the ER and P3 states shift more
than those of the P1 and P2 states, cells in the ER and P3 states are more 
likely to have errors.  These errors (ER~$\leftrightarrow$~P1 and
P2~$\leftrightarrow$~P3) affect the MSB of the cell.}

\begin{figure}[h]
\centering
\includegraphics[trim=0 0 0 0,clip,width=\figscale\linewidth]
{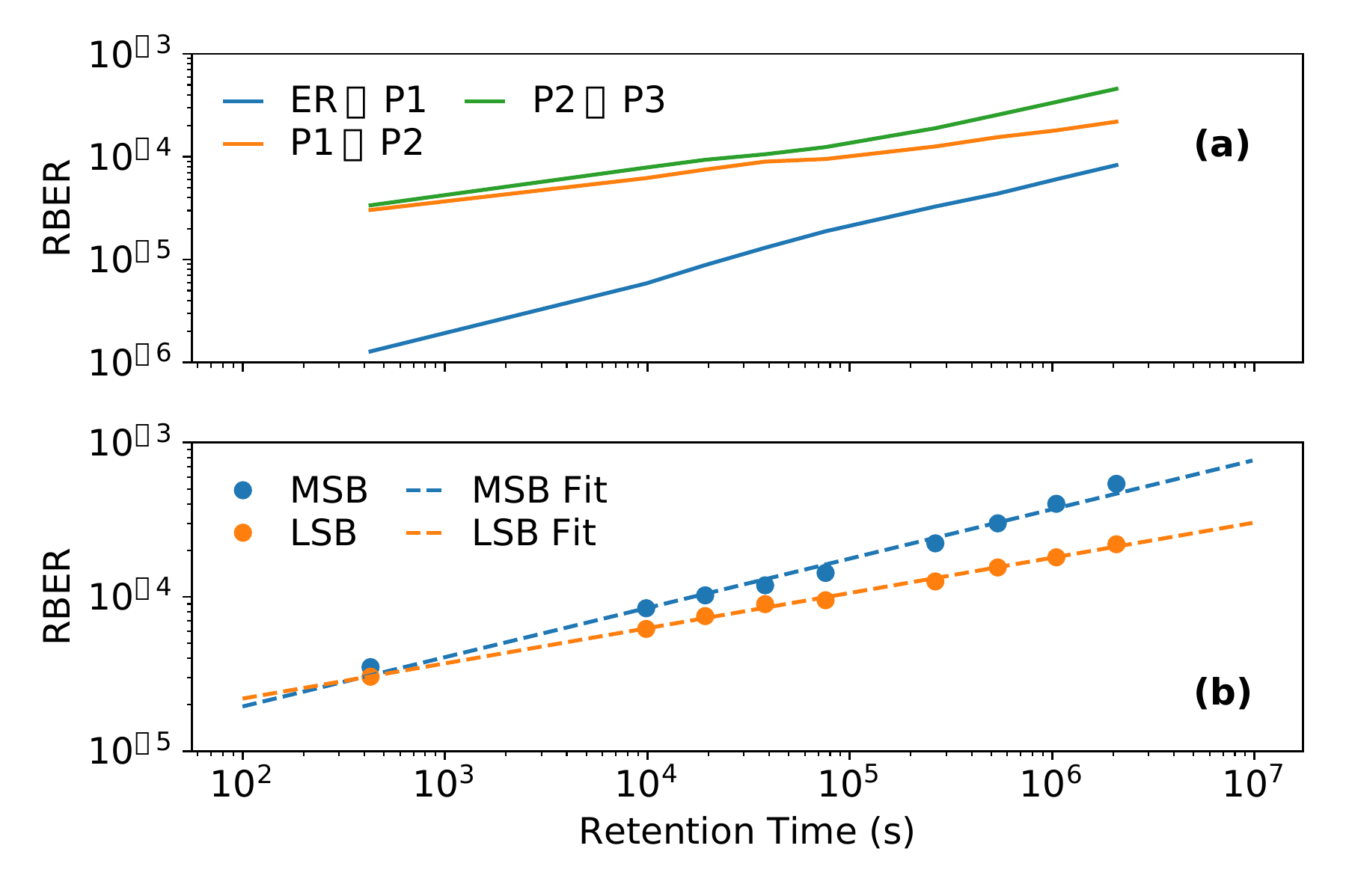}
\caption{RBER \chVI{vs.} retention time, broken down by \chVI{(a)~the state transition
of each flash cell, and (b)~MSB or LSB page}.}
\label{fig:retention-opterr}
\end{figure}

\textbf{Insights.} We compare the errors due to retention loss in 3D NAND \chX{flash memory to those} in planar
NAND flash memory, as reported in prior work~\cite{mielke.irps08, cai.hpca15,
cai.iccd12}. We find another major \chVI{difference} in 3D NAND \chVI{flash
memory} in terms of threshold
voltage \chX{distribution,} in addition to those
discussed in Section~\ref{sec:retention}. We find that \chX{the}
retention loss \chX{phenomenon we observe} in 3D NAND \chX{flash memory
(1)~}shifts the threshold voltage distributions of the P1, P2 and \chX{P3 states} lower, and 
\chX{(2)~}has little effect on the width of the distribution of each state. In contrast, the
retention loss \chX{phenomenon observed in} planar NAND flash memory 
\chX{(1)~}does \emph{not} shift the P1 and P2 state
distributions by much, and 
\chX{(2)~}increases the width of each state's distribution 
\emph{significantly}~\cite{cai.hpca15}. 
This indicates that a mechanism that
adjusts the optimal read reference voltage to the threshold voltage shift caused
by retention \chX{loss} can be more effective on 3D NAND \chIX{flash memory} than on
planar NAND \chIX{flash memory}, \chXI{because the distributions shift by a 
greater amount (indicating a greater need for voltage adjustment) \chXII{with}
a smaller amount of overlap between two threshold voltage distributions 
(reducing the number of read errors when the optimal read reference voltage is used)}. 
\chIX{We conclude that, due to \chX{the} early
retention loss phenomenon \chX{we observe in 3D NAND flash memory}, 
\chX{the threshold voltage of a flash cell}
changes quickly within several hours after programming, leading to
significant changes in RBER and optimal read reference \chX{voltage values}.}


\subsection{Read-Induced Errors}
\label{sec:read}

\chX{In this section, we} analyze how each type of read-induced error affects the
RBER and the threshold voltage distribution of 3D NAND flash
memory.

\subsubsection{Read Errors}
\label{sec:read:variation}

\chX{A read error is a type of read-induced error where two reads to a
flash cell may return different data values if the read reference voltage used
to read the cell is close to the cell's threshold voltage\chXI{~\cite{joe.ted11, 
compagnoni.edl09, ghetti.ted09}} (see Section~\ref{sec:background:errors}).}
\chXI{A read error adds uncertainty to the outcome of \emph{every} 
read operation performed by the SSD controller.  However, despite the potential
for widespread impact, read} errors are \chX{\emph{not}} well-studied by prior work. 

To quantify read errors, we use the data \chIX{we collected} in
Section~\ref{sec:retention}. \chX{For each cell, we see if the \emph{actual}} read outcome
\chX{(i.e., the bit value output by the flash controller after a read operation)
matches the \emph{expected}} read outcome \chX{(i.e., the
value that the read should have returned based on the current voltage of the
flash cell).  We determine the expected read outcome} by comparing $V_{ref}$
with $V_{th}$ (i.e., we expect to read 1 if $V_{th} < V_{ref}$,
\chX{because $V_{ref}$ is high enough that it should turn on the cell}).
\chX{We obtain $V_{th}$} by combining the outcomes of multiple reads when sweeping the read
reference voltage, thus \chXI{we expect that the combined output
eliminates the impact of read errors and is thus accurate.}
\chX{We say that a read error occurs if the actual read outcome and the
expected read outcome do \chXI{\emph{not}} match.}

\textbf{Observations.}
Figure~\ref{fig:read-error-offset} shows \chIX{how \chX{the} read
error rate changes \chX{as a function of the} \emph{read offset} (i.e., $V_{ref} - V_{th}$).}
We observe that\chIX{,} as the \chIX{absolute value of} \chX{the} read offset \chIX
{increases, \chX{the} read} error rate decreases exponentially. \chIX{This is likely
because \chX{when} $V_{ref}$ is closer to $V_{th}$ (i.e., \chXI{when} $V_{ref} - V_{th}$ has a
smaller absolute value), the amount of noise (i.e., voltage fluctuations) in
the sense amplifier increases exponentially\chXI{~\cite{compagnoni.edl09, ghetti.ted09}}.  The
\chXI{larger amount of} noise increases the
likelihood that the sense amplifier incorrectly detects whether the cell 
turns on, which leads to a \chXI{larger} probability that a read error occurs.}

\begin{figure}[h]
\centering
\includegraphics[trim=10 10 10 10,clip,width=.6\linewidth]
{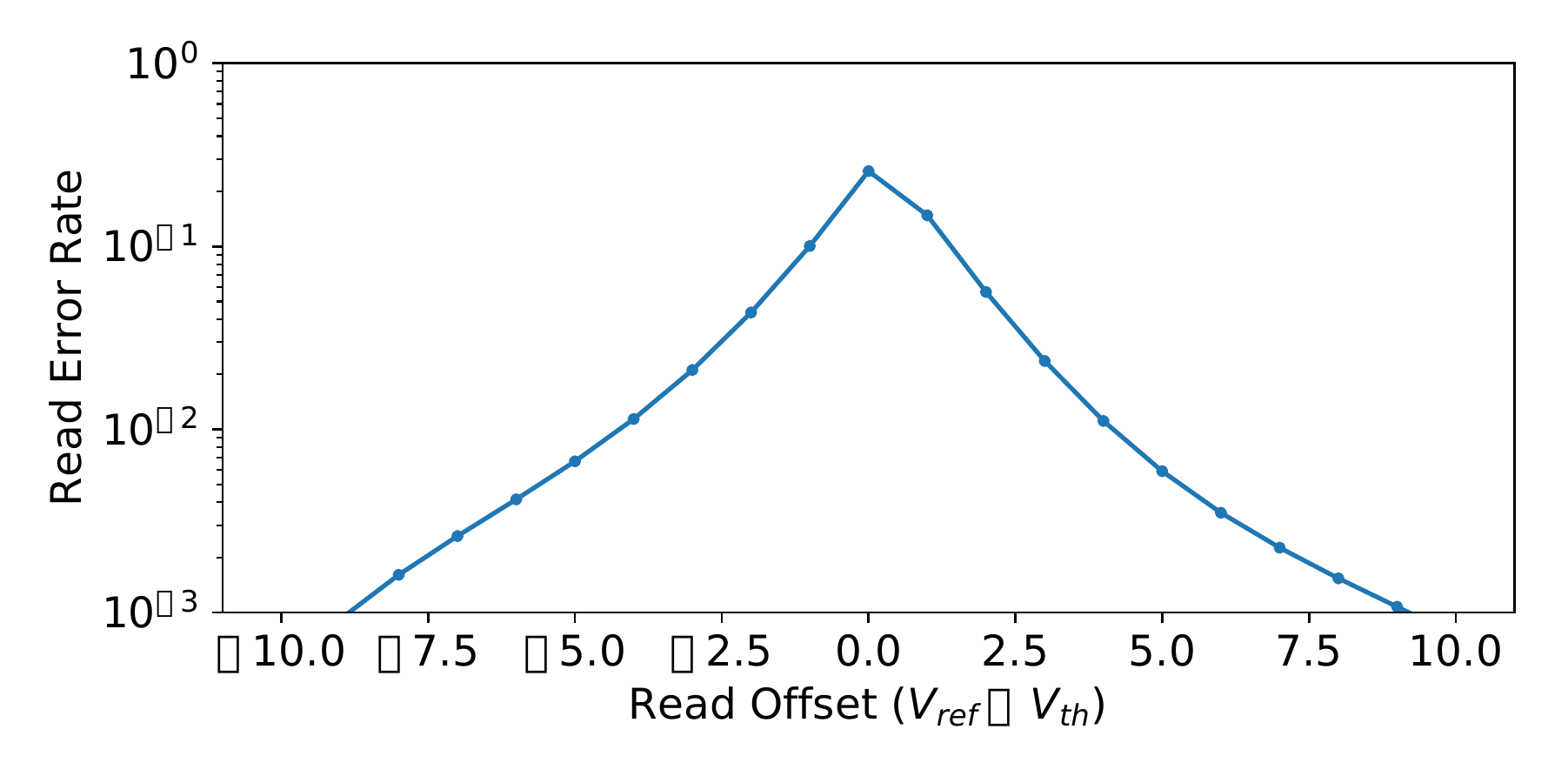}
\caption{Read error \chXI{rate} vs.\ read offset \chXI{($V_{ref} - V_{th}$)}.}
\label{fig:read-error-offset}
\end{figure}

Figure~\ref{fig:read-error-ratio} shows the correlation
between the read error rate and the total RBER in a flash page.
We observe that \chXI{the} read error rate is linearly correlated with the overall
RBER\@. This is because, when the RBER is high, the 
threshold voltage distributions of neighboring states overlap with
each other by a greater amount.  This causes a larger number of cells to
be close to the read reference voltage value, increasing the probability
that a read error occurs (see Figure~\ref{fig:read-error-offset}).

\begin{figure}[h]
\centering
\includegraphics[trim=0 10 0 10,clip,width=\figscale\linewidth]
{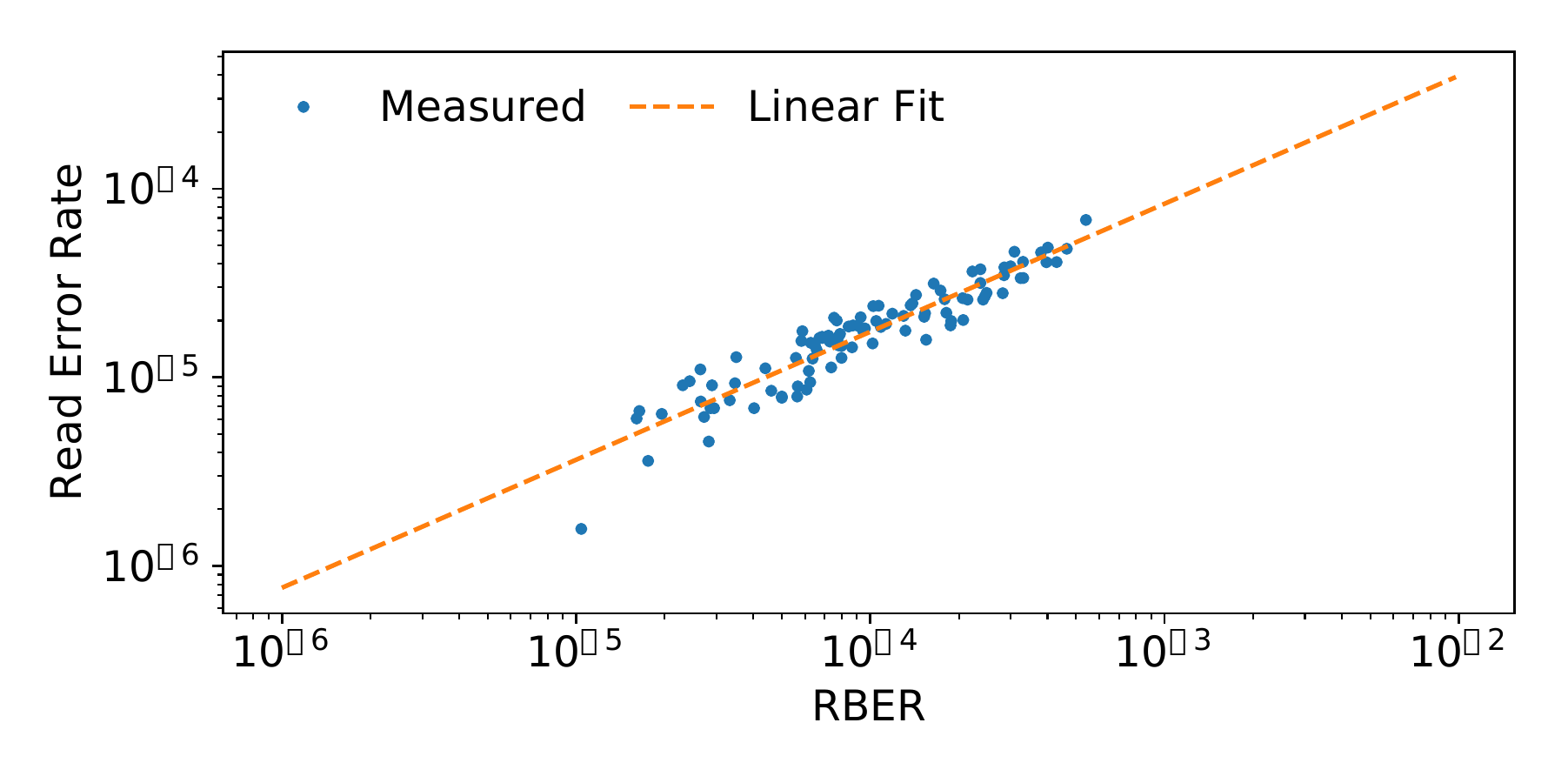}
\caption{\chXI{Relationship between the read error rate and} the RBER.}
\label{fig:read-error-ratio}
\end{figure}

\textbf{Insights.}
We are the first to discover and quantify the extent of read errors,
and to show the
correlation of these errors with the RBER and with the read reference voltage.
\chIX{We conclude that read errors are correlated with the read offset 
\chX{(i.e., $V_{ref} - V_{th}$)} and
the overall RBER of the flash page.}


\subsubsection{Read Disturb Errors}
\label{sec:read:disturb}

\chXI{Read disturb errors occur when a read operation to one page in a 
flash block may introduce errors in \emph{other, unread} pages in the
same block~\cite{cai.dsn15, papandreou.glsvlsi14}}
(see Section~\ref{sec:background:errors}). Read disturb
errors are caused by the high pass-through voltage applied \chXI{to cells
in the unread pages}.

To characterize read disturb errors, we first randomly select 11 flash blocks
and wear out each block to 10K P/E cycles \chVI{by repeatedly erasing and
programming pseudorandomly generated data} \chX{into each page of each 
block}. Then, we program
\chX{pseudorandomly-generated} data to \chX{each page of} each flash block. To minimize the
impact of other errors,
especially retention errors due to early retention loss, we wait until the
data has \chVI{a} 2-day retention time before inducing read disturb. This ensures
that, according to our results in Section~\ref{sec:retention}, \chVI{after
2 days,} retention loss \chVI{has slowed down and} can
only shift the threshold voltage by at most 1 voltage step during the \chVI
{relatively short} characterization process \chVI{($\sim$\SI{9}{\hour})}. To
induce read disturb in the flash block, we
repeatedly read from a wordline within the block for up to 900K times (i.e.,
up to 900K read disturbs). During this process, to characterize \chX{the} read disturb
effect, we obtain the RBER and threshold voltage distribution at ten different read
disturb counts from 0 to 900K.

\textbf{Observations.}
\chX{Figure~\ref{fig:read-mean-var}} shows how the
mean and standard deviation \chVI{of the threshold voltage distribution} change with
read disturb count.
\chVI{Each subfigure in the top row \chX{shows} the mean for a
different state; each subfigure in the bottom row \chX{shows} the standard
deviation for a different state. The blue dots shows the measured data; each
orange line shows a linear \chX{trend line} fitted to the measured data. The x-axis
shows the P/E cycle count; the y-axis shows the distribution parameters in
voltage steps.}
We make three observations from \chVI{this figure}. \chVI{First}, \chX{the} read
disturb effect increases the \chVI{mean threshold voltage of \chX{the} ER state
significantly, by $\sim$8 voltage steps after 900K read disturbs. In contrast,
the mean threshold \chX{voltages} of the programmed states change by \chX{only a small amount}
($<$3 voltage steps).} The increase in the \chVI{mean threshold voltage} is lower for
a higher $V_{th}$ state. 
\chX{This is because the impact of read disturb is correlated with the 
difference between the pass-through voltage (see
Section~\ref{sec:background:flash}) and the threshold voltage of a
cell.  When the difference is larger (i.e., when the threshold voltage of a
cell is lower), the impact of read disturb increases.}
\chXI{In fact, we observe that the threshold voltage distribution of the P3 state
even shifts to slightly \emph{lower} voltage values during the experiment,
because read disturb has little effect on cells in the P3~state, and the impact
of retention \chXII{loss dominates}.}
\chVI{Second}, the distribution width of each
state (i.e., standard deviation) \emph{decreases} slightly \chVI{as \chX{the} read disturb count
increases, by $<$0.2 voltage steps after 900K read disturbs}. 
\chVI{Third}, the change in \chVI{each} distribution
\chVI{parameter} can be
modeled as a linear function of \chXI{the} read disturb \chVI{count} \chX{(as shown by the orange
dotted lines)}. This shows that read disturb in 3D NAND flash memory follows a
similar linear trend as that observed in planar NAND flash memory by prior
work~\cite{cai.dsn15}.

\begin{figure}[h]
\centering
\includegraphics[trim=10 10 10 10,clip,width=\linewidth]
{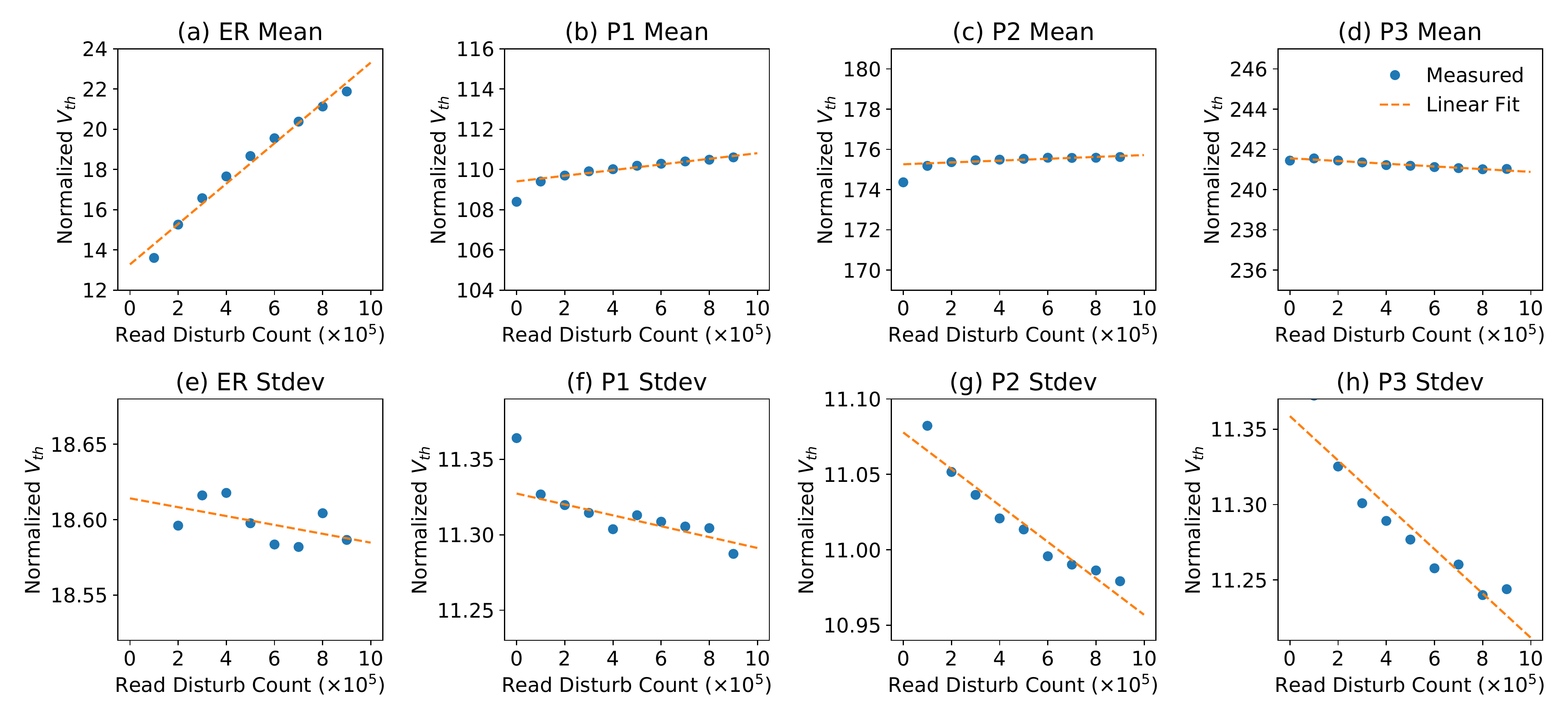}
\caption{\chVI{Mean and standard deviation of threshold voltage distribution
of each state}, \chX{vs.\ }read disturb count.}
\label{fig:read-mean-var}
\end{figure}

Figure~\ref{fig:read-opterr} plots how RBER increases \chX{with}
read disturb \chVI{count for a flash block \chXI{that has endured} 10K P/E cycles}. The top graph
breaks down the errors according to the change in cell state as a
result of the errors; the bottom graph breaks down
the errors into \chX{MSB and LSB} errors.  We make three
observations from Figure~\ref{fig:read-opterr}.
\chIX{First, ER$\leftrightarrow$P1 errors increase significantly with read
disturb count, whereas P1$\leftrightarrow$P2 and P2$\leftrightarrow$P3 errors
do not. This is because \chX{the} ER state threshold voltage distribution \chX{shifts}
significantly with read disturb count (see Figure~\ref{fig:read-mean-var}),
reducing
the threshold voltage difference between \chXI{the ER and P1 states}. Second, MSB errors
increase much faster than LSB errors with read disturb count because
ER$\leftrightarrow$P1 errors are a type of MSB \chX{error, and they increase}
significantly \chXI{with read disturb count}. Third, the increase in RBER with read disturb count follows a
linear trend (\chX{as shown by} the dotted line in Figure~\ref{fig:read-opterr}b), which
is similar to the observation \chX{made for} planar NAND flash memory by prior
work~\cite{cai.dsn15}.}

\begin{figure}[h]
\centering
\includegraphics[trim=0 10 0 10,clip,width=\figscale\linewidth]
{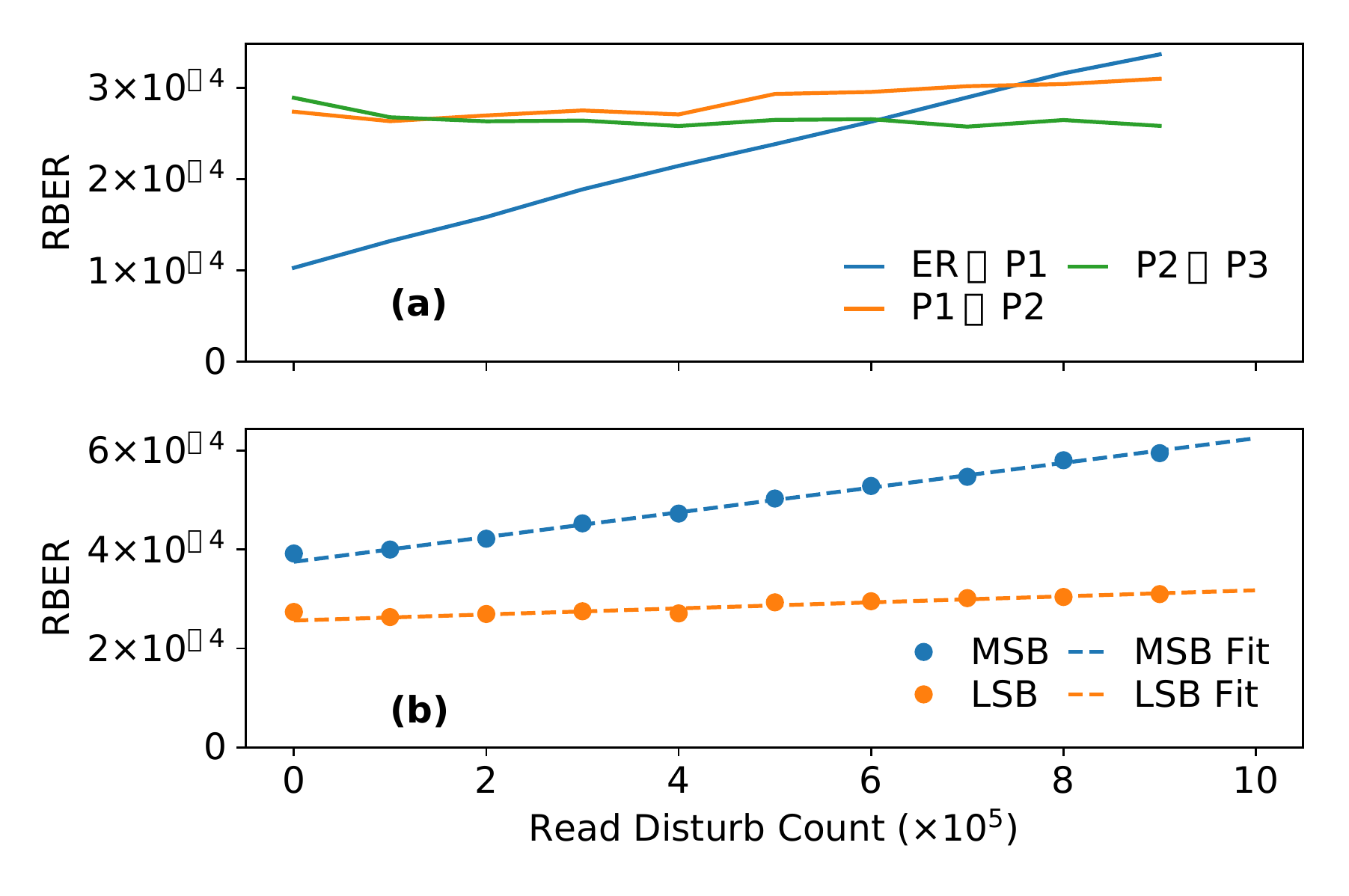}
\caption{RBER vs.\ read disturb count, broken down by (a)~the state
transition of each flash cell, and (b)~MSB or LSB page.}
\label{fig:read-opterr}
\end{figure}

Figure~\ref{fig:read-optvrefs} shows how the optimal read reference voltages
change \chIX{with} read disturb \chIX{count}.
\chX{The three} subfigures show the optimal voltages for $V_a$, $V_b$, and $V_c$. We
make two observations from this figure. \chIX{First, the optimal voltages
for $V_b$ and $V_c$ \chXI{change by relatively little} as \chXI{the} read disturb
count increases ($<$3 voltage steps after 900K read disturbs), whereas the
optimal $V_a$ \chX{changes more with the} read disturb count.
\chXI{This is because read disturb causes the threshold voltage distributions of
lower-voltage states to change by a greater amount, which requires the
read reference voltages separating the lower-voltage states (e.g., $V_a$) to
change more.}
Second, the increase in the optimal $V_a$ follows a linear trend with read
disturb count, because the ER state threshold voltage distribution shifts
linearly (as we see from Figure~\ref{fig:read-mean-var}).}

\begin{figure}[h]
\centering
\includegraphics[trim=10 10 10 10,clip,width=.8\linewidth]{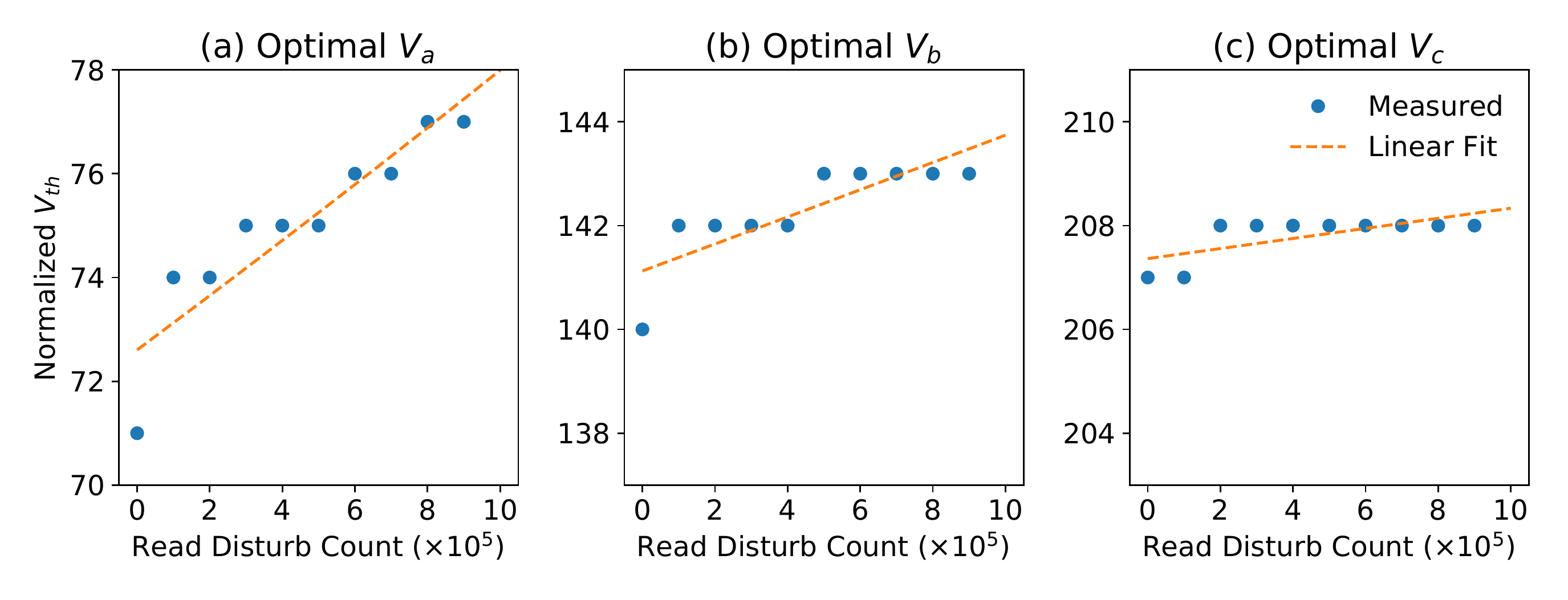}
\caption{Optimal read reference voltages vs.\ read disturb \chXII{count}.}
\label{fig:read-optvrefs}
\end{figure}

\textbf{Insights.} We compare the read disturb effect \chX{that we observe in
3D NAND flash memory to that observed in
planar NAND flash memory} \chXI{by prior} work~\cite{cai.dsn15}. We make the
observation that, although RBER \chX{increases} linearly \chIX{with read disturb count} in
both 3D NAND and planar NAND \chX{flash memory}, the slope of \chXI{the} increase (i.e., the sensitivity \chX{of \chXI{the} RBER} to
read disturb) at 10K P/E cycles is 96.7\% \chXI{\emph{lower}} in 3D NAND \chX{flash memory} than \chX{that} in planar
NAND \chX{flash memory}~\cite{cai.dsn15}. We believe that this difference in \chIX{the} sensitivity to
read disturb \chIX{effect} is due to the use of a larger process technology
\chIX{node} \chX{(\SIrange{30}{40}{\nano\meter})}
in \chX{current} 3D NAND \chIX{flash memory}. 
\chX{The comparable planar NAND flash memory results} from prior work \chX{are
collected on} \SIrange{20}{24}{\nano\meter} planar NAND \chIX{flash memory} \chX{devices~\cite{cai.dsn15}}. 
\chX{We}
expect the \chIX{read disturb effect} in 3D NAND \chX{flash memory} to increase in the future as
\chX{the process technology node size shrinks}.
\chIX{We conclude that the 96.7\% reduction in the read disturb effect
we observe in 3D NAND flash memory compared to planar NAND flash memory is
mainly caused by the difference in manufacturing process technology \chX{nodes}
\chXI{of the two types of NAND flash memories}.}


\subsection{Layer-to-Layer Process Variation}
\label{sec:appendix:variation}

In this \chX{section}, we present \chX{new results and analyses} of \chXI{the layer-to-layer}
process variation \chXI{phenomenon} in
3D NAND \chX{flash memory,} in addition to the key findings 
\chX{we already presented} in Section~\ref{sec:variation}. We use
the same methodology as \chIX{we describe} in Section~\ref{sec:variation}.

\chIX{Figure~\ref{fig:variation-wlmean-var} shows how the}
threshold voltage distribution mean and standard deviation of each state
\chIX{changes with layer number}, for a flash block \chX{that has endured} 10K
P/E cycles. \chIX{Each subfigure in the top row \chX{shows} the mean for a
different state; each subfigure in the bottom row \chX{shows} the standard
deviation for a different state.} We make two observations from \chIX{this
figure}.
First, the ER state \chIX{threshold voltage} \chIX{increases} by as much as
25 voltage steps \chIX{as \chX{the layer number changes}}, while the \chIX{mean
threshold voltages} of the other three states do not \chIX{vary} by much.
\chX{This is because \chXI{the threshold voltage of a cell in ER~state
is} set after
an erase \chXII{operation, and the value it is set to} is a function of
manufacturing process variation and of wearout.
In contrast, \chXI{the threshold voltage of} a cell in one of the other states
(P1, P2, or \chXI{P3) is} set to a \chXII{\emph{fixed}} target voltage value \chXII{\emph{regardless}} of
process variation~\cite{mielke.irps08, bez.procieee03, suh.jssc95, wang.ics14} (see
Section~\ref{sec:background:flash}).}
\chX{Since only the voltage of the ER~state is affected by layer-to-layer
process variation, only one of the read reference voltages, $V_a$, changes
with the layer number, \chXI{as we already observed} in
Figure~\ref{fig:variation-wloptvrefs}.}
\chIX{Second}, the distribution \chIX{widths of ER and P1 states} \chIX{(i.e.,
\chX{their standard
deviations}) increase} in the \chXI{top layers}, and decrease in the
bottom \chXI{layers}. \chIX{This pattern is similar to the pattern of how \chX{the}
RBER changes
with layer number, which we show in Figure~\ref{fig:variation-wlopterr}
(Section~\ref{sec:variation}).
\chX{A} wider threshold voltage distribution increases the overlap of
neighboring distributions, leading to more errors \chX{in the top layer}.} However, the
distribution widths of the P2 and P3 states \chXI{mainly} \chIX
{decrease as layer number increases.} \chIX{Unfortunately, we are unable to
\chX{completely} explain why mean threshold voltage and distribution width change differently
with layer number \chXI{for different states} because we do not have \chX
{exact} circuit-level information about
layer-to-layer process variation.}

\begin{figure}[h]
\centering
\includegraphics[trim=10 10 10 10,clip,width=\linewidth]
{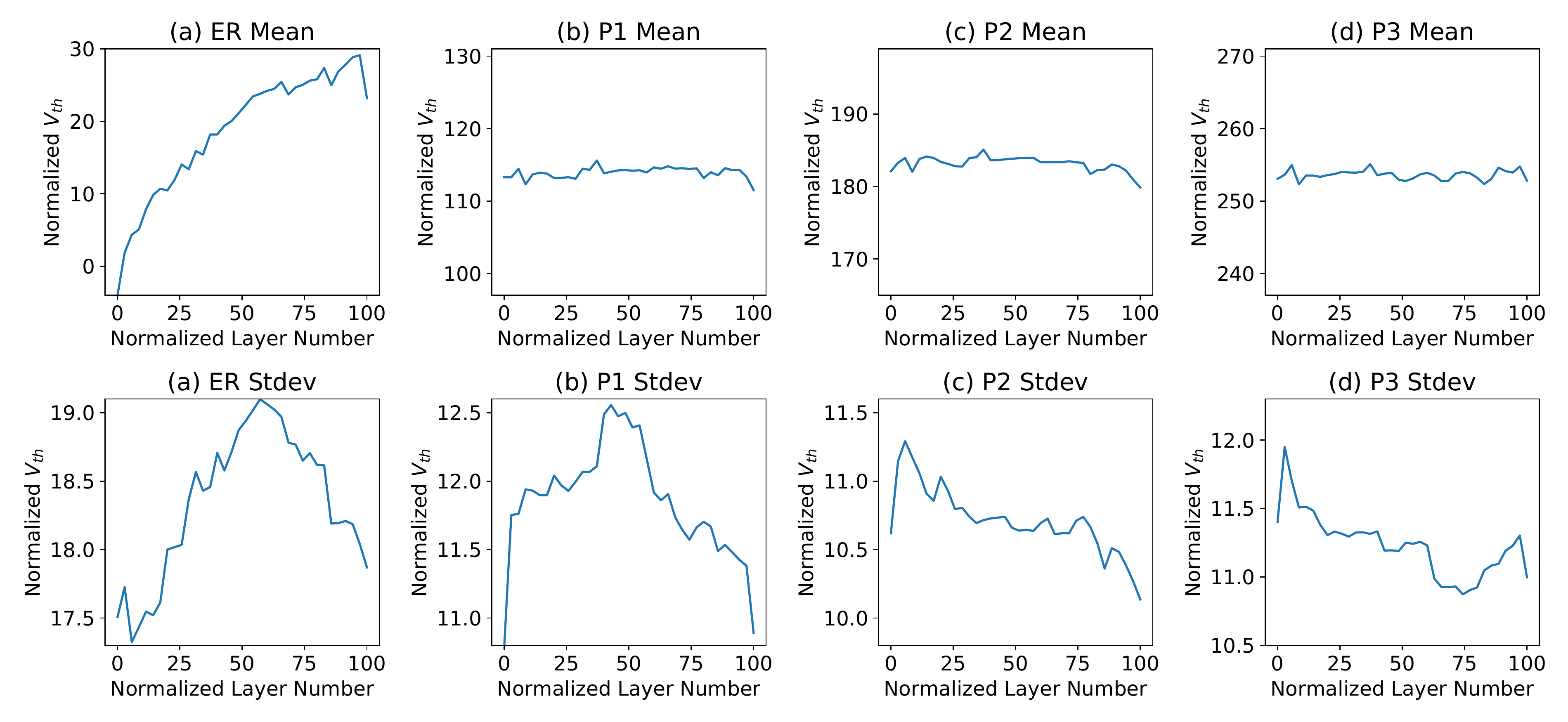}
\caption{\chIX{Mean and standard deviation of our Gaussian threshold voltage
distribution model of each state, versus layer number.}}
\label{fig:variation-wlmean-var}
\end{figure}

\chIX{We conclude that layer-to-layer process
variation significantly impacts the threshold voltage distribution and leads
to large variations in RBER and optimal read reference voltages across
layers.}

\subsection{Bitline-to-Bitline Process Variation}
\label{sec:bitline-variation}

We perform \chX{an analysis} \chXI{of} the variation of RBER and threshold
voltage distribution along the y-axis (i.e., across groups of bitlines) for a
flash block \chXI{that has endured} 10K P/E cycles.  \chX{We use a similar
methodology
to our layer-to-layer process variation experiments  (see 
Section~\ref{sec:variation}).}

\chIX{Figure}~\ref{fig:variation-blmean-var} shows how the
threshold voltage distribution mean and standard deviation of each state
changes with layer number, for a flash block \chX{that has endured} 10K
P/E cycles. Each subfigure in the top row \chX{shows} the mean for a
different state; each subfigure in the bottom row \chX{shows} the standard
deviation for a different state. Note that we normalize the number of bitlines
from 0 to 100, by multiplying the actual bitline number
with a constant, to maintain the anonymity of the chip vendors. We
make two observations from this figure.  First, the variations in mean
threshold voltage and the distribution width (i.e., standard deviation) are
much smaller in this figure compared to that observed in
Figure~\ref{fig:variation-wlmean-var} \chX{for layer-to-layer variation} (Appendix~\ref{sec:appendix:variation}).
This indicates that bitline-to-bitline process variation is much smaller
compared to layer-to-layer process variation in 3D NAND flash memory. Second,
\chX{we observe that the pattern of the mean threshold voltage repeats
periodically, for every 25 bitlines.  We believe that this indicates a
repetitive architecture in the way that the 3D NAND flash memory chip is
organized (for example, each block may be made up of four arrays of
flash cells that \chXII{are connected together}).  Unfortunately, we cannot
completely explain this behavior without access to circuit-level design
information that is proprietary to NAND flash memory vendors.}

\begin{figure}[h]
\centering
\includegraphics[trim=10 10 10 10,clip,width=\linewidth]
{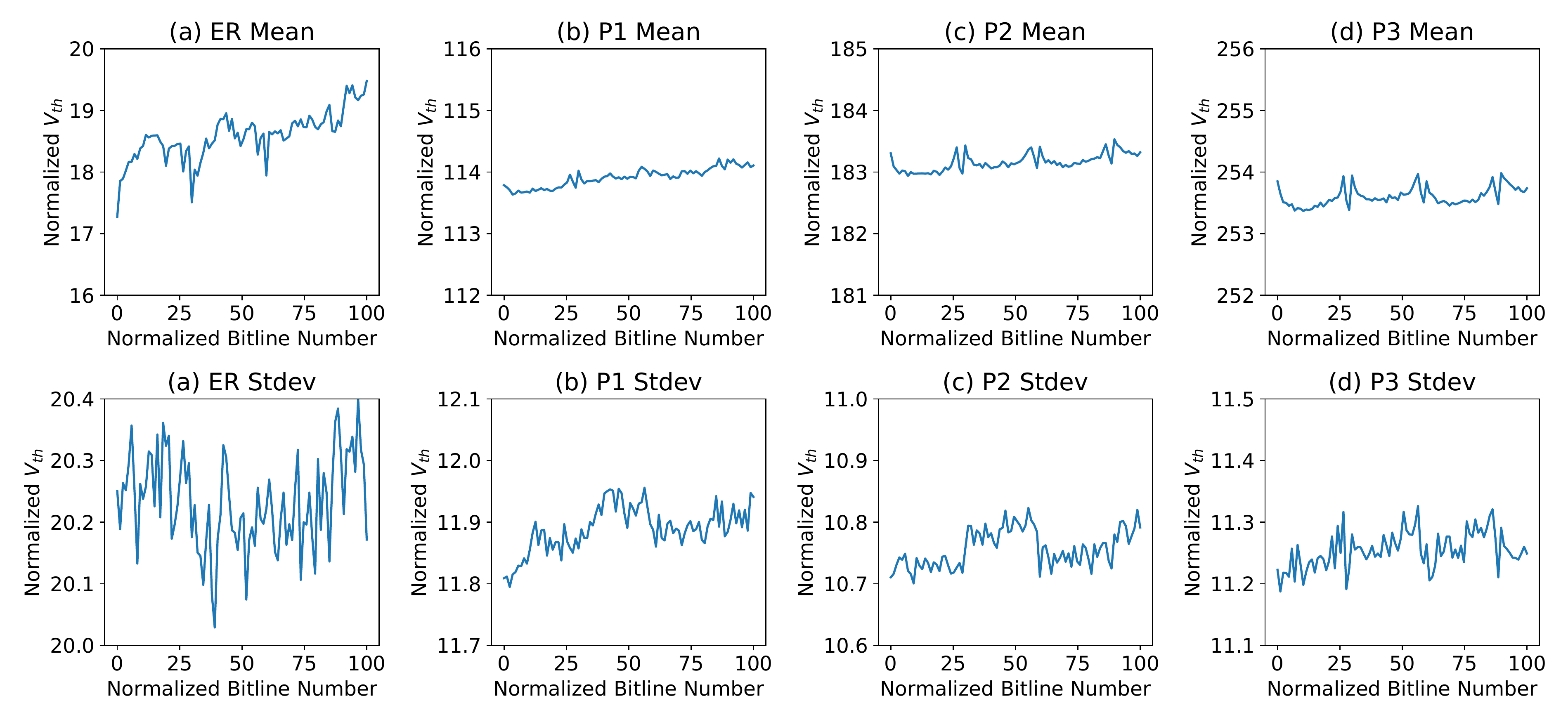}
\caption{\chXI{Mean and standard deviation of our Gaussian threshold voltage distribution model of each state,
versus bitline number.}}
\label{fig:variation-blmean-var}
\end{figure}

\chIX{\chX{Figures}~\ref{fig:variation-blopterr} and \ref{fig:variation-bloptvrefs} show
how \chX{the RBER and} optimal read reference voltages change with bitline number,
for a flash block \chXI{that has endured} 10K P/E cycles. We observe that
neither RBER nor the
optimal read reference voltages change by much across bitlines. This indicates
that the \chX{changes that} we observe in Figure~\ref{fig:variation-blmean-var} may not be
significant enough to \chX{lead to variation in the reliability of different} bitlines. We conclude that
bitline-to-bitline process variation is much smaller than layer-to-layer
process variation in 3D NAND flash memory.}

\begin{figure}[h]
\centering
\includegraphics[trim=10 10 10 10,clip,width=\figscale\linewidth]
{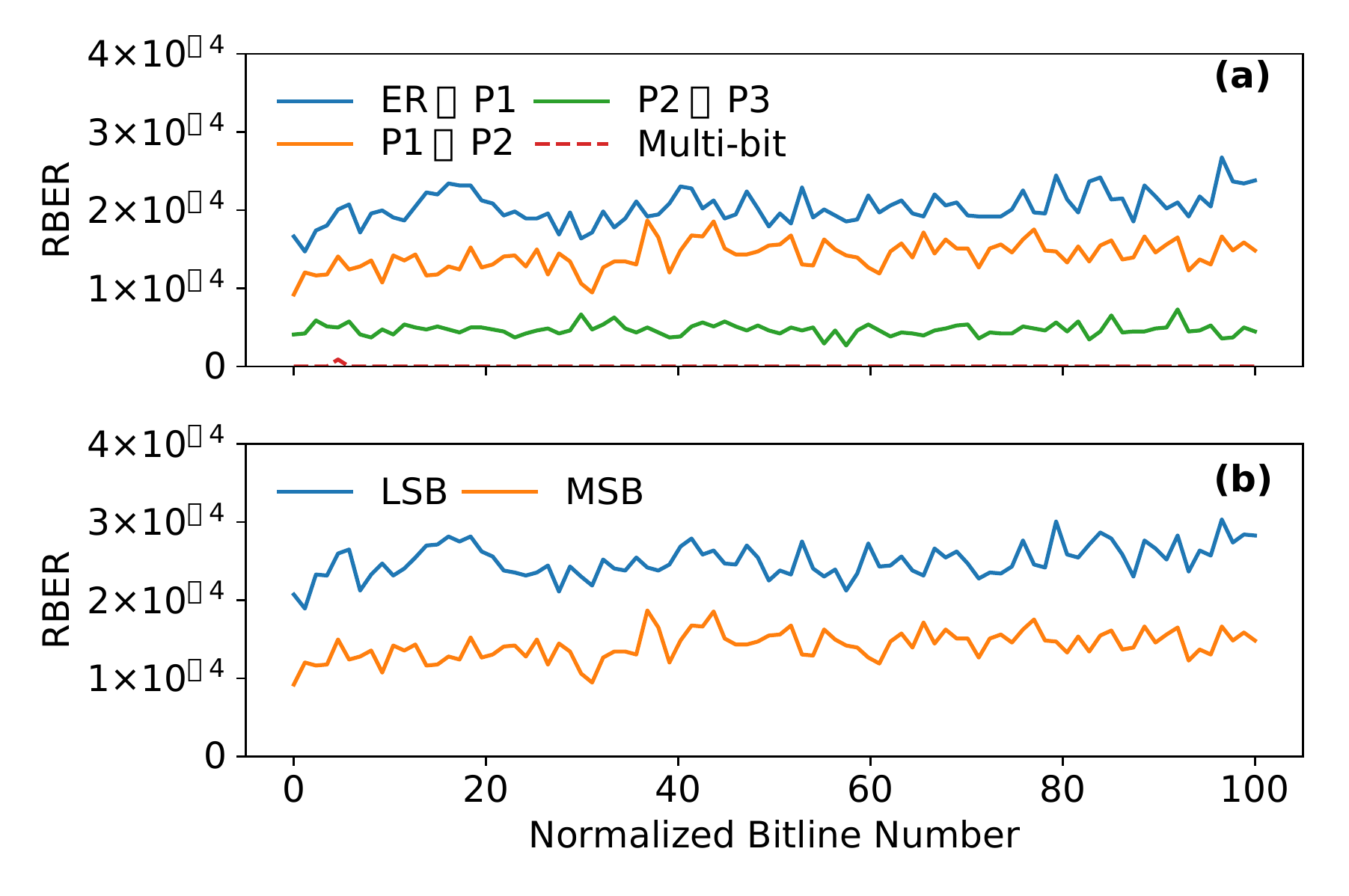}
\caption{\chXI{RBER vs. bitline number, broken down by (a)~the state transition of
each flash cell, and (b)~MSB or LSB page.}}
\label{fig:variation-blopterr}
\end{figure}

\begin{figure}[h]
\centering
\includegraphics[trim=10 10 10 10,clip,width=.8\linewidth]
{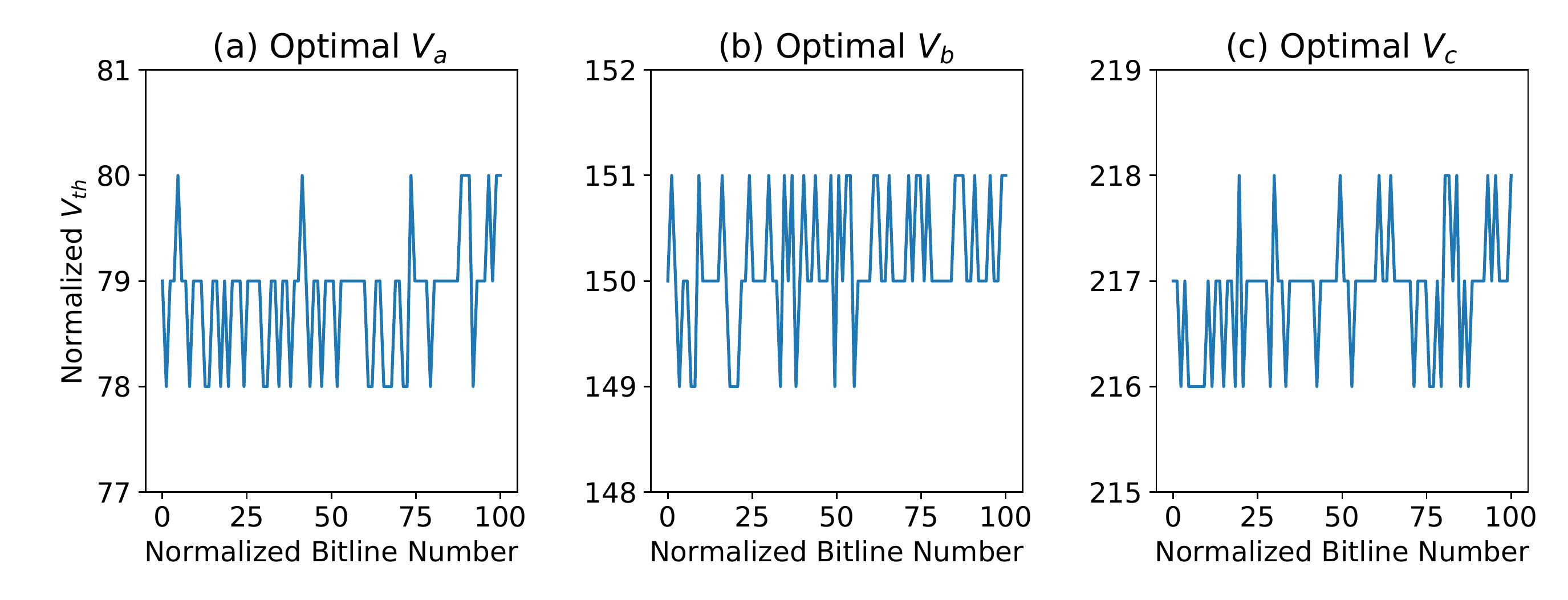}
\caption{\chXI{Optimal read reference voltages vs. bitline number.}}
\label{fig:variation-bloptvrefs}
\end{figure}

\end{document}